\documentclass[12pt]{article}
\usepackage{a4wide,latexsym,graphicx,epsfig,psfrag,here}
\usepackage{amsmath} 
\usepackage{longtable}
\usepackage{amssymb} 
\usepackage{bbm}
\usepackage{cite}  

\def\slashchar#1{\setbox0=\hbox{$#1$}\dimen0=\wd0%
\setbox1=\hbox{/}\dimen1=\wd1%
\ifdim\dimen0>\dimen1%
\rlap{\hbox to
\dimen0{\hfil/\hfil}}#1\else
\rlap{\hbox to \dimen1{\hfil$#1$\hfil}}/\fi}
\newcommand{\no}{\nonumber\\}
\newcommand{\non}{\nonumber}
\newcommand{\co}{\; ,}
\newcommand{\fs}{\; .}
\newcommand{\extraline}{$\\ &$}

\newcommand{\LNCinf}{{\cal L}_{\infty}}

\newcommand{\FV}{F_V}
\newcommand{\GV}{G_V}
\newcommand{\MV}{M_V}
\newcommand{\FA}{F_A}
\newcommand{\MA}{M_A}
\newcommand{\cd}{c_d}
\newcommand{\cm}{c_m}
\newcommand{\MS}{M_S}
\newcommand{\dm}{d_m}
\newcommand{\MP}{M_P}

\newcommand{\rS}{r_S}
\newcommand{\rV}{r_V}

\newcommand{\Cs}{C_{28}^{\cal R}}
\newcommand{\Css}{C_{32}^{\cal R}}

\newcommand{\lVA}{\overline{\lambda}^{VA}}
\newcommand{\lb}[1]{\overline{\lambda}^{#1}}

\newcommand{\bea}{\begin{eqnarray}}
\newcommand{\eea}{\end{eqnarray}}
\newcommand{\beq}{\begin{equation}}
\newcommand{\eeq}{\end{equation}}

\newcommand{\cO}{{\cal O}}
\newcommand{\beqa}{\begin{eqnarray}}
\newcommand{\eeqa}{\end{eqnarray}}
\newcommand{\vp}{\varphi}

\newcommand{\cL}{{\cal L}}
\newcommand{\dg}{\dagger}
\newcommand{\nn}{\nonumber \\}
\newcommand{\lgl}{\langle}
\newcommand{\rgl}{\rangle}

\newcommand{\nc}{N_C}

\begin{document}
\parskip=3pt plus 1pt

\begin{titlepage}
\begin{flushright} 
{
CALTECH MAP-319 \\
CPT-P05-2006 \\
FTUV/06-0319 \\
IFIC/06-02\\
UWThPh-2005-19
}
\end{flushright}
\vskip 0cm

\setcounter{footnote}{0}
\renewcommand{\thefootnote}{\fnsymbol{footnote}}

\begin{center} 
{\LARGE \bf Towards a consistent estimate of} \\[15pt]
{\LARGE \bf the chiral low-energy
constants\footnote{Work supported in part by HPRN-CT2002-00311 (EURIDICE) and by
Acciones Integradas, Project No. 19/2003 (Austria),
HU2002-0044 (MCYT, Spain).}} 
\\[30pt] 

{\normalsize \bf  V. Cirigliano$^{1}$, 
G. Ecker$^{2}$, 
M. Eidem\"uller$^{3}$,} \\[10pt]

{\normalsize \bf 
R. Kaiser$^{4}$, 
A. Pich$^{3}$ and 
J. Portol\'es$^{3}$} 

\vspace{1cm}
${}^{1)}$ California Institute of Technology \\ 
Pasadena, California 91125, USA  \\[10pt] 

${}^{2)}$ Institut f\"ur Theoretische Physik, Universit\"at 
Wien\\ Boltzmanngasse 5, A-1090 Vienna, Austria \\[10pt] 

${}^{3)}$ Departament de F\'{\i}sica Te\`orica, IFIC, CSIC --- 
Universitat de Val\`encia \\ 
Edifici d'Instituts de Paterna, Apt. Correus 22085, E-46071 
Val\`encia, Spain \\[10pt]

${}^{4)}$ Centre de Physique Th\'{e}orique\footnote{Unit\'{e} mixte de
recherche (UMR 6207) du CNRS et des Universit\'{e}s Aix-Marseille I,
Aix-Marseille II, et du Sud Toulon-Var; laboratoire affili\'{e} \`{a}
la FRUMAM (FR 2291).}, CNRS-Luminy, Case 907,\\ F-13288
Marseille Cedex 9, France
\\[0pt]
\end{center} 

\setcounter{footnote}{0}
\renewcommand{\thefootnote}{\arabic{footnote}} 

\vfill

\begin{abstract}
\noindent 
Guided by the large-$N_C$ limit of QCD, we construct the most general
chiral resonance Lagrangian that can generate chiral low-energy
constants up to ${\cal O}(p^6)$. By integrating out the resonance
fields, the low-energy constants are parametrized 
in terms of resonance masses and couplings. Information on those
couplings and on the low-energy constants can be extracted by 
analysing QCD Green functions of currents both for large and small
momenta. The chiral resonance theory generates 
Green functions that interpolate between QCD and
chiral perturbation theory. As specific examples we consider
the $\langle VA\,P \rangle$ and $\langle S\,P\,P \rangle$ Green
functions.    
\end{abstract}

PACS~: 11.15.Pg, 12.38.-t, 12.39.Fe \\ \hspace*{0.5cm}
Keywords~: Chiral Lagrangians, $1/N_C$ expansion, QCD.
\vfill

\end{titlepage} 
\newpage

\tableofcontents

\newpage

\section{Introduction} 
\label{sec:intro}
\renewcommand{\theequation}{\arabic{section}.\arabic{equation}}
\setcounter{equation}{0}

Chiral perturbation theory~\cite{Weinberg:1978kz,Gasser:1983yg,
Gasser:1984gg} ($\chi$PT) is the effective theory of QCD 
at small external momenta. In the low-energy regime, the leading
singularities of QCD Green functions of quark currents are generated
by the octet of light pseudoscalar mesons ($\pi, K, \eta$), the
explicit degrees of freedom in the effective theory.   
$\chi$PT is constructed by exploiting the chiral symmetry of QCD in the
limit of massless light quarks (we consider here the three-flavour
case with $m_u=m_d=m_s = 0$), its spontaneous
symmetry breaking according to the pattern $SU(3)_L \times SU(3)_R
\rightarrow SU(3)_V$  and its explicit breaking due to nonvanishing 
quark masses. 

The structure of the effective Lagrangian is determined by chiral symmetry
and the discrete symmetries of QCD. It is organized as an expansion
in derivatives of the Goldstone fields  and in powers of the light quark
masses ($m_q$). In the standard scenario 
the two expansions are related ($m_q \sim \cO(M^2) \sim \cO(p^2)$) 
and the mesonic effective chiral Lagrangian takes the form
\begin{equation}
 {\cal L}_{\rm eff} = \sum_{n\ge 1} \ {\cal L}_{2n}^{\rm \chi PT} \ ,  
\qquad  \qquad \qquad 
{\cal L}_{2n}^{\rm \chi PT} \sim   \cO(p^{2n}) \,. 
\label{eq:intro1}
\end{equation}
The intrinsic scale $\Lambda_\chi$ of this expansion is set by the 
lightest mesonic non-Goldstone states ($\Lambda_{\chi} \sim M_V \sim
1$ GeV). 
The effective Lagrangian depends on a number of low-energy constants
(LECs), which are not determined by symmetry considerations, encoding 
the underlying QCD dynamics.  
Applications of current interest require working to NNLO ($\cO(p^6)$)
\cite{Bijnens:2004pk}.
Since ${\cal L}_{6}^{\rm \chi PT}$ involves 90
LECs for three light flavours 
\cite{Fearing:1994ga,Bijnens:1999sh,Bijnens:1999hw}, 
a theoretical assessment of 
the size of those couplings is mandatory for phenomenology.
\par
Determining the LECs from QCD is a difficult nonperturbative problem.
However, both empirical evidence and theoretical arguments suggest
that the most important contributions to the LECs in the strong chiral 
Lagrangian come from physics at the scale $\Lambda_\chi$, i.e. the 
physics of low-lying resonances.
In general, the LECs can be characterized as coefficients of the Taylor
expansion of QCD correlators around zero momentum, once the
non-expandable singularities due to Goldstone modes have been
removed.  
If the appropriate correlators are order parameters of spontaneous 
chiral symmetry breaking (vanishing to all orders in QCD perturbation
theory), they fall off at high momenta with an inverse power 
determined by the operator product expansion (OPE).  Therefore, the
LECs are expected to be sensitive to the intermediate-momentum
region where the low-lying hadronic resonances turn the polynomial
behaviour of the correlator into an inverse-power behaviour, as required 
by the QCD short-distance constraints. 
\par
The natural framework to incorporate systematically the above
considerations is provided by the $1/N_C$ expansion of
QCD~\cite{'tHooft:1973jz,Witten:1979kh}.  Earlier
studies of resonance saturation for $\cO(p^4)$ couplings
~\cite{Ecker:1988te,Ecker:1989yg,donoghue89} can be embedded in this
framework~\cite{Peris:1998nj,Pich:2002xy,deRafael:2002tj}. A
number of more recent works has already applied large-$N_C$ techniques 
to estimate
subsets of the $\cO(p^6)$ LECs~\cite{Moussallam:1997xx,
Knecht:2001xc,Ruiz-Femenia:2003hm,
Bijnens:2003rc,Cirigliano:2004ue,Cirigliano:2005xn}.
A few studies at next-to-leading order in the $1/N_C$
expansion have also been performed \cite{Rosell:2004mn}.
In this work we aim to study systematically resonance
contributions to the full ${\cal L}_6^{\rm \chi PT}$ to leading order 
in $1/N_C$. We first recall the salient features common to most 
procedures based on the $1/N_C$ expansion, describing along the way 
several approximations used in this work.  We then comment on the
specific aspects that characterize the interaction between Goldstone 
modes and resonance fields. 

\vspace{1cm} 
\noindent
\underline{\bf Matching at large $N_C$}
\par 
In principle, the matching of QCD with $\chi$PT is straightforward in 
the $N_C \rightarrow \infty$ limit (QCD$_{\infty}$).  
To leading order in the $1/N_C$ expansion, any correlator of quark
bilinears $\langle J_1... J_n \rangle$ is given by a sum of tree-level 
diagrams involving interactions of an infinite tower of
narrow meson states with appropriate quantum numbers
\footnote{Crossing and unitarity imply that the correlators 
can be obtained by tree-level insertions of an appropriate 
hadronic Lagrangian~\cite{Witten:1979kh}.}.
Hadron masses and couplings are adjusted so as to satisfy chiral
Ward identities and to match the QCD asymptotic behaviour at large
momenta.  The Taylor expansion of the correlator around vanishing
momenta, after removing the Goldstone poles, allows one to read off
the corresponding chiral LECs. 
\par
In practice, since a solution to QCD$_\infty$ is not available,
one has to make a set of approximations in implementing the matching
outlined above.  The main approximations involve truncating the
hadronic spectrum to a finite number of states and choosing the
appropriate set of short-distance constraints to determine the
hadronic parameters.  In this work we truncate the spectrum to the
lowest-lying resonance multiplets with given $J^{PC}$. We consider
explicitly the channels V($1^{--}$), A ($1^{++}$), S ($0^{++}$),
P($0^{-+}$).  This choice has to be considered a working hypothesis that
can be extended if needed. It is based on the observations that (i)
the low-lying hadronic spectrum has the largest impact on the LECs;
(ii) the QCD asymptotic behaviour sets in at energies $E \sim 1.5$ GeV
(for correlators that are order parameters of spontaneous chiral
symmetry breaking the fall-off is well reproduced by a few hadronic
states \cite{Peris:1998nj}); (iii) retaining only lowest-lying states
 leads to a successful phenomenology for $\cO(p^4)$
couplings~\cite{Ecker:1988te}.  
\par
In this work we disregard the lightest P($0^{-+}$) singlet because
the $\eta^\prime$ meson plays a special role in the large-$N_C$
counting \cite{Gasser:1984gg,Peris:1994dh,Leutwyler:1997yr}. 
The contributions of $\eta^\prime$
exchange to the LECs of $\cO(p^6)$ are worked out in an accompanying
paper \cite{RK}. 
\par
Concerning the short-distance constraints, the minimal requirement is
that the Green functions obey the asymptotic behaviour dictated by QCD
to leading power in the inverse large momenta. In addition, although 
not derived from first
principles, it is heuristically inferred~\cite{Lepage:1979zb} and 
phenomenologically supported that form factors of QCD currents should 
vanish smoothly at large momenta.
It should be kept in mind, however, that there are intrinsic
limitations of the matching program when only a finite number of
resonance multiplets are included (e.g., 
Refs.~\cite{Bijnens:2003rc,Cirigliano:2005xn}). 

\noindent
\underline{\bf Lagrangian formulation}
\par
The matching strategy outlined above can be pursued within or without
a Lagrangian description of the chiral invariant Goldstone-resonance
interactions.
One useful aspect of the Lagrangian formulation is that, within a
given set of assumptions on the large-$N_C$ spectrum, it provides a
common framework to study many observables or correlators at the same
time as opposed to constructing different hadronic ans\"atze on a 
case-by-case basis. Another important feature of the Lagrangian
approach is 
that the resonance fields can be integrated out at the level of the
generating functional as opposed to expanding resonance propagators in 
individual Green functions.  This allows one to obtain all 
resonance contributions to the $\cO(p^6)$ chiral LECs once and for
all, even before specific values of the resonance couplings have been
determined by the short-distance analysis.
\par
In the strict large-$N_C$ limit the hadronic Lagrangian $ \LNCinf$ has 
to be used at tree level only. Its couplings are
determined in such a way that the corresponding Green functions 
reproduce the asymptotic behaviour of QCD$_{\infty}$.  
Although the construction of $ \LNCinf$ is a formidable task,
progress can be made if one considers a limited set of hadronic states
and if one focuses on the subset of $ \LNCinf$ that, upon
integrating out heavy fields at tree level, contributes to the
chiral Lagrangian up to a given chiral order only. This limits both
the number of resonance fields and the chiral order of the respective
terms in the resonance Lagrangian.
The first systematic studies in this direction go back to
Refs.~\cite{Ecker:1988te,Ecker:1989yg} where the most general 
$ \LNCinf$ contributing to ${\cal L}_4^{\rm \chi PT}$ was constructed
and the equivalence of different representations for spin-one fields was
demonstrated, once the Green functions generated by $ \LNCinf$
are forced to satisfy the correct asymptotic behaviour dictated by QCD.  
\par
In this paper we perform the first steps towards a systematic 
study of resonance contributions to ${\cal L}_{6}^{\rm \chi PT}$ 
within the truncated large-$N_C$ matching described above.  
The program involves several tasks: 
\begin{enumerate}
\item[i.] Construct the most general chiral invariant Lagrangian 
describing the interactions of Goldstone modes with V, A, S, P meson 
resonance fields
that
contributes to ${\cal L}_6^{\rm \chi PT}$ after integrating out 
the resonance fields.  
Apart from kinetic and mass terms for the resonance fields, it has the 
following structure:
\begin{equation}
 \LNCinf \ =  {\cal L}^{\rm GB}_{(2)+(4)+(6)} \ + \  
   {\cal L}^{R_i}_{(2)}  +  {\cal L}^{R_i}_{(4)} \ + \
{\cal L}^{ R_i R_j}_{(2)} \ + \ 
{\cal L}^{ R_i R_j R_k}_{(0)}  \ , 
\label{eq:intro2}
\end{equation}
where ${\cal L}^{\rm GB}_{(2)+(4)+(6)} = {\cal L}_{(2)}^{\rm GB} + 
{\cal L}_{(4)}^{\rm GB}+{\cal L}_{(6)}^{\rm GB}$ is the Goldstone 
chiral Lagrangian
up to $\cO(p^6)$.  $ {\cal L}^{[...]}_{(n)} $
is a term of chiral order $p^n$ involving the resonances
specified in $[...]$, i.e. up to cubic terms in resonance fields. 
Higher-derivative operators can be
added to the Lagrangian. Although they cannot contribute to 
${\cal L}_6^{\rm \chi PT}$ such operators  may be required in 
order to satisfy short-distance constraints \cite{Moussallam:1994at}.  
By using field redefinitions, we identify the minimal set of resonance
couplings contributing to ${\cal L}_6^{\rm \chi PT}$. 
It is important to distinguish between 
${\cal L}_{2n}^{\rm \chi PT}$ and ${\cal L}_{(2n)}^{\rm GB}$ for $n> 1$
(${\cal L}_2^{\rm \chi PT} \equiv  {\cal L}_{(2)}^{\rm GB}$)
although both have the same structure and operators. ${\cal
  L}_{2n}^{\rm \chi PT}$ denotes the full chiral Lagrangian whereas
${\cal L}_{(2n)}^{\rm GB}$ is part of the large-$N_C$ inspired Lagrangian
(\ref{eq:intro2}) where the meson resonances are still active degrees 
of freedom. 
\par
We adopt here the antisymmetric tensor 
representation for spin-one mesons~\cite{Gasser:1983yg,Ecker:1988te}.
Although we do not explicitly prove the equivalence with the
Proca formalism to $\cO(p^6)$, we expect that our results are
representation independent once the Lagrangian couplings are forced
to satisfy QCD short-distance constraints.
In this way we reproduce the well-known results for the low-energy
constants of ${\cal O}(p^4)$: the antisymmetric tensor representation
provides the simplest possible framework where short-distance
constraints imply~\cite{Ecker:1989yg} the absence of 
${\cal L}^{\rm GB}_{(4)}$ in Eq.~(\ref{eq:intro2}). At 
${\cal O}(p^6)$ this issue remains to be clarified
(see also subsection~\ref{subsec:Proca}). 

\item[ii.]  Integrate out resonance fields and express the 
${\cal O}(p^6)$ LECs in terms of resonance masses and couplings. 
\item[iii.]  Determine or at least constrain the resonance couplings
  by enforcing the correct asymptotic behaviour of appropriate QCD
  Green  functions.
\end{enumerate}
\par
In this work we complete the first two goals outlined above. In Sec.~2 
we explain how chiral symmetry constrains the structure of the 
Lagrangian describing the interaction between Goldstone bosons and 
resonance fields. We proceed in Sec.~3 to eliminate the resonance 
fields with the help of their equations of motion to obtain a 
parametrization of the LECs of ${\cal O}(p^6)$. Sec.~4 is committed 
to explore the consequences of short-distance constraints for the 
resonance couplings. We incorporate known constraints from all relevant
two-point functions. In Sec.~5, we reanalyse the three-point functions 
$\langle VA\,P \rangle $ and $\langle S\,P\,P  \rangle$ 
\cite{Cirigliano:2004ue,Cirigliano:2005xn} within the present scheme.
In Sec.~6 we collect our conclusions. Several appendices complement 
the results achieved in this  article.

\section{Chiral resonance Lagrangian}
\label{sec:lagrangianR}
\renewcommand{\theequation}{\arabic{section}.\arabic{equation}}
\setcounter{equation}{0}

In this section we construct the most general Lagrangian
$ \LNCinf$ of Goldstone and resonance fields consistent 
with $SU(3)_L \times SU(3)_R$ 
chiral symmetry, parity (P), charge conjugation (C) and 
the $N_C \rightarrow \infty$ limit. 
We only retain those operators in $ \LNCinf$
that contribute to chiral LECs of up to $\cO(p^6)$
after integrating out the resonance fields. 
Since chiral symmetry plays a major role in constraining the
structure of the Lagrangian we shortly review the formalism of 
broken chiral symmetry and nonlinear realizations of the chiral group
below. 

\subsection{Building blocks}
With massless light quarks ($q^\top = (u,d,s)$), the QCD
Lagrangian (omitting the heavy-quark part)
\beq \label{eq:qcd}
\cL_{\rm QCD}^0 =
- \displaystyle\frac{1}{4}\, G^a_{\mu\nu} G^{\mu\nu}_a
 + i \,\bar{q}_L \gamma^\mu D_\mu q_L  
 + i \,\bar{q}_R \gamma^\mu D_\mu q_R 
\eeq
is invariant under global $SU(3)_L\times SU(3)_R$ transformations of
the left- and right-handed quarks in flavour space: $q_{L,R} \to
g_{L,R} \; q_{L,R} \,$, $g_{L,R} \in SU(3)_{L,R}$. The chiral group
$G=SU(3)_L\times SU(3)_R$ is spontaneously broken to the diagonal
subgroup $SU(3)_V$. According to Goldstone's theorem~\cite{GO:61},
eight pseudoscalar massless bosons appear in the theory. 
\par
The Goldstone fields $\vp$ parametrize the elements $u(\vp)$ 
of the coset space $SU(3)_L\times SU(3)_R/$ $SU(3)_V$, transforming as 
\begin{equation} \label{eq:uphi}
u(\vp) \to u(\vp')=g_R u(\vp) h(g,\vp)^{-1} 
= h(g,\vp) u(\vp) g_L^{-1}  
\end{equation} 
under a general chiral rotation $g=(g_L,g_R) \in G$ 
in terms of the $SU(3)_V$ compensator field $h(g,\vp)$.  
An explicit parametrization of $u(\vp)$ is given by 
\begin{equation}
u(\vp) = \exp{\left\{\frac{i}{\sqrt{2} F} \Phi \right\}} \ , 
\end{equation}
with 
$$
\Phi =  
\left(
\begin{array}{ccc}
 \displaystyle\frac{1}{\sqrt 2}\,\pi^0 + \displaystyle\frac{1}{\sqrt
 6}\,\eta_8 
& \pi^+ & K^+ \\
\pi^- & - \displaystyle\frac{1}{\sqrt 2}\,\pi^0 + 
\displaystyle\frac{1}{\sqrt 6}\,\eta_8 
& K^0 \\
 K^- & \bar{K}^0 & - \displaystyle\frac{2}{\sqrt 6}\,\eta_8 
\end{array}
\right)
\ .
$$
\par
The nonlinear realization of $G$ on massive non-Goldstone fields
depends on their transformation properties under the unbroken subgroup
$SU(3)_V$ \cite{Coleman:1969sm}.
In this work we consider massive states transforming as
octets ($R_8$) or singlets ($R_0$):
\beq
R_8 \rightarrow h(g,\vp) \,   R_8  \, h(g,\vp)^{-1} \ , \qquad 
\qquad R_0 \rightarrow   R_0 \ ,   
\eeq 
with the notation
$R_8 = 1/\sqrt{2} \sum_i \lambda_i R_i$. 
In the large-$N_C$ limit, octet and singlet become degenerate in the 
chiral limit (with common mass $M_R$), and we collect them in a
nonet field 
\beq \label{eq:antisy}
R =  \sum_i \lambda_i  R_i/\sqrt{2}  + R_0/\sqrt{3} ~\mathbbm{1}  \ . 
\eeq
In order to calculate Green functions of vector, axial-vector, scalar
and pseudoscalar densities, it is convenient to include in the QCD
Lagrangian external hermitian sources $\ell_{\mu}, r_\mu, s, p$:
\beq
\cL_{\rm QCD} =  \cL_{\rm QCD}^0 + 
\bar{q}_L \gamma^\mu \ell_{\mu}\, q_L
+ \bar{q}_R \gamma^\mu r_\mu\, q_R -
\bar{q}_L (s - i p)\, q_R -  
\bar{q}_R (s + i p)\, q_L   \ . 
\label{eq:qcd2}
\eeq
The extended Lagrangian is invariant under  local 
$G$ transformations, with external sources 
transforming  as
\beqa 
\ell_{\mu} &\longrightarrow & 
g_L \ell_{\mu} g_L^\dagger + i g_L \partial_\mu g_L^\dagger  \; , \nn 
r_\mu &\longrightarrow & 
g_R r_\mu g_R^\dagger + i g_R \partial_\mu g_R^\dagger \; ,  \nn 
s + i p &\longrightarrow & 
g_R (s + i p) g_L^\dagger \; .
\eeqa
Given the fundamental building blocks $u(\vp)$, $R$, $\ell_{\mu}, r_\mu, s,
p$, the hadronic Lagrangian is given by the most general set of
monomials invariant under Lorentz, chiral, P and C transformations.
Invariant monomials to leading order 
in $1/N_C$ can be constructed by
taking single traces of products of chiral operators $X$ that either
transform as
\begin{equation} \label{eq:hXh}
X \to h(g,\vp) \, X \, h(g,\vp)^{-1} 
\end{equation} 
or remain invariant under chiral transformations. The possible
occurrence of multiple-trace terms will be discussed in 
subsection~\ref{sec:Lres} and App.~\ref{sec:multiple}. 
\par
The building blocks can be labeled according to chiral power counting.
Booking as usual $u(\vp)$ and $R$ as $\cO(1)$,  $\partial_\mu$, $\ell_{\mu}$,
$r_\mu$ as $\cO(p)$, and $s$, $p$ as $\cO(p^2)$, the 
independent building blocks $X$ of lowest dimension are:
\begin{eqnarray} 
\label{eq:ingr}
u_\mu &=& i \{ u^\dg(\partial_\mu - i r_\mu)u - 
u(\partial_\mu - i \ell_\mu) u^\dg\} 
\qquad \qquad   [\cO(p)] \; , 
\nn
\chi_\pm &=&  u^\dg \chi u^\dg \pm u \chi^\dg u   
\qquad \qquad \qquad \qquad \qquad \qquad 
  [\cO(p^2)] \; , 
\nn
f_\pm^{\mu\nu} &=& u F_L^{\mu\nu} u^\dg \pm u^\dg F_R^{\mu\nu} u   
\qquad \qquad \qquad \qquad \qquad  \ 
 [\cO(p^2)]  \; , 
\nn
h_{\mu\nu} &=& \nabla_\mu u_\nu + \nabla_\nu u_\mu 
\qquad  \qquad \qquad \qquad \qquad \qquad    [\cO(p^2)] \; , 
\end{eqnarray}
with $\chi= 2 B (s+ip)$ and non-Abelian field strengths 
$F_R^{\mu\nu} = \partial^\mu r^\nu - \partial^\nu r^\mu -
i[r^\mu,r^\nu] $, $F_L^{\mu\nu} = \partial^\mu \ell^\nu - \partial^\nu
\ell^\mu - i [\ell^\mu,\ell^\nu]$. The covariant derivative is 
defined by
\begin{equation} 
\nabla_\mu X = \partial_\mu X + [\Gamma_\mu,X] 
\end{equation} 
in terms of the chiral connection
$\Gamma_\mu = \{ u^\dg (\partial_\mu - i r_\mu)u + 
u (\partial_\mu - i \ell_\mu) u^\dg \}/2$ 
for any operator $X$ transforming as in Eq.~(\ref{eq:hXh}).
Higher-order chiral tensors can be obtained by taking products of
lower-dimensional building blocks or by acting on them with the
covariant derivative.

\subsection{Constructing $ \LNCinf$}
\label{sec:Lres} 
Having identified the building blocks and accounting for their behaviour 
under P, C and
chiral transformations, one can proceed with the construction of
$ \LNCinf$.  In the Goldstone sector one recovers the usual 
$\chi$PT effective Lagrangian 
\cite{Gasser:1984gg,Fearing:1994ga,Bijnens:1999sh}, excluding
operators subleading in $1/N_C$. 
To distinguish  $\LNCinf$ from the $\chi$PT Lagrangian, we use the notation
${\cal L}_{(2n)}^{\rm GB}$ for the Goldstone Lagrangian of
$\cO(p^{2n})$  instead of ${\cal L}_{(2n)}^{\rm \chi PT}$.
\par
The interactions of resonances can be classified by (i)
the number of massive fields and (ii) the number of derivatives and
quark mass insertions in a given monomial.  Since we are only
interested in resonance contributions  
to the chiral Lagrangian up to $\cO(p^6)$ only a few types of
monomials can occur. One way to see which terms can occur is as 
follows: solving the resonance equations of motion in an expansion in
the resonance masses, the fields $R_i$ are expressed as a series of 
chiral monomials times inverse powers of $M_{R_i}$, with chiral 
monomials starting at $\cO(p^2)$. Therefore, for bookkeeping purposes, 
we can book the resonance fields as $\cO(p^2)$ and construct chiral 
Lagrangians with resonance fields up to $\cO(p^6)$. 
These considerations lead us to write 
\begin{eqnarray}\label{eq:lag1}
 \LNCinf \ &=&  {\cal L}^{\rm GB}_{(2)}
 \ + \  {\cal L}^{\rm GB}_{(4)} \ + {\cal L}^{\rm GB}_{(6)} 
\nn 
&&  + \
{\cal L}^{R}_{\rm kin}  \ + \
{\cal L}^{R}_{(2)}  \ + \
{\cal L}^{R}_{(4)} \ + \
{\cal L}^{ R R}_{(2)} \ + \ 
{\cal L}^{ R R R}_{(0)}  \ , 
\end{eqnarray}
where ${\cal L}^{\rm GB}_{(n)}$ is the Goldstone chiral Lagrangian
of $\cO(p^{n})$, 
${\cal L}_{\rm kin}^{R}$ is the resonance kinetic term, and 
${\cal L}^{[...]}_{(n)} $ is a sum of monomials involving the number of
resonances specified in $[...]$ with chiral building blocks of order
$p^n$.  In general, higher-derivative operators can
be added to the Lagrangian. They do not contribute to ${\cal
L}_6^{\rm \chi PT}$, but may be required in order to satisfy
short-distance constraints \cite{Moussallam:1994at}. 
\par
The Lagrangian (\ref{eq:lag1}) brings up the question of double
counting. As in every effective field theory, the LECs carry
information about physics at higher scales. Since the low-lying
resonances are represented as explicit fields in the Lagrangian 
(\ref{eq:lag1}) the LECs in ${\cal L}^{\rm GB}_{(4)+(6)}$ should only
be sensitive to even higher scales beyond the lightest meson
resonances. Within the approximation of including only the lightest
resonance multiplets in the analysis of QCD Green functions, one may
even expect those truncated LECs to be negligible. At ${\cal O}(p^4)$ 
it could actually be shown \cite{Ecker:1989yg} that all local terms in 
${\cal L}^{\rm GB}_{(4)}$ (using the antisymmetric tensor representation 
for spin-one mesons) have to vanish in order not to upset the
asymptotic behaviour of QCD correlators. A corresponding result  at 
${\cal O}(p^6)$ still has to be achieved.
In the following we do not consider nonvanishing 
contributions from ${\cal L}^{\rm GB}_{(6)}$ explicitly (see, however,
subsection~\ref{subsec:Proca}).

Using the antisymmetric tensor formalism for spin-one fields, 
the kinetic terms for resonances read
\begin{equation}
{\cal L}^{R}_{\rm kin}  =  
\sum_{R=S,P} \frac{1}{2}\langle \nabla^{\mu}R \nabla_{\mu}R 
- M_{R}^{2} R^2 \rangle \, 
- \ \sum_{R=V,A} \frac{1}{2} \langle \nabla^\lambda R_{\lambda \mu} \,
\nabla_\nu R^{\nu\mu} -\frac{M_R^2}{2}R_{\mu\nu}R^{\mu\nu}\rangle \, \; . 
\end{equation} 
The Lagrangian ${\cal L}_{(2)}^{R}$ for resonance nonets of the type
$R_i=V(1^{--}), A(1^{++}),S(0^{++}),P(0^{-+})$ is of the form 
\cite{Ecker:1988te}
\begin{eqnarray}
\label{eq:R_int}
\cL_{(2)}^V   &\; =\; &  \displaystyle\frac{F_{V}}{2\sqrt{2}}\;
   \langle V^{\mu\nu} f_{+ \, \mu\nu}\rangle\, +\,
   \displaystyle\frac{i\, G_{V}}{\sqrt{2}} \,\,\langle V^{\mu\nu} 
u_\mu u_\nu\rangle   \, , 
\nn
\cL_{(2)}^A & = &  \displaystyle\frac{F_{A}}{2\sqrt{2}} \;
   \langle A^{\mu\nu} f_{- \, \mu\nu} \rangle\, , \nn
\cL_{(2)}^{S} & = &  c_{d} \; \langle S\, u^\mu
u_\mu\rangle\, +\, c_{m} \; \langle S\, \chi_+ \rangle \, ,
\nn
\cL_{(2)}^P &=& i\, d_{m}\;\langle P\, \chi_- \rangle\, ,  
\end{eqnarray}
where we have adopted the standard notation
for the resonance couplings $F_{V}$, $G_{V}$, $F_{A}$, $c_{d}$,
$c_{m}$ and $d_{m}$. 
The Lagrangian ${\cal L}^{R}_{\rm kin} +{\cal L}^{R}_{(2)}$ 
is sufficient to describe all resonance contributions to 
${\cal L}_{(4)}^{\rm \chi PT}$~\cite{Ecker:1988te}.
\par
In order to obtain the resonance contributions to ${\cal L}_{(6)}^{\rm
\chi PT}$, we have worked out the operators contributing to $
{\cal L}^{R}_{(4)}$ (70 monomials), ${\cal L}^{ R R}_{(2)}$ (38
monomials), and ${\cal L}^{ R R R}_{(0)}$ (7 monomials).  In
the construction of this basis we have eliminated redundant operators by use
of:
\begin{itemize}
\item Partial integration;  
\item Equations of motion (EOM) for the 
lowest-order Goldstone Lagrangian:
\begin{equation} \label{eq:eom2}
\nabla^\mu u_\mu = \frac{i}{2}\left(\chi_- - \frac{1}{N_F}\lgl \chi_-
\rgl \right) \; , 
\end{equation}  with $N_F$ the number of light flavours ($N_F=3$ in
our case); 
\item  The identity
\begin{equation} \label{eq:iden1}
\nabla^\mu h_{\mu \nu} \ =  \ \nabla_{\nu} h^{\mu}_{\mu} \  + 
\  \left[ \, u^{\mu} \, , \, 
i \, f_{+ \mu \nu} - \frac{1}{2} [u_\mu , u_\nu] \, \right]  
\ - \ \nabla^\mu f_{- \mu \nu}  \  ; 
\end{equation}
\item The Bianchi identity 
\begin{equation} \label{eq:Bianchi}
\nabla_\mu \Gamma_{\nu\rho}+\nabla_\nu \Gamma_{\rho\mu}+
\nabla_\rho \Gamma_{\mu\nu} = 0 \ , \qquad \qquad \qquad 
\Gamma_{\mu\nu} = \frac{1}{4}[u_\mu,u_\nu]-\frac{i}{2}f_{+\mu\nu} \ . 
\end{equation}
\end{itemize}
For the Lagrangian density linear in resonance fields we find 
a total of 70 independent operators: 
\begin{equation} \label{eq:l4r}
{\cal L}_{(4)}^{R} = \sum_{i=1}^{22} \, \lambda_i^{V} \, {\cal O}^V_i
+ \sum_{i=1}^{17} \, \lambda_i^{A} \, {\cal O}^A_i
+ \sum_{i=1}^{18} \, \lambda_i^{S} \, {\cal O}^S_i
+ \sum_{i=1}^{13} \, \lambda_i^{P} \, {\cal O}^P_i \ . 
\end{equation}
The corresponding monomials ${\cal O}_i^{R}$ are reported in
Tables~\ref{tab:lagV} --
\ref{tab:lagP}. 
\par
For the Lagrangian quadratic in the resonance fields
we find a total of 38 operators: 
\begin{equation} \label{eq:l2rr}
{\cal L}_{(2)}^{RR} = \sum_{(i j) n} \, \lambda_n^{R_i R_j} 
\ {\cal O}^{R_i R_j}_n  \ , 
\end{equation}
with $R_i R_j= VV,AA,SS,PP,SA,SP,SV,PV,PA,VA$. 
The operators are listed in Tables~\ref{tab:VVSSPP}, \ref{tab:SPSVSA},
\ref{tab:PVPAVA}.        
\par
Finally, for the Lagrangian cubic in resonance fields there are
7 independent operators (see Table~\ref{tab:cubic}):  
\begin{equation} \label{eq:l0rrr}
{\cal L}_{(0)}^{RRR} = \sum_{( i j k )}  \, \lambda^{R_i R_j R_k }
\  {\cal O}^{R_i R_j R_k } \ .
\end{equation}

The form of the EOM (\ref{eq:eom2}) implies that the number of traces
is not conserved over the course of constructing the effective
Lagrangian. In principle, this could be circumvented at the cost of 
ignoring the EOM and writing the Lagrangian in terms of 
$\nabla^\mu u_\mu $ \footnote{Indeed, one recovers our Lagrangian when 
first considering the most general expression involving exclusively
terms with single traces and only afterwards using the equation of
 motion.}. More fundamentally, this circumstance reflects 
the fact that the counting of traces in the effective theory is in
general not in direct correspondence with the order in $1/\nc$ of a 
term. A well-known example is the term $L_7  \langle \chi_- \rangle^2$ 
that receives contributions from the exchange of the $\eta'$ 
\cite{Gasser:1984gg, Peris:1994dh, Ecker:1988te, Kaiser:2000gs, 
Kaiser:2005eu}. 

As stated above, we do not include this particle explicitly in our 
Lagrangian. However, we devote App.~\ref{sec:multiple} to the 
clarification of the role of the multiple-trace terms (see also 
Ref.~\cite{RK}). The analysis shows that 4 of the 7 multiple-trace 
terms need not be considered because they lead to subleading 
contributions in $1/N_C$. The same is true of a possible contribution 
$\propto \langle P \rangle \langle \chi_- \rangle$ to $\cL_{(2)}^P$.  


\begin{table}[!t]
\begin{center}
\renewcommand{\arraystretch}{1.5}
\begin{tabular}{|c|c|||c|c|} 
\hline
\multicolumn{1}{|c|}{i} &
\multicolumn{1}{|c|||}{Operator ${\cal O}^V_i$} & 
\multicolumn{1}{|c|}{i} &
\multicolumn{1}{|c|}{Operator ${\cal O}^V_i$}  \\
\hline
\hline
 1  & $i \, \langle \, V_{\mu \nu} \,  u^{\mu} u_{\alpha} u^{\alpha} 
u^{\nu} \, 
 \rangle $
 & $12^*$ & $ \langle \, V_{\mu \nu} \, u_{\alpha} \,  f_{+}^{\mu \nu} \, 
 u^{\alpha} \, \rangle $\\
\hline
 2 & $i \, \langle \,  V_{\mu \nu} \,  u^{\alpha} u^{\mu} u^{\nu} 
 u_{\alpha} \, \rangle $
& $13^*$ & $ \langle \, V_{\mu \nu} \, ( \, u^{\mu} \, f_{+}^{\nu \alpha} \, 
u_{\alpha} \, + \, u_{\alpha} \, f_{+}^{\nu \alpha} \, u^{\mu} \, ) \, 
\rangle $\\
\hline
3 & $i \, \langle \, V_{\mu \nu} \, \{ \,  u^{\alpha} , u^{\mu} u_{\alpha} 
u^{\nu} \, \} \, 
\rangle $ 
&$14^*$& $ \langle \, V_{\mu \nu} \, ( \, u^{\mu} u_{\alpha} \, 
f_{+}^{\alpha \nu} \, + \, f_{+}^{\alpha \nu} \, u_{\alpha} u^{\mu} \, ) \,
\rangle $\\
\hline
4  & $i \, \langle \, V_{\mu \nu} \, \{ \, u^{\mu} u^{\nu},  u^{\alpha} 
u_{\alpha} \, \} \, 
\rangle $ 
& $15^*$ & $\langle \, V_{\mu \nu} \, ( \, u_{\alpha} u^{\mu} \, 
f_{+}^{\alpha \nu} \, + \, f_{+}^{\alpha \nu} \, u^{\mu} u_{\alpha} \, ) \,
\rangle $\\
\hline
$5^*$ & $i \, \langle \, V_{\mu \nu} \, f_{-}^{\mu \alpha} \, 
f_{-}^{\nu \beta} \, \rangle \, g_{\alpha \beta} $
&$16^*$& $ i \, \langle \, V_{\mu \nu} \, [ \, \nabla^{\mu} 
f_{-}^{\nu \alpha} \, 
, \, u_{\alpha} \, ] \, \rangle$ \\
\hline
$6^*$ & $ \langle \, V_{\mu \nu} \, \{ \, f_{+}^{\mu \nu} \, , \, 
\chi_{+} \, \} \, \rangle $ 
&$17^*$& $ i \, \langle \, V_{\mu \nu} \, [ \, \nabla_{\alpha} 
f_{-}^{\mu \nu } \, 
, \, u^{\alpha} \, ] \, \rangle$\\
\hline
$7^*$ & $i \, \langle \, V_{\mu \nu} \, f_{+}^{\mu \alpha} \, 
f_{+}^{\nu \beta}  \, \rangle \, g_{\alpha \beta} $
&$18^*$ & $ i \, \langle \, V_{\mu \nu} \, [ \, \nabla_{\alpha} 
f_{-}^{\alpha \mu} \, 
, \, u^{\nu} \, ] \, \rangle$\\
\hline
$8^*$ & $ i \, \langle \, V_{\mu \nu} \, \{ \, \chi_{+} \, , \,
u^{\mu} u^{\nu} \, 
\}
 \, \rangle $  
&$19^*$& $ i \, \langle \, V_{\mu \nu} \, [ \, f_{-}^{\mu \alpha} \, , \, 
h^{\nu}_{\alpha} \, ] \, \rangle  $\\
\hline
$9^*$ & $ i \, \langle \, V_{\mu \nu} \, u^{\mu} \, \chi_{+} \, u^{\nu} \, 
\rangle $ 
&$20^*$&  $ \langle \, V_{\mu \nu} \, [ \, f_{-}^{\mu \nu} \, , \, 
\chi_{-} \, ] \, \rangle $ \\
\hline
$10^*$ & $ \langle \, V_{\mu \nu} \, [ \, u^{\mu} \, , \, \nabla^{\nu} 
\chi_{-} \, ] \, \rangle $
&$21^\dagger$& $i \, \left\langle \, V_{\mu\nu} \, \nabla_{\alpha} 
\nabla^{\alpha} \,
\left( u^{\mu} \, u^{\nu} \right) \, \right\rangle $\\
\hline
$11^*$ & $\langle \, V_{\mu \nu} \, \{ \, f_{+}^{\mu \nu} \, , \, u^{\alpha}
u_{\alpha} \, \} \, \rangle  $
&$22^\dagger$& $ \langle \, V_{\mu \nu} \, \nabla_{\alpha} \nabla^{\alpha} \, 
f_{+}^{\mu \nu} \, \rangle $\\
\hline
\multicolumn{4}{c}{}
\end{tabular} 
\end{center}
\vspace*{-1.cm}
\caption{\label{tab:lagV}
Monomials contributing to ${\cal L}_{(4)}^{V}$. Operators that can be 
dismissed on the basis of field redefinitions and the OPE are marked
with $*$ and $\dagger$, respectively (see 
Sec.~\ref{sec:minimal}).}
\end{table}

\begin{table}[!h]
\begin{center}
\renewcommand{\arraystretch}{1.5}
\begin{tabular}{|c|c|||c|c|} 
\hline
\multicolumn{1}{|c|}{$i$} &
\multicolumn{1}{|c|||}{Operator ${\cal O}^A_i$} & 
\multicolumn{1}{|c|}{$i$} &
\multicolumn{1}{|c|}{Operator ${\cal O}^A_i$}  \\
\hline
\hline
 1 & $ \langle \, A_{\mu \nu} \, ( \, u^{\mu} u_{\alpha} \, h^{\nu \alpha}
 \, + \, h^{\nu \alpha} \, u_{\alpha} u^{\mu} \, )  \, 
 \rangle $
& $9^*$ & $ \langle \, A_{\mu \nu} \, ( \, u^{\mu} u_{\alpha} \, 
f_{-}^{\nu \alpha}
\, + \, f_{-}^{\nu \alpha} \, u_{\alpha} u^{\mu} \, )  \, 
\rangle $ \\
\hline
 2  & $ \langle \,  A_{\mu \nu} \, ( \, u_{\alpha} u^{\mu} \, 
 h^{\nu \alpha} \, + \, h^{\nu \alpha} \, u^{\mu} u_{\alpha} \, ) \, 
\rangle $
&$10^*$ & $ \langle \, A_{\mu \nu} \, ( \, u_{\alpha} u^{\mu} \, 
f_{-}^{\nu \alpha}
\, + \, f_{-}^{\nu \alpha} \, u^{\mu} u_{\alpha} \, )  \, \rangle $\\
\hline
3 & 
$ \langle \, A_{\mu \nu} \, (  \, u^{\mu} \, h^{\nu \alpha} \, u_{\alpha}
\, + \, u_{\alpha} \, h^{\nu \alpha} \, u^{\mu} \, ) \, 
\rangle $ 
& $11^*$ & $\langle \, A_{\mu \nu} \, ( \, u^{\mu} \, f_{-}^{\nu \alpha} \,
u_{\alpha} \, + \, u_{\alpha} \, f_{-}^{\nu \alpha} \, u^{\mu} \, ) 
 \, \rangle  $\\
\hline
$4^*$ & $ \langle \, A_{\mu \nu} \, [ \, f_{+}^{\mu \nu} \, , \, \chi_{-} \, 
] \, \rangle$
 & $12^*$ & $i \, \langle \, A_{\mu \nu} \, [ \, f_{+}^{\mu \alpha} \, , \, 
 h^{\nu}_{\alpha} \, ] \, \rangle $ \\
\hline
$5^*$ & $ i \, \langle \, A_{\mu \nu} \, [ \, \chi_{-} \, , \, u^{\mu} 
u^{\nu} \,] \, \rangle $  
& $13^*$ & $ i \, \langle \, A_{\mu \nu} \, [ \, \nabla^{\alpha} 
f_{+}^{\mu \nu} \,
 , \, u_{\alpha} \, ] \,  \rangle $ \\
\hline
$6^*$ & $  \langle \, A_{\mu \nu} \, \{ \, u^{\mu} \, , \, 
\nabla^{\nu} \chi_{+} \, \} \, \rangle $ 
& $14^*$ & $i \,  \langle \, A_{\mu \nu} \, [ \, f_{+}^{\mu \alpha} \, , 
\, f_{-}^{\nu \beta} \, ]  \,  \rangle \, g_{\alpha \beta}$\\
\hline
$7^*$ & $ \langle \, A_{\mu \nu} \, \{ \, f_{-}^{\mu \nu} \, , \, 
u_{\alpha} u^{\alpha} \, \} \, \rangle $ 
&$15^*$ & $i \, \langle \, A_{\mu \nu} \, [ \, \nabla_{\alpha} 
f_{+}^{\nu \alpha} \,
 , \, u^{\mu} \, ] \,  \rangle  $\\
\hline
$8^*$ & $\langle \, A_{\mu \nu} \, u_{\alpha} \, f_{-}^{\mu \nu} \, 
u^{\alpha} \, \rangle $  
&$16^*$& $ \langle \, A_{\mu \nu} \, \{ \, f_{-}^{\mu \nu} \, , \, 
\chi_{+} \, \} \, \rangle$\\
\hline
 & & $17^\dagger$ & $\langle \, A_{\mu \nu} \, \nabla_{\alpha} 
\nabla^{\alpha} \, f_{-}^{\mu \nu} \, \rangle $  \\
\hline
\multicolumn{4}{c}{}
\end{tabular} 
\end{center}
\vspace*{-1.cm}
\caption{\label{tab:lagA}
Monomials contributing to ${\cal L}_{(4)}^{A}$. Operators that can be 
dismissed on the basis of field redefinitions and the OPE are marked 
with $*$ and $\dagger$, respectively (see 
Sec.~\ref{sec:minimal}).}
\end{table}

\begin{table}[!h]
\begin{center}
\renewcommand{\arraystretch}{1.5}
\begin{tabular}{|c|c|||c|c|} 
\hline
\multicolumn{1}{|c|}{$i$} &
\multicolumn{1}{|c|||}{Operator ${\cal O}^S_i$} & 
\multicolumn{1}{|c|}{$i$} &
\multicolumn{1}{|c|}{Operator ${\cal O}^S_i$}  \\
\hline
\hline
 1 & $ \langle S \, u_{\mu} u^{\mu} \, u_{\nu} u^{\nu}  \, 
 \rangle $
&$10^*$ & $i \,  \langle \, S \, \{ \, f_{+}^{\mu \nu} \, , \, 
u_{\mu} u_{\nu} \, \}  \, \rangle $ \\
\hline
 2 & $ \langle \, S \, u_{\mu} \, u_{\nu} u^{\nu} \, u^{\mu} \, \rangle $
&$11^*$ & $i \, \langle \, S \, u_{\mu} \, f_{+}^{\mu \nu} \, u_{\nu}  
 \, \rangle  $\\
\hline
3 & $ \langle \, S \, u_{\mu} u_{\nu} u^{\mu} u^{\nu}  \, 
\rangle $ 
 & $12^*$ & $ \langle \, S \, \{ \, \nabla_{\alpha} \, 
f_{-}^{\mu \alpha} \, , \,
 u_{\mu} \, \} \, \rangle $\\
\hline
4 & $ i \, \langle \, S \, u^{\mu} \, \rangle \, \langle \, 
\nabla_{\mu}  \,\chi_-
\, \rangle $ 
& $13^*$ & $ \langle \, S \, \chi_{+} \, \chi_{+} \, \rangle $\\
\hline
5 & $\langle \, S \, \chi_- \, \rangle \, \langle \,\chi_- \, \rangle$ 
& $14^*$ & $ \langle \, S \, \chi_{-} \, \chi_{-} \, \rangle $ \\
\hline
$6^*$ & $  \langle \, S \, \{ \, \chi_{+} \, , \, u^{\mu} u_{\mu} \, \} \, 
\rangle $ 
&$15^*$ & $\langle \, S \, f_{+ \mu \nu} \, f_{+}^{\mu \nu} \, \rangle $\\
\hline
$7^*$ & $ \langle \, S \, u_{\mu} \, \chi_{+} \, u^{\mu}  \, \rangle $ 
&$16^*$& $\langle \, S \, f_{- \mu \nu} \, f_{-}^{\mu \nu} \, \rangle $ \\
\hline
$8^*$ & $i \, \langle \, S \, \{ \, u^{\mu} \, , \, \nabla_{\mu} \, 
\chi_{-} \, \} 
 \, \rangle $  
& $17^\dagger$ & $\left\langle  \, S \, \nabla_{\alpha}
\nabla^{\alpha} \, \left( u_{\mu} \,
u^{\mu} \right) \, \right\rangle $ \\
\hline
$9^\ddag$ & $ \langle \, S \, \rangle \, \langle \, \chi_- \, \rangle
\, \langle \, \chi_- \, \rangle $
&$18^\dagger$ & $ \langle \, S \, \nabla_{\mu} \nabla^{\mu} \,\chi_{+} \, 
\rangle $ \\
\hline
\multicolumn{4}{c}{}
\end{tabular} 
\end{center}
\vspace*{-1.cm}
\caption{\label{tab:lagS}
Monomials contributing to ${\cal L}_{(4)}^{S}$.
Operators that can be dismissed on the basis of field redefinitions, 
the OPE and large $\nc$ are marked with $*$, $\dagger$ and $\ddag$, 
respectively (see Sec.~\ref{sec:minimal} and App.~\ref{sec:multiple}).}
\end{table}

\begin{table}[!h]
\begin{center}
\renewcommand{\arraystretch}{1.5}
\begin{tabular}{|c|c|||c|c|} 
\hline
\multicolumn{1}{|c|}{$i$} &
\multicolumn{1}{|c|||}{Operator ${\cal O}^P_i$} &
\multicolumn{1}{|c|}{$i$} &
\multicolumn{1}{|c|}{Operator ${\cal O}^P_i$} \\
\hline
\hline
 1 & $ \langle \, P \, \{ \, h^{\mu \nu} \, , \, u_{\mu} u_{\nu} \, \}  \, 
 \rangle $ 
&$7^{*,\ddag}$ & $i \, \langle \, P \, u_{\mu} u^{\mu} \, \rangle \, \langle 
\, \chi_- \, 
\rangle$\\
\hline
 2 & $ \langle \, P \, u_{\mu} \, h^{\mu \nu} \, u_{\nu} \, \rangle $ 
& $8^*$ & $ \langle \, P \, [ \, f_{-}^{\mu \nu} \, , \, u_{\mu} u_{\nu} \, ]  
 \, \rangle $  
\\
\hline
3  &  $i \, \langle \, P \, \chi_+ \, \rangle \, \langle \, 
\chi_- \, \rangle$
&$9^*$ & $i \, \langle \, P \, [ \, \nabla_{\mu} \, f_{+}^{\mu  \nu} \, , \, 
u_{\nu} \, ] \,  \rangle $ \\
\hline
$4^*$ & $ i \, \langle \, P \, \{ \, \chi_{-} ,\, u_{\mu} u^{\mu} \, \}
\, \rangle $ 
&$10^*$ & $i \,  \langle \, P \, \{ \, \chi_{+}  \, , \, \chi_{-}  \, \}
  \, \rangle $\\
\hline
$5^*$ & $i \, \langle \, P \, u_{\mu} \, \chi_{-} \, u^{\mu}
\, \rangle $ 
&$11^*$ & $i \, \langle \, P \, [ \, f_{+}^{\mu \nu} \, , \, f_{- \mu
\nu} \, ]  
 \, \rangle  $\\
\hline
$6^*$ & $  \langle \, P \, \{ \, \nabla^{\mu} \, \chi_{+} \, , \,
u_{\mu} \, \}
 \, \rangle $ 
 & $12^\ddag$ &  $i \, \langle \, P \, \rangle \, \langle \, \nabla_{\mu} 
\nabla^{\mu} \, \chi_- \, \rangle $\\
\hline
&
&$13^\dagger$ & $ i \, \langle \, P \, \nabla_{\mu} \nabla^{\mu} \, \chi_{-}  
\, \rangle $  \\
\hline
\multicolumn{4}{c}{}
\end{tabular} 
\end{center}
\vspace*{-1.0cm}
\caption{\label{tab:lagP}
Monomials contributing to ${\cal L}_{(4)}^{P}$.
Operators that can be dismissed on the basis of field redefinitions,
the OPE and large $\nc$ are marked with $*$, $\dagger$ and $\ddag$, 
respectively (see Sec.~\ref{sec:minimal} and App.~\ref{sec:multiple}). 
}
\end{table}

\begin{table}[!h]
\begin{center}
\renewcommand{\arraystretch}{1.5}
\begin{tabular}{|c|c|c|c|} 
\hline
\multicolumn{1}{|c|}{$i$} &
\multicolumn{1}{|c|}{Operator ${\cal O}^{RR}_i$, $R=V,A$}  &
\multicolumn{1}{|c|}{Operator ${\cal O}^{SS}_i$} &
\multicolumn{1}{|c|}{Operator ${\cal O}^{PP}_i$} \\
\hline
\hline
 1 & $\langle \, R_{\mu \nu} R^{\mu \nu} \, u^{\alpha} u_{\alpha} \, 
 \rangle $ & $ \langle \, S \, S  \, u_{\mu} u^{\mu}   \, 
 \rangle $ & $\langle \, P \, P \, u_{\mu} u^{\mu}  \, 
 \rangle $\\
\hline
 2 & $\langle \, R_{\mu \nu} \, u^{\alpha} \, R^{\mu \nu} \, u_{\alpha}
  \,  \rangle $ & $ \langle \, S \, u_{\mu} \, S \, u^{\mu}  \, \rangle $
  & $ \langle \, P \, u_{\mu} \, P \, u^{\mu} 
  \,  \rangle $ \\
\hline
 3 & $ \langle \, R_{\mu \alpha} \,   R^{\nu \alpha} \, 
  u^{\mu} \, u_{\nu} \,  \rangle $ & $ \langle \, S \, S \, \chi_{+} \, 
\rangle $& $ \langle \, P \, P \, \chi_{+}
  \,  \rangle $  \\
\hline
 4 & $ \langle \, R_{\mu \alpha} \,   R^{\nu \alpha} \, 
  u_{\nu} \, u^{\mu} \,  \rangle $ & & \\
\hline
 5 & $ \langle \, R_{\mu \alpha} \, ( \,  u^{\alpha} \, R^{\mu \beta} \, 
 u_{\beta} \, + \, u_{\beta} \, R^{\mu \beta} \, u^{\alpha} \, ) 
  \,  \rangle $ & & \\
\hline
 6 & $ \langle \, R_{\mu \nu} \,  R^{\mu \nu} \, \chi_{+}   
  \,  \rangle $ & & \\
\hline
 7 & $i \,  \langle \,  R_{\mu \alpha}  \, R^{\alpha \nu} \, 
  f_{+ \beta \nu}   
  \,  \rangle \, g^{\beta \mu}$ & & \\
\hline
\multicolumn{4}{c}{}
\end{tabular} 
\end{center}
\vspace*{-1.3cm}
\caption{
\label{tab:VVSSPP}
Independent monomials of type $\cO^{VV}_i$, $\cO^{AA}_i$, $\cO^{SS}_i$ 
and $\cO^{PP}_i$.
}
\end{table}

\begin{table}[!h]
\begin{center}
\renewcommand{\arraystretch}{1.5}
\begin{tabular}{|c|c|c|c|} 
\hline
\multicolumn{1}{|c|}{$i$} &
\multicolumn{1}{|c|}{Operator ${\cal O}^{SP}_i$} &
\multicolumn{1}{|c|}{Operator ${\cal O}^{SV}_i$}  &
\multicolumn{1}{|c|}{Operator ${\cal O}^{SA}_i$}  \\
\hline
\hline
 1 & $ \langle \, \{ \, \nabla_{\mu} \,S \, , \,  P \, \}  \, u^{\mu}    \, 
 \rangle $& $i \, \langle \, \{ \, S \, , \, V_{\mu \nu} \, \} \,
 u^{\mu} u^{\nu}  \, 
 \rangle $ & $\langle \, \{ \, \nabla_{\mu} \, S \, , \, A^{\mu \nu} \, \} 
 \,  u_{\nu}  \, 
 \rangle $ \\
\hline
 2 & $i \,  \langle \, \{ \, S \, , \, P \, \} \, \chi_{-}  \, \rangle $
  & $i \,  \langle \, S \, u_{\mu} \, V^{\mu \nu} \, u_{\nu} \,  \rangle $
 & $ \langle \, \{ S \, , \,   A_{\mu \nu} \, \} \, f_{-}^{\mu \nu} 
  \,  \rangle $  \\
\hline
3 & $ i \, \langle \, S \, P \, \rangle \, \langle \, \chi_- \, 
\rangle^\ddag $  
& $\langle \, \{ \, S \, , \, V_{\mu \nu} \, \} \, f_{+}^{\mu \nu} \, 
 \rangle $ & \\
\hline
\multicolumn{4}{c}{}
\end{tabular} 
\end{center}
\vspace*{-1.3cm}
\caption{
\label{tab:SPSVSA}
Independent monomials of type $\cO^{SP}_i$, $\cO^{SV}_i$ and 
$\cO^{SA}_i$. The operator $\cO^{SP}_3$ (marked with~$\ddag$) can be 
dismissed on the basis of large $\nc$ (see App.~\ref{sec:multiple}). 
}
\end{table}

\begin{table}[!h]
\begin{center}
\renewcommand{\arraystretch}{1.5}
\begin{tabular}{|c|c|c|c|} 
\hline
\multicolumn{1}{|c|}{$i$} &
\multicolumn{1}{|c|}{Operator ${\cal O}^{VA}_i$} &
\multicolumn{1}{|c|}{Operator ${\cal O}^{PA}_i$} &
\multicolumn{1}{|c|}{Operator ${\cal O}^{PV}_i$}  \\
\hline
\hline
 1 &$\langle \, [ \, V^{\mu \nu} \, , \, A_{\mu \nu} \, ] \, \chi_{-} \,  
 \rangle $ & $i \,  \langle \, [ \, P \, , \, A_{\mu \nu} \, ] \, 
 f_{+}^{\mu \nu} 
  \,  \rangle $ & $i \, \langle \, [ \, \nabla^{\mu} \, P \, , \, 
  V_{\mu \nu} \, ] \,
  u^{\nu}   \, 
 \rangle $ \\
\hline
 2 & $i \, \langle \,  [ \, V^{\mu \nu} \, , \, A_{\nu \alpha} \, ] \, 
 h_{\mu}^{\alpha}
  \,  \rangle $ & $\langle \, [ \, P \, , \, A_{\mu \nu} \, ] \, 
  u^{\mu} u^{\nu}  \, 
 \rangle $ & $i \,  \langle \, [ \, P \, , \, V_{\mu \nu} \, ] \, 
  f_{-}^{\mu \nu} 
  \,  \rangle $ \\
\hline
 3 & $i \,  \langle \, [ \, \nabla^{\mu} \, V_{\mu \nu} \, , \, A^{\nu \alpha}
 \, ] \, u_{\alpha} 
  \,  \rangle $ & & \\
\hline
 4 & $i \,  \langle \, [ \, \nabla_{\alpha} \, V_{\mu \nu} \, , \, 
 A^{\alpha \nu}
 \, ] \, u^{\mu} 
  \,  \rangle $ & & \\
\hline
 5 & $ i \, \langle \,[ \, \nabla_{\alpha} \, V_{\mu \nu} \, , \, A^{\mu \nu}
 \, ] \, u^{\alpha}  
  \,  \rangle $ & & \\
\hline
 6 & $ i \, \langle \, [ \, V^{\mu \nu} \, , \, A_{\mu \alpha} \,] \, 
 f_{- \beta \nu}   
  \,  \rangle \, g^{\alpha \beta}$ & & \\
\hline
\multicolumn{4}{c}{}
\end{tabular} 
\end{center}
\vspace*{-1.3cm}
\caption{
\label{tab:PVPAVA}
Independent monomials of type $\cO^{VA}_i$, $\cO^{PA}_i$ and $\cO^{PV}_i$.
}
\end{table}

\begin{table}[!h]
\begin{center}
\renewcommand{\arraystretch}{1.5}
\begin{tabular}{|c|c|} 
\hline
\multicolumn{1}{|c|}{$R_1 R_2 R_3$} &
\multicolumn{1}{|c|}{Operator $\cO^{(R_1,R_2,R_3)}$}  \\
\hline
\hline
 $SVV$ & $\langle \, S \,  V_{\mu \nu} V^{\mu \nu}  
\, \rangle $ \\
\hline
 $SAA$ & 
$\langle \, S \,  A_{\mu \nu} A^{\mu \nu}  
  \,  \rangle $ \\
\hline
 $SSS$ & $ \langle \, S \, S \, S \,  \rangle $ \\
\hline
 $SPP$ & $ \langle \, S \, P \, P \,  \rangle $ \\
\hline
 $VVV$ & $ i \, \langle \, 
V_{\mu \nu} \, V^{\mu \rho} \, V^{\nu \sigma} \,  \rangle \, 
g_{\rho \sigma}$ \\
\hline
 $VAP$ & $ i \, \langle  \, \left[ \, V_{\mu \nu} \, , \,   A^{\mu \nu} \, 
 \right] \, P \, \rangle $ \\
\hline
 $VAA$ & $i \,  \langle \,
V_{\mu \nu} \, A^{\mu \rho}  \, A^{\nu \sigma} 
\,  \rangle \, g_{\rho \sigma}$ \\
\hline
\multicolumn{2}{c}{}
\end{tabular} 
\end{center}
\vspace*{-1.3cm}
\caption{
\label{tab:cubic}
Independent monomials of type $\cO^{RRR}$.
}
\end{table}

\subsection{Minimal operator basis for the resonance Lagrangian}
\label{sec:minimal}

The operator basis constructed in subsection~\ref{sec:Lres} is still
redundant in the following sense. Many of the resonance couplings 
contribute to the LECs of $\cO(p^6)$ in certain combinations
only. The number of those combinations turns out to be
considerably smaller than the number of original couplings.
An elegant way to identify and to eliminate this redundancy is to make
use of field redefinitions.  The idea behind field redefinitions is
very simple: when considering the generating functional of Green
functions, the fields $R_i$ are nothing but integration variables.
Therefore, any change of variables, consistent with the symmetries  
and the spectrum of the original field theory, does not affect the 
Green functions generated by functional integration.
\par
Here we will see that redefinitions of the resonance fields will
simplify the content of $ \LNCinf$ enormously.
We will be able to take advantage of these shifts to discard
most of the operators of ${\cal L}_{(4)}^R$, without generating new
contributions to the ${\cal O}(p^6)$ chiral Lagrangian. However, since
we explicitly ignore operators generated by the field redefinitions 
that do not contribute to the ${\cal O}(p^6)$ chiral Lagrangian, Green 
functions on the basis of the simplified Lagrangian $ \LNCinf$ are in 
principle different from those produced by the full resonance Lagrangian.

\subsubsection{Linear field redefinitions}

We consider here transformations of the type 
\begin{equation}\label{eq:redef1}
R_i  \longrightarrow  R_i \ + \ g_{ij} \ F_{(2)} (R_j)   \ , 
\end{equation}
where the $g_{ij}$ are arbitrary constants and $F_{(2)} (R_j)$ are chiral
monomials of $\cO(p^2)$, linear in the resonance field $R_j$, and with
the same Lorentz, C, P and hermiticity properties as $R_i$.
Applying these field redefinitions to a given monomial in the
Lagrangian increases its chiral order. Many terms in
$ \LNCinf$ in Eq.~(\ref{eq:lag1}) therefore generate
monomials that can only influence LECs of $\cO(p^8)$ or higher.
The exceptions are the mass terms and ${\cal L}_{(2)}^{R}$, for which one 
has schematically: 
\beqa\label{eq:redef2}
-\frac{1}{2} M_{R}^2  \langle R^2  \rangle  
& \longrightarrow & -\frac{1}{2} M_{R}^2  \langle R^2  \rangle  
\ + \  {\cal L}_{(2)}^{R R} \; , 
\nn 
{\cal L}_{(2)}^{R}   & \longrightarrow &  
{\cal L}_{(2)}^{R}  \ + \ {\cal L}_{(4)}^{R} \; .
\eeqa
By appropriate choices of the constants $g_{ij}$ in
Eq.~(\ref{eq:redef1}) we can therefore
eliminate monomials belonging to ${\cal L}_{(4)}^{R}$, while
redefining some of the couplings appearing in ${\cal L}_{(2)}^{RR}$.
\par
We have found 18 possible redefinitions for vector, 17 for
axial-vector, 11 for
scalar and 10 for pseudoscalar nonet fields. The complete list is
reported in App.~\ref{app:lfr}.  Using the above 56  field 
transformations we can eliminate 47 of the 70 operators of
the type 
${\cal O}_i^R$. There are 9 monomials that do not appear at all 
in the above field transformations: 
\beq \label{eq:hec}
\cO^V_{21}, \cO^V_{22}, 
\cO^A_{17}, 
\cO^S_{17}, \cO^S_{18}, 
\cO^P_{13} 
\quad \rm{and} \quad   
\cO^S_4, 
\cO^S_{9}, \cO^P_{12}. 
\eeq
Therefore, they can certainly not be transformed away.
Those of the first group in Eq.~(\ref{eq:hec}) 
can, however, all be discarded due to the bad high-energy behaviour 
they generate (see Sec.~\ref{sec:shortd}). Of the three remaining 
(multiple-trace) terms in Eq.~(\ref{eq:hec}) it is shown in 
App.~\ref{sec:multiple} that $\cO^S_4$ (along with $\cO^S_5$ and 
$\cO^P_3$) should be retained while the other two ($\cO^S_{9}$, 
$\cO^P_{12}$) only lead to contributions subleading in $1/\nc$ and 
can therefore be dismissed.
\par
For the remaining terms one has to make a choice. We have adopted the 
strategy to eliminate preferentially terms with $\chi$'s and leave those 
with many derivatives in the list because the latter (some will not even 
contribute to ``simple" Green functions) have the worst possible 
high-energy behaviour and should therefore be more easily eliminated
with the  help of high-energy constraints. 
In our analysis, we have kept the following 15 operators
(70 - 47 - 8 = 15) in the Lagrangian ${\cal L}_{(4)}^{R}$:
\beq
\cO^V_1, \cO^V_2, \cO^V_3,\cO^V_4, \cO^A_1, 
\cO^A_2,\cO^A_3,\cO^S_1, \cO^S_2, \cO^S_3, \cO^S_4, \cO^S_5, 
\cO^P_1, \cO^P_2, \cO^P_3 \ . 
\eeq

\subsubsection{Nonlinear field redefinitions}

We may also consider transformations of the type 
\begin{equation}\label{eq:redef3}
R_i  \longrightarrow  R_i \ + \ g_{ijk} \ F_{(0)} (R_j R_k)   \ , 
\end{equation}
where $g_{ijk}$ are again arbitrary constants. The $F_{(0)} (R_j R_k)$ 
are chiral monomials of $\cO(p^0)$ involving the fields $R_j$, $R_k$ 
and with the same Lorentz, C, P and hermiticity properties as $R_i$.
The relevant transformations of monomials in $ \LNCinf$ are 
\beqa\label{eq:redef4}
-\frac{1}{2} M_{R}^2  \langle R^2  \rangle  
& \longrightarrow & -\frac{1}{2} M_{R}^2  \langle R^2  \rangle  
\ + \  {\cal L}_{(0)}^{R R R} \; , 
\nn 
{\cal L}_{(2)}^{R}   & \longrightarrow &  
{\cal L}_{(2)}^{R}  \ + \ {\cal L}_{(2)}^{R R} \; , 
\eeqa
thus allowing one to remove either bilinear or trilinear couplings.
We have found 13 independent transformations of the
type~(\ref{eq:redef4}). They could in principle be used 
to eliminate all cubic operators (seven) and six out of the 38 
bilinear operators. However, in contrast to the previous case of
linear field transformations both the
bilinear and trilinear terms are of leading chiral order. We have
therefore chosen to keep all monomials in ${\cal L}_{(2)}^{R R}$ and $
{\cal L}_{(0)}^{R R R}$ for the time being. However, the redundancy
will manifest itself in certain combinations of the 
$\lambda_{i}^{R_i R_j}$ and $\lambda^{R_i R_j R_k}$ that always occur 
together in the LECs of $\cO(p^6)$ (see App.~\ref{app:resultsCI}).

In summary, this leaves us with a minimal Lagrangian of the form given 
in Eq.~(\ref{eq:lag1}) where ${\cal L}^{\rm GB}_{(4,6)}$, 
${\cal L}_{(4)}^{R}$ and ${\cal L}_{(2)}^{SP}$ are replaced with their 
minimal versions. Explicitly,
\begin{align}\label{eq:lagminimal}
 {\cal L}^{\rm GB}_{(4,6)}  |_{\rm minimal} & = 0 \quad 
 ({\rm see,~however,~subsection~\ref{subsec:Proca}}) \; , 
\no 
{\cal L}_{(4)}^{R}|_{\rm minimal} & = \sum_{i=1}^{4} \, \lambda_i^{V} \, 
{\cal O}^V_i
+ \sum_{i=1}^{3} \, \lambda_i^{A} \, {\cal O}^A_i
+ \sum_{i=1}^{5} \, \lambda_i^{S} \, {\cal O}^S_i
+ \sum_{i=1}^{3} \, \lambda_i^{P} \, {\cal O}^P_i  
\co \\
{\cal L}_{(2)}^{SP}|_{\rm minimal} & = \sum_{i=1}^2 \, \lambda_i^{SP} 
\ {\cal O}^{SP}_i   
\fs \non 
\end{align}
Note that $\cO_3^{SP} $ has been dismissed on the basis of large $\nc$ 
(cf. App.~\ref{sec:multiple}). 

\section{The chiral Lagrangian from resonance exchange}
\label{sec:integratingout}
\renewcommand{\theequation}{\arabic{section}.\arabic{equation}}
\setcounter{equation}{0}

In this section we sketch the derivation of resonance exchange
contributions to the chiral Lagrangian up to $\cO(p^6)$, deferring most
definitions and results to App.~\ref{app:defintout}. The mesonic
chiral Lagrangian in the notation of Eq.~(\ref{eq:intro1}) takes the form
\begin{equation}
{\cal L}_{\chi PT} \, = \, {\cal L}_{2}^{\rm \chi PT} \, + \, 
{\cal L}_{4}^{\rm \chi PT} \, + \, {\cal L}_{6}^{\rm \chi PT} \,+ ...
\end{equation}
The leading-order term 
\begin{equation} 
{\cal L}_2^{\rm \chi PT}= \frac{F^2}{4} \left\lgl u_\mu u^\mu+
 \chi_+\right\rgl
\label{eq:L2}
\end{equation}
contains only two LECs, the meson decay constant in the chiral limit
$F$ and the constant $B$ in $\chi= 2 B (s+ip)$ that is related to the
quark condensate. These parameters characterize the spontaneous
breaking of chiral symmetry and they are insensitive to physics at
shorter distances.  

Higher orders in the chiral expansion bring in information from 
higher energy scales that have been integrated out by evolving
down to low energies. This information is encoded in the LECs, the
coupling constants of the higher-order Lagrangians:
\begin{equation}
{\cal L}_4^{\rm \chi PT} \, = \, \sum_{i=1}^{10} \, L_i \; {\cal O}_i^{(4)}
\; \; \; \; \qquad , \; \; \; \; \qquad 
{\cal L}_6^{\rm \chi PT} \, = \, \sum_{i=1}^{90} \, C_i \; {\cal O}_i^{(6)}
\; . 
\end{equation}
The numbering refers to $N_F = 3$ light flavours and we have omitted
in the sums the contact terms involving external fields only. Explicit 
expressions for the operators ${\cal O}_i^{(4)}$ and ${\cal O}_i^{(6)}$  
can be found in Refs.~\cite{Gasser:1984gg,Bijnens:1999sh}.
\par
The chiral expansion scale $\Lambda_{\chi} \sim M_V$ indicates that
the LECs receive contributions from energies at or above $M_V$.
It is therefore natural to expect that
the most important contributions to the LECs will come from the
lightest meson resonances. This was in fact confirmed for the LECs of
${\cal O}(p^4)$ that appear to be saturated by resonance exchange 
\cite{Ecker:1988te}. A qualification is in order here. To
leading order in $1/N_C$ (tree-level exchange only), no scale
dependence is generated for the LECs. The saturation by resonance
exchange at ${\cal O}(p^4)$ appears to be valid for a renormalization
scale between 0.5 and 1 GeV. Here we are going to perform the integration
of the resonance fields up to ${\cal O}(p^6)$ in the chiral expansion
assuming that a similar saturation holds up to this order.
\par
To make the expressions more compact, we rewrite the resonance
couplings in the Lagrangian $ \LNCinf$ in Eq.~(\ref{eq:lag1}) as
\begin{equation}
{\cal L}_{(2)}^{R} \, + \, {\cal L}_{(4)}^{R} \, = \,  
 \sum_{R=S,P} \left\langle R \, (g_2^R + g_4^R) 
\right\rangle +
\sum_{R=V,A} \left\langle R_{\mu\nu} \, (g_2^R + g_4^R)^{\mu\nu} 
\right\rangle \; ,
\end{equation}
and analogously for the part ${\cal L}_{(2)}^{RR}$ bilinear in
resonance fields (see App.~\ref{app:defintout}
for the precise definitions). 

Integrating out the resonance fields at tree level amounts to 
solving the EOM of the fields perturbatively up to the requested
order. Inserting the solutions
into $ \LNCinf$ and keeping only those pieces contributing 
up to ${\cal O}(p^6)$ in the chiral expansion, we write the final
results in the form 
\begin{equation} \label{eq:intout}
{\cal L}_{(4+6)}^{R-{\rm exchange}} = 
{\cal L}_{SP} + {\cal L}_{VA} + {\cal L}_{SPVA}
  \; . 
\end{equation}
The Lagrangians ${\cal L}_{SP}$, ${\cal L}_{VA}$ and ${\cal L}_{SPVA}$
are also given in App.~\ref{app:defintout}.

The Lagrangian (\ref{eq:intout}) is of course of the general form
${\cal L}_4^{\rm \chi PT}+{\cal L}_{6}^{\rm \chi PT}$ and 
we can identify the expressions for the LECs $L_i$ and $C_i$ in terms 
of resonance masses and couplings. The results for $L_i$ are
well known \cite{Ecker:1988te} and will not be reproduced here. The
main result of this paper are the resonance exchange contributions to
the LECs $C_i$ of ${\cal O}(p^6)$ collected in
App.~\ref{app:resultsCI}.
\par
The final results for the LECs $C_i$ display the dependence on
resonance couplings and masses. At this point, the information is
still rather limited. On the one hand, Table~\ref{tab:RESCi} shows
which LECs are not sensitive at all to resonance exchange and
may therefore be expected to be negligible in the spirit of our
approach. 
Closer inspection of Table~\ref{tab:RESCi} reveals that there are in
addition several linear relations among the LECs, e.g., $C_{20}= -3
C_{21}=C_{32}= C_{35}/6=C_{94}/8$, $C_{24}=6 C_{28}= 3 C_{30}$, etc. Finally,
some of the LECs of $\cO(p^6)$ are found to depend only on
resonance couplings that already occur at  $\cO(p^4)$
\cite{Ecker:1988te}. In fact, neglecting the contact terms, there are 
only three of them which have already been analysed
\cite{Cirigliano:2004ue,Cirigliano:2005xn}: $C_{12}, C_{38},
C_{87}$. With some more effort, one finds that the same applies
also to $C_1 +4 C_3$, $3C_1-4 C_4$, $C_1/12-C_{28}+r_S C_{32}$, 
$C_{88}-C_{90}$ \cite{Cirigliano:2005xn}, $C_{91}$, $C_{93}$, etc. 
\par
The information in Table~\ref{tab:RESCi} comes from the matching of
chiral resonance theory to $\chi$PT. Still missing is the matching of
chiral resonance theory to QCD that will give us information on the
resonance couplings and then in turn on the LECs. This matching
procedure is the subject of Sec.~\ref{sec:shortd}.

\subsection{Proca fields}
\label{subsec:Proca}
In this paper we are only concerned with the chiral Lagrangian of even
intrinsic parity. In the resonance Lagrangian $ \LNCinf$ in 
Eq.~(\ref{eq:lag1}) we have tacitly assumed that also there only
even-intrinsic-parity couplings are relevant. This is however not the
full story because resonance exchange with two odd-intrinsic-parity
vertices can also produce contributions to LECs in the
even-intrinsic-parity sector. At ${\cal O}(p^6)$ only spin-1 exchange
can contribute here but only with Proca vector fields instead of
antisymmetric tensor fields \cite{Ecker:1990in}. The relevant
Lagrangian with Proca fields $V^{\mu}$, $A^{\mu}$ consists of three
terms only \footnote{Nonlinear resonance couplings and those involving 
spin-0 resonance fields start to contribute at ${\cal O}(p^8)$ only.}~:
\begin{eqnarray} \label{eq:epsilon}
{\cal L}_{\varepsilon} \, & = & \, g_1^V \, 
{\cal O}_{V1}^{\varepsilon} \, + \,
 g_2^V \, {\cal O}_{V2}^{\varepsilon} \, + 
 \, g_1^A \, {\cal O}_{A}^{\varepsilon} \; , 
\end{eqnarray}
where
\begin{eqnarray}
{\cal O}_{V1}^{\varepsilon} \, & = & \, i \, 
\varepsilon_{\mu\nu\rho\sigma} \, 
\left\langle \, V^{\mu} \, u^{\nu} \, u^{\rho} 
\, u^{\sigma} \, \right\rangle \; ,  \nonumber \\ 
{\cal O}_{V2}^{\varepsilon} \, & = & \, 
\varepsilon_{\mu\nu\rho\sigma} \, 
\left\langle \, V^{\mu} \, \left\lbrace \, u^{\nu} 
\, , \, f_{+}^{\rho\sigma} \, \right\rbrace 
\right\rangle \; , \nonumber \\
{\cal O}_{A}^{\varepsilon} \, & = & \, 
\varepsilon_{\mu\nu\rho\sigma} \, 
\left\langle \, A^{\mu} \, \left\lbrace \, 
u^{\nu} \, , \, f_{-}^{\rho\sigma} \, \right\rbrace 
\right\rangle \; .
\end{eqnarray} 
Integrating out the vector and axial-vector Proca fields 
produces an additional contribution to the resonance-induced chiral
Lagrangian of $\cO(p^6)$. In the standard basis for $N_F=3$
\cite{Bijnens:1999sh} we find~:
\begin{eqnarray}  \label{eq:ies}
{\cal L}_{(6)}^{{\rm odd}\times{\rm odd}} \, & = & \, 
\frac{(g_1^V)^2}{2 \, M_V^2} \, \left[ \,
\cO^{(6)}_{42}\, - \, 2 \, \cO^{(6)}_{44} \, - \, \cO^{(6)}_{46} \, + 
\, 2 \, \cO^{(6)}_{47} \, \right] 
\nonumber \\
& & \, + \, \frac{2 \, g_1^V \, g_2^V}{M_V^2} \, \left[ \,
\cO^{(6)}_{48} \, + \, 2 \, \cO^{(6)}_{50} \, - \, \cO^{(6)}_{52} 
\, \right] \nonumber \\
& & \, + \, \frac{2 \, (g_2^V)^2}{M_V^2} \, \left[ \,
\cO^{(6)}_{53} \, + \, \cO^{(6)}_{55} \, - \, 2 \, \cO^{(6)}_{56} 
\, - \, \cO^{(6)}_{59} \ \right] 
\nonumber \\
& & \, + \, \frac{2 \, (g_1^A)^2}{M_A^2} \, \left[ \, \cO^{(6)}_{70} 
\, + \, \cO^{(6)}_{72} \, - \,
2 \, \cO^{(6)}_{73} \, - \, \cO^{(6)}_{76} \, \right] \; .
\end{eqnarray}
The corresponding contributions have been added to the LECs $C_i$ in
Table~\ref{tab:RESCi}.
\par
Some additional remarks are in order here. A short-distance analysis
would be required to investigate whether the Proca-type couplings
$g_1^V$, $g_2^V$ and $g_1^A$ are actually nonzero. In fact, general
quantum field theory (the Froissart theorem applied to $\pi^0$ Compton
scattering \cite{Ecker:1990in}) does indeed require $g_2^V \neq 0$.
The coupling $g_2^V$ can be estimated from $V \to P \gamma$ decays and
it was in fact included in the ${\cal O}(p^6)$ analysis of $\gamma 
\gamma \to \pi^0 \pi^0$ \cite{Bellucci:1994eb}. As already noted, 
antisymmetric tensor field exchange cannot produce the terms proportional
to $(g_2^V)^2$ in Eq.~(\ref{eq:ies}) for purely kinematical reasons. 
In other words, adopting the antisymmetric tensor fields everywhere would
require the explicit addition of the terms $\sim (g_2^V)^2$ to the
resonance Lagrangian. With the role of vector and antisymmetric tensor
fields interchanged, an analogous situation occurs at ${\cal O}(p^4)$
\cite{Ecker:1989yg}. A corresponding short-distance analysis is not yet
available for $g_1^V$ and $g_1^A$.

\section{Short-distance constraints on the resonance couplings}
\label{sec:shortd}
\renewcommand{\theequation}{\arabic{section}.\arabic{equation}}
\setcounter{equation}{0}

QCD imposes severe constraints on both couplings and operators of the 
effective field theory that implement
the strong interactions in the nonperturbative low-energy region.
The chiral symmetry of massless QCD, for instance,
determines the structure of the operators both in $\chi$PT 
and in resonance chiral theory, as we have seen in 
Sec.~\ref{sec:lagrangianR}. To determine the couplings themselves is 
much more involved as it would amount to solve the theory in the 
nonperturbative regime. On the other hand, we know how to describe
QCD at high energies, in its perturbative domain.
As the spectral functions of both vector and axial-vector
current correlators show, the perturbative continuum describes them
reasonably well above the resonance region. Hence we can 
conclude that for $E \gtrsim 2 \, \mbox{GeV}$ we know how to handle, 
both qualitatively and quantitatively, the strong interaction.
\par
Our large-$N_C$ Lagrangian $ \LNCinf$ in Eq.~(\ref{eq:lag1})
is intended to describe the strong interactions in the energy region
of the light-flavour resonances ($M_V \lesssim E \lesssim 2 \, 
\mbox{GeV}$). It is true that the
phenomenologically known spectrum of resonances in this domain is only 
partially represented in $ \LNCinf$, as we are 
neglecting the $\eta'$ (see Ref.~\cite{RK}) and
include the 
lightest nonets of resonances only. However, the inclusion of a more 
complete set of states is a systematic procedure that can be carried 
out in successive steps. In addition, heavier degrees of freedom tend 
to be suppressed by inverse powers of their mass,
for instance in their contributions to the LECs.

\par
In the last years there has been increasing interest in the development
of a hadronic description of the energy region where resonances are 
active degrees of freedom, with different goals in mind. The 
{\em Minimal Hadronic Ansatz} \cite{Peris:1998nj} has unveiled
interesting aspects of the strong dynamics and its role in the 
determination of matrix elements of operators of the effective 
electroweak Hamiltonian.
Studies within two-, three- \cite{Moussallam:1997xx,Knecht:2001xc,
Ruiz-Femenia:2003hm,
Bijnens:2003rc,Cirigliano:2004ue,Cirigliano:2005xn} or even four-point 
functions \cite{Ananthanarayan:2004qk} have
allowed to implement QCD dynamics in different approaches. The common
features of these techniques are, on one side, the use of 
large-$N_C$ ideas and on the other hand, performing a matching
procedure between the resonance region and
the perturbative regime of QCD. The promising phenomenological
results achieved so far encourage us to focus on these two aspects.

Green functions are the 
fundamental objects of any quantum field theory. Here we are only 
interested in the colour-singlet Green functions of QCD currents,
whose short-distance behaviour can be determined within QCD. More
specifically, we concentrate on the spectral functions of two-current 
correlators and on three-point Green functions that we discuss in turn.

\subsection{Spectral functions of two-current correlators}
\label{sec:2current}

Within perturbative QCD the leading-order behaviour of the spectral
functions of two-current correlators is well known (see 
Ref.~\cite{Gasser:1983yg} and references therein). This knowledge
allows, through the use of dispersion relations, the construction
of low-energy theorems in terms of sum rules.
\par
Those spectral functions have also been employed from 
another point of view. In the large-$N_C$ limit,
the QCD result is saturated with an infinite number
of intermediate hadronic states. Hence, from the QCD behaviour one can
extract information on general aspects of the
individual contributions at large momenta and, in consequence,
on the high-energy behaviour of form factors of QCD currents. As an 
example of this last procedure, let us consider the spectral function
of the isovector component of the vector-vector current 
correlator ${\rm Im} \, F_V^{(1)}(t)$.
At leading order in QCD it is known to behave like a constant~:
 ${\rm Im} \, F_V^{(1)}(t) = N_C / 12 \, \pi$ \cite{Floratos:1978jb}.
To recover this result, each of the infinite number
of hadronic contributions to this spectral function is expected to
vanish at high $t$, in particular the two-pion contribution in terms
of the vector form factor of the pion $F_{\pi}(t)$. 
\par
This result for the high-energy behaviour of form factors of QCD 
currents is also known as the Brodsky-Lepage condition \cite{Lepage:1979zb}
although their reasoning involves 
parton dynamics. As a general statement it says that form factors of 
QCD currents should vanish at high momentum transfer. 
This result is well supported phenomenologically and it has been widely
applied \cite{Ecker:1989yg,Pich:2002xy,Amoros:2001gf,GomezDumm:2003ku}. 
Here a question arises concerning form factors with resonances
as asymptotic states. Of course, such form factors are not observable 
quantities. Consequently, phenomenology does not give us any
information about their asymptotic behaviour. On the other hand,
large $N_C$ suggests that these form factors should be
treated on the same footing as those with pseudo-Goldstone bosons 
in the final states because at leading order in the $1/N_C$ 
expansion resonances are stable. However, at this order there should 
also be an infinite number of stable resonances in the theory. Since
we will always limit the number of resonances to a few we will 
adopt the pragmatic point-of-view that the Brodsky-Lepage condition
must be satisfied for form factors with actual asymptotic states but
not necessarily for form factors of resonances. One exception is the 
case of the $\langle SPP \rangle$ Green function where we consider 
also pion-to-resonance transition form factors.    
\par
Using these and analogous ideas on the high-energy behaviour of 
form factors and scattering amplitudes we now come back to 
Eq.~(\ref{eq:hec}) and explain why the couplings of those operators
must vanish. 
\begin{itemize}
\item The contribution of $\cO^S_{17}$ to the scalar form factor 
$F^{i j}_S (t)$ (where $i,j$ are flavour indices)
grows linearly with $t$ for large $t$. This is
inconsistent with the quark counting rules unless $\cO^S_{17}$ is absent.
Moreover, by analysing the correlator $\Pi_{SS - PP}^{ij} (t)$,
one sees that both $\cO^S_{18}$ and 
$\cO^P_{13}$ contribute, for high $t$, terms linear in $t$ and constant. 
The OPE implies that the correlator goes like $1/t^2$ for high
$t$. Setting to zero the linear and constant terms gives conditions
that are satisfied only if both operators $\cO^S_{18}$ and
$\cO^P_{13}$ are absent. 
\item The terms $\cO^V_{21}, \cO^V_{22}$ and $\cO^A_{17}$ can be
discarded by using exactly the same high-energy constraints as in
Ref.~\cite{Ecker:1989yg}. Starting with elastic meson-meson
scattering in the forward direction, $\cO^V_{21}$ is inconsistent with
a once-subtracted dispersion relation. Once this term is dropped,
the vector pion form factor requires $\cO^V_{22}$ to be absent.
Finally, the unsubtracted dispersion relation for the left-right 
two-point function then eliminates $\cO^A_{17}$. 
\end{itemize}
Altogether we therefore have
\begin{eqnarray} \label{eq:fffff}
\lambda_{13}^P & = & 0 \, , \nonumber \\
\lambda_{17}^S = \lambda_{18}^S & = & 0 \, , \nonumber \\
\lambda_{17}^A & = & 0 \, , \nonumber \\
\lambda_{21}^V = \lambda_{22}^V & = & 0 \, . 
\end{eqnarray}
For the parameters of the leading-order resonance Lagrangian the same type of requirements leads to the well-known conditions
\cite{Ecker:1989yg, Pich:2002xy,Weinberg:1967kj,Golterman:1999au,Jamin:2000wn,
Jamin:2001zq} 
\begin{eqnarray}
\label{eq:after fffff}
F_V \, G_V =  F^2 \, ,  & \hspace*{3cm} & 
F_V^2 \, - \, F_A^2  =  F^2 \, ,\no
F_V^2 \, M_V^2  =  F_A^2 \, M_A^2 \, ,  & \hspace*{3cm} &
 \\
4 c_d c_m=  F^2 \, ,  & \hspace*{3cm} & 
8 (c_m ^2- d_m^2)   =  F^2 \fs \non 
\end{eqnarray}

\subsection{Three-point functions of QCD currents}

\begin{figure}
\begin{center}
\includegraphics[scale=1]{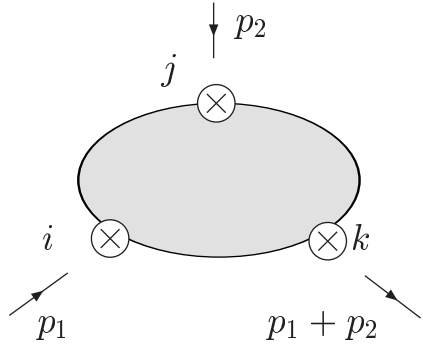}
\caption[]{\label{fig:ijk}
Momentum configuration for the Green functions $\Pi_{123}^{ijk}$.}
\end{center}
\end{figure}

The Green functions of interest are 
\begin{equation}
\Pi_{123}^{ijk}(p_1,p_2) \, = \, i^2 \, \int \, d^4x \, d^4y \, 
e^{i \, \left( p_1 \cdot x \, + \, p_2 \cdot y \right)} \, 
\left\langle \, 0 \, \left| \, T \,\left\{ \, J_1^i(x) \, 
 J_2^j(y) \, J_3^k(0) \, \right\} \, \right| \, 0 \, 
\right\rangle \; , 
\end{equation}
where the QCD currents $J_a^i$ are defined as
\begin{equation}
S^i \,  =   \, \overline{q} \, \lambda^i \, q \; \; \; , \qquad \;   
P^i \,  =   \, \overline{q} \, i \, \gamma_5 \, \lambda^i \, q \; \; \; ,
\qquad \;  
V_{\mu}^i \,  =  \, \overline{q} \, \gamma_{\mu} \, \frac{\lambda^i}{2} \, 
q  \; \; \; , \qquad \; 
  A_{\mu}^i \,  =  \, \overline{q} \, \gamma_{\mu} \, 
\gamma_5 \, \frac{\lambda^i}{2}\, q \; ,
\end{equation}
with the normalization $\langle \lambda^i \lambda^j \rangle \, = 
\, 2 \, \delta^{ij}$. The momenta are assigned as shown in 
Fig.~\ref{fig:ijk}.
We are particularly interested in those Green functions that are 
order parameters of the spontaneous breaking of chiral symmetry. They 
do not receive contributions from perturbative QCD at large momentum
transfer in the chiral limit. As a consequence, their behaviour at
short distances is smoother than expected on purely dimensional
grounds. 
\par
Chiral Ward identities, discrete and $SU(3)$ symmetries constrain
the structure of these Green functions \cite{Gerstein:1969cx}. Their 
short-distance behaviour can be determined within perturbative QCD 
in terms of an OPE for different kinematical regimes, 
namely $\Pi(\lambda p_1,p_2)$, $\Pi(p_1, \lambda p_2)$,
$\Pi( \lambda p_1, \lambda p_2)$ and $\Pi(\lambda p_1, p_2 - \lambda p_1)$,
for large $\lambda$. We will only consider the leading orders both 
in $1/\lambda$ and in the perturbative expansion of QCD. Consequently, 
our results hold up to ${\cal O}(\alpha_S)$ corrections.

The procedure is then straightforward. We first compute
the corresponding three-point Green functions within our 
approach based on the Lagrangian $ \LNCinf$.
For the different kinematical
regimes specified above, we then match the results with those of the
OPE. In this way we obtain information on the couplings of the 
Lagrangian and therefore on the LECs $C_i$. The specific details of 
the matching have to be worked out in each case 
\cite{Moussallam:1997xx,Knecht:2001xc,Ruiz-Femenia:2003hm,
Cirigliano:2004ue,Cirigliano:2005xn}.

\section{Resonance exchange for $\mathbf{\langle V A P \rangle}$ and 
  $\mathbf{\langle S P P \rangle}$ correlators}
\label{sec:Rex}
\renewcommand{\theequation}{\arabic{section}.\arabic{equation}}
\setcounter{equation}{0}
In this section we reanalyse the three-point functions
$\langle V A P \rangle$ and  $\langle S P P \rangle $ in the present
framework. In previous treatments, the chiral resonance approach was
either not used at all \cite{Cirigliano:2005xn} or with a selected
set of couplings only \cite{Cirigliano:2004ue}. 

A virtue of the resonance Lagrangian framework lies in the fact that 
chiral symmetry is built in from the start. As a consequence the
chiral symmetry relations arising at special kinematical points hold 
automatically and, moreover, the corresponding high-energy behaviour
is also inherited. An example is the relation between the $\langle SPP  
\rangle$ correlator and the $\langle SS- PP  \rangle$ two point
function discussed in \cite{Cirigliano:2005xn}. In the framework of
Ref.~\cite{Cirigliano:2005xn}, the relation between $\langle SPP  
\rangle$ and $\langle SS- PP  \rangle$ leads to independent constraints 
on the chosen ansatz. Here, not only is the relation  
automatically satisfied but also the high-energy behaviour is correct
once the proper high-energy behaviour of the two-point function has
been implemented. The relation between the $\langle VA |\pi   \rangle$ 
matrix element and the two-point function $\langle VV-AA \rangle$ is a
similar case \cite{Moussallam:1997xx,Knecht:2001xc}. This should also
be of great advantage for the study of four- and higher-point
functions. For this reason the relations in Eqs.~(\ref{eq:fffff}) and 
(\ref{eq:after fffff}) will be used throughout the present section. 

The role of the terms that have been removed by use of field 
redefinitions is less clear a priori. However, in the examples 
considered below we demonstrate explicitly that the omission of these 
terms can be justified by the asymptotic constraints.

\subsection{$\langle V A P \rangle $ Green function}

 Chiral Ward identities, $SU(3)_V$, parity and time reversal 
\cite{Moussallam:1997xx,Knecht:2001xc} provide the general expression
for the $\langle V A P \rangle $ Green function~\footnote{Note that 
our convention for the correlator differs by a factor of (-2) compared 
to Refs.~\cite{Knecht:2001xc,Cirigliano:2004ue}, whereas the 
definitions of the  functions ${\cal F}$ and ${\cal G}$ coincide.}:
\begin{eqnarray}
\label{eq:GeneralSolution}
(\Pi_{V\!AP})_{\mu\nu}^{ijk}(p,q) & = & (-2)
  f^{ijk}\,\Bigg\{ \, \langle{\overline\psi}\psi\rangle_0 
\left[\frac{(p+2q)_\mu q_\nu}{q^2 (p+q)^2}
 - \frac{g_{\mu\nu}}{(p+q)^2}  \right] 
\nonumber \\  
&&  + \, P_{\mu\nu}(p,q) \, {\cal F}(p^2,q^2,(p+q)^2) \, 
+ \, Q_{\mu\nu}(p,q) \,  {\cal G}(p^2,q^2,(p+q)^2) \Bigg\} \, ,
\end{eqnarray}
where the transverse tensors $P_{\mu \nu}$ and $Q_{\mu \nu}$ are 
defined by~:
\begin{eqnarray}
\label{eq:PQdef}
P_{\mu\nu}(p,q) &=& q_\mu p_\nu - (p \cdot q) g_{\mu\nu} \, , \nonumber \\ 
Q_{\mu\nu}(p,q) &=& p^2 q_\mu q_\nu + q^2 p_\mu p_\nu - (p \cdot q)
p_\mu q_\nu - p^2 q^2 g_{\mu\nu} \, .  
\end{eqnarray}

The $\langle V A P \rangle$ correlator was studied in 
Ref.~\cite{Cirigliano:2004ue} in a resonance
Lagrangian framework. In this context the functions $ \cal F $ and 
$\cal G$ were found to be of the general structure 
\begin{align}
{\cal F}(p^2,q^2,(p+q)^2) & = 
\frac{ \left\langle \overline{\psi} 
\psi\right\rangle _{0}}{(p^2-M_V^2)(q^2-M_A^2)} \, 
\left[  a_0 \, + \frac{b_1 \, + \, 
b_2 \,p^2 + \, b_3 \, q^2}{(p+q)^2}  + \frac{c_1\, + \, c_2 \, p^2 
\, + \, c_3 \, q^2}{(p+q)^2-M_P^2} \right] \,  , 
\nonumber \\ 
{\cal G}((p^2,q^2,(p+q)^2)) & = \frac{\left\langle \overline{\psi} 
\psi\right\rangle _{0}}{(p^2-M_V^2)q^2}
\, \left[ \frac{d_1 \, + \, d_2\,q^2}{(p+q)^2 \, (q^2- M_A^2)} 
\, + \, \frac{f}{(p+q)^2 - M_P^2}  \right] \, .
\end{align}
\begin{figure}
\begin{center}
\includegraphics[scale=0.85]{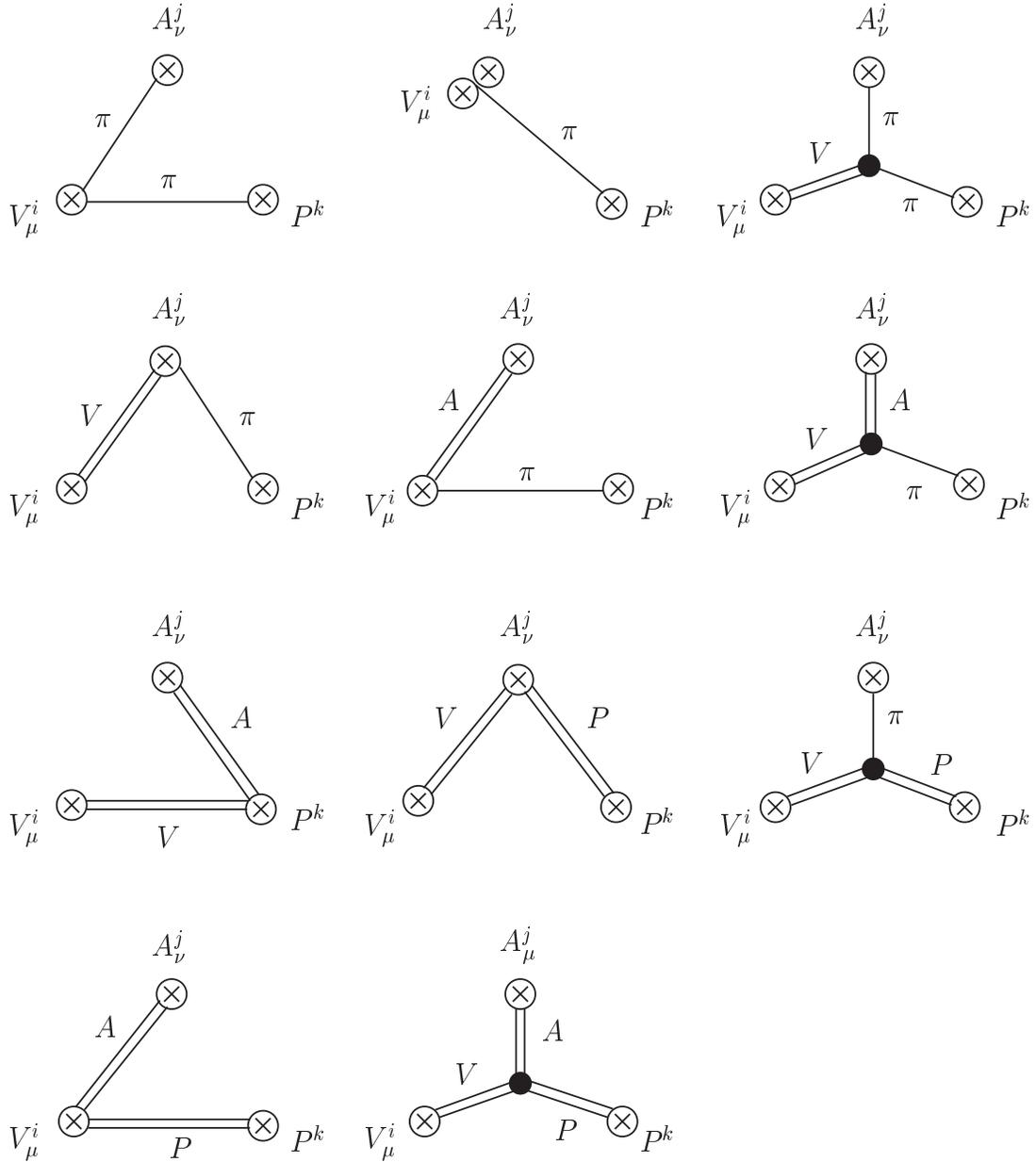}
\caption[]{\label{fig:vapr}
Feynman diagrams contributing to $\Pi_{VAP}$ in the minimal resonance
Lagrangian. $\pi$ stands for a Goldstone boson, $V$, $A$ and $P$
denote vector, axial-vector and pseudoscalar resonances, respectively.}
\end{center}
\end{figure}
The comparison with the present resonance Lagrangian in its minimal
form according to subsection~\ref{sec:minimal} shows that the analysis
of Ref.~\cite{Cirigliano:2004ue} included all relevant terms with
the exception of the three-resonance coupling $ \lambda^{VAP} $. 
In Fig.~\ref{fig:vapr} we show the Feynman diagrams derived from the
resonance Lagrangian of Sec.~\ref{sec:lagrangianR} in its minimal form.
Since we are also using the same notation for the couplings in the
resonance Lagrangian, the results for the coefficients 
$ a_0 , \ldots,  f $  may simply be taken over from
Ref.~\cite{Cirigliano:2004ue} except in the case of 
$c_1$ which receives an additional contribution proportional
to $\lambda^{VAP}$,
\begin{align}
c_1 & =  -c_2  M_V^2 -  c_3  M_A^2\, + \, 8 \, \lambda^{VAP} \, 
{F_V F_A d_m }/{F^2} \; . 
\end{align} 
The various short-distance conditions discussed in 
Ref.~\cite{Cirigliano:2004ue} impose constraints on all but precisely 
this coefficient, 
\begin{align}
\label{eq:abcdef}
2 a_0  = - b_2 = 2 c_2 = 2c_3 = f = -1  
\co  \quad 
b_3 = d_2 = 0 
\co \quad 
b_1 = \MA^2 -  \MV^2 
\co \quad 
d_1 = 2 \MA^2 ~,
\end{align}
implying that the conclusions of Ref.~\cite{Cirigliano:2004ue} 
remain unaffected. The predictions for the chiral LECs
may be expressed in terms of the coefficients $ a_0 , 
\ldots,  f$. Therefore they coincide with the ones of Ref.~\cite{Cirigliano:2004ue}, with the exception of the 
coupling constant $C_{82}$ that receives a contribution from 
$\lambda^{VAP}$: 
\begin{align}
C_{82} = - \frac{F^2 (5 \MV^2 +4 \MA^2)}{32 \MV^4 \MA^2} - 
\frac{F^2}{32 \MA^2 \MP^2} - \frac{\FV \FA d_m}{2 \MV^2\MA^2 \MP^2}\, 
\lambda^{VAP} \fs
\end{align}
The predictions for this and the other LECs are of course contained in 
Table~\ref{tab:RESCi} of the present work. 
Taking into account the
restrictions on the resonance couplings imposed by Eq.~(\ref{eq:abcdef}), we find,
as in Ref.~\cite{Cirigliano:2004ue}:
\begin{align}
\sqrt{2} \lambda_0  =-4\lambda_1^{VA}-\lambda_2^{VA} -
\frac{\lambda_4^{VA}}{2}-\lambda_5^{VA} & =
\frac{1}{2\sqrt{2}}(\lambda' + \lambda'') 
\co \\
\sqrt{2} \lambda'  =  \lambda_2^{VA}-\lambda_3^{VA} +
\frac{\lambda_4^{VA}}{2}+\lambda_5^{VA} &= \frac{M_A}{2 M_V}
\co \no
\sqrt{2} \lambda''  =  \lambda_2^{VA} -\frac{\lambda_4^{VA}}{2}-
\lambda_5^{VA}& = \frac{M_A^2 - 2 M_V^2}{2 M_V M_A}
\co \no
 \lambda_1^{PV} = - 4 \lambda_2^{PV} = -\frac{F 
\sqrt{M_A^2 - M_V^2}}{4 \sqrt{2} d_m M_A} \co \quad  \lambda_1^{PA} 
&=  \frac{F \sqrt{M_A^2 - M_V^2}}{16 \sqrt{2} d_m  M_V} 
 \co \non
\end{align}
where the relations in Eq.~(\ref{eq:after fffff}) have been used. 
Inserting these relations in Table~\ref{tab:RESCi} one recovers the 
predictions for the coupling constants $C_{78}$, $C_{87}$,   $C_{88}$, 
$C_{89}$ and $C_{90}$ in Ref.~\cite{Cirigliano:2004ue}.      

One may ask the question what would have become of these results if 
instead one had performed the calculation with the full resonance 
Lagrangian before the field redefinitions. We have in fact performed 
this calculation and the answer to the question is simple: the above 
result for the $\langle VAP \rangle$ correlator remains valid also in 
this case. The reason is that the potential additional 
contributions also lead to conflicts with the OPE and are thus
required to vanish.

\subsection{$\langle S P P \rangle$ Green function}

In Ref.~\cite{Cirigliano:2005xn} the $\langle S P P  \rangle$ Green
function was analysed with a general meromorphic ansatz to comply
with the large-$N_C$ limit of QCD. 
From $SU(3)_V$ and C invariance, the $\langle S P P \rangle$ Green
function is given in terms of a single scalar function:
\begin{equation}
\Pi_{SPP}^{ijk}(p,q)  \, = \, d^{ijk} \, 
\Pi_{SPP}\left(p^2, q^2, (p+q)^2\right) \; . 
\end{equation}
Bose symmetry implies that the function is symmetric in its second
and third arguments. According to the results of Ref.~\cite{Cirigliano:2005xn} 
it is of the form 
\begin{align}\label{SPP-model}
\Pi_{SPP}(s,t,u)  = 8 B^3 F^2 \MS^2 \MP^4 
\frac{1+P_1+P_2+P_3+P_4}{[\MS^2 -s][-t][-u][\MP^2-t][\MP^2-u]} \; ,
\end{align}
where the $P_n$ are polynomials of degree $n$ in the variables $s$,
$t$ and $u$
\begin{align}
 \label{eq:poly}
P_1 & = c_{010}(t+u) \; , \\
P_2 & = c_{011}tu  \; ,  \no 
P_3 & = [c_{111}s+c_{021}(t+u)]tu \; ,  \no
P_4 & = [c_{211}(s^2-(t-u)^2)+c_{121}(s(t+u)-(t-u)^2)]tu \; .  \nonumber 
\end{align}
As discussed in Ref.~\cite{Cirigliano:2005xn}, the restrictions on the 
form of the polynomials arise from the requirements that the
asymptotic behaviour of the function $\Pi_{SPP}$ in the various limits 
be no worse than what follows from the OPE and that the scalar and 
pseudoscalar (transition) form factors with at least one pion vanish 
asymptotically. 
\begin{figure}
\begin{center}
\includegraphics[scale=0.85]{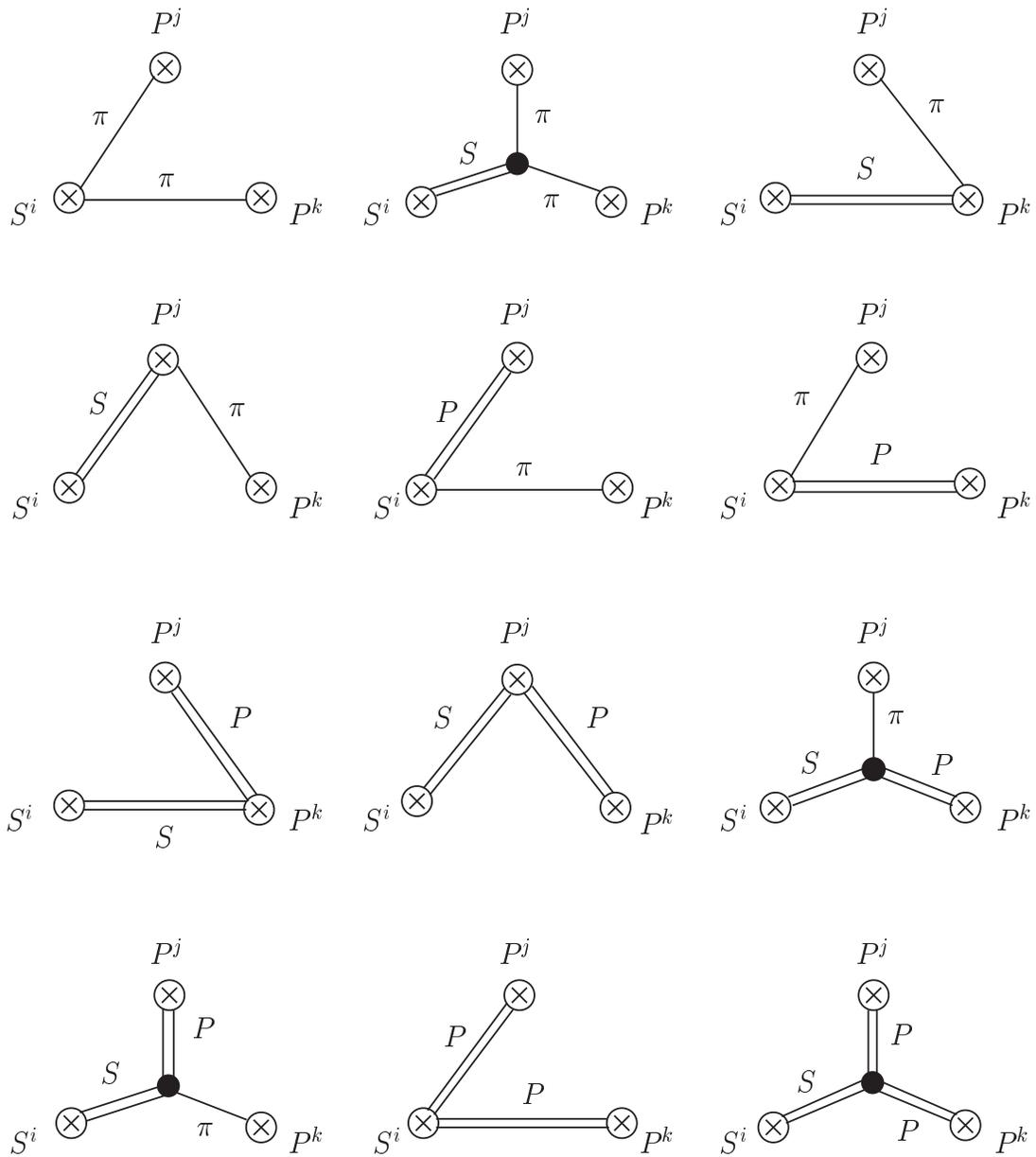}
\caption[]{\label{fig:sppr}
Feynman diagrams contributing to $\Pi_{SPP}$ in the minimal resonance
Lagrangian. $\pi$ stands for a Goldstone boson, $S$ and $P$
denote scalar and pseudoscalar resonances, respectively.}
\end{center}
\end{figure}
The computation within the chiral resonance framework in its minimal 
version according to subsection~\ref{sec:minimal}
is straightforward. The relevant diagrams are given
in Fig.~\ref{fig:sppr} and they yield the result
\begin{eqnarray}\label{eq:sppreso}
\Pi_{SPP}(p_1^2,p_2^2, q^2)  &= & -  \, 64 \,  B^3  \,  \Bigg\{ \, 
\frac{- F^2}{8 \, q^2 \, p_2^2} 
 + \frac{c_m^2}{p_1^2 - M_S^2} \left(\frac{1}{q^2} + \frac{1}{p_2^2} \right)
\nonumber \\[.2cm]  
&  & - \, d_m^2 \left( \frac{1}{q^2 \left(p_2^2 - M_P^2 \right)}  + 
\frac{1}{p_2^2 \left(q^2 - M_P^2 \right)}  
\right) - \frac{c_m \, c_d \,q \cdot p_2 }{q^2  p_2^2 
 \left(p_1^2 - M_S^2 \right)} 
\nonumber \\[.2cm] 
&  &  + \frac{2 \, c_m \, d_m \, \lambda^{SP}_1}{p_1^2 - M_S^2} 
\left( - \frac{q \cdot p_1}{q^2 \left(p_2^2 - M_P^2 \right)} 
+ \frac{p_1 \cdot p_2}{p_2^2 \left(q^2 - M_P^2 \right)} \right)  
\nonumber \\[.2cm] 
&  & + \frac{4 c_m \,d_m \, \lambda^{SP}_2 }{p_1^2 - M_S^2} 
\left(\frac{1}{q^2 - M_P^2} + 
\frac{1}{p_2^2 - M_P^2} \right) 
\nonumber \\[.2cm] 
&  &  + \frac{4 \,d_m^2 \, \lambda^{PP}_3}{\left( q^2 - M_P^2 \right) 
\left(p_2^2 - M_P^2 \right)} 
- \frac{4 \, c_m \, d_m^2 \, \lambda^{SPP}}{\left( p_1^2 - M_S^2
\right)  \left( q^2 - M_P^2 \right) \left(p_2^2 - M_P^2 \right)}  
  \Bigg\} \, . 
\end{eqnarray}
In addition to the couplings from the Lagrangian ${\cal L}_{(2)}^{R}$
in Eq.~(\ref{eq:R_int}), $\Pi_{SPP} $ is
found to depend on the resonance couplings $\lambda_i^{SP}~(i=1,2)$,
$\lambda_3^{PP}$ and $\lambda^{SPP}$. Inspection reveals that this 
result is of the desired form, up to a contribution $\propto  
(\lambda^{SP}_1 +d_m/c_m) (s(t+u)-t^2 -u^2) $ to $P_2$ that is in 
conflict with demanding that the scalar form factor
involving a Goldstone boson and a pseudoscalar resonance fades away
at high momenta. The conflict is resolved by the condition 
\begin{equation} 
\lambda_1^{SP} = - \displaystyle\frac{d_m}{c_m}~. 
\end{equation} 
Inserting this relation in Table~\ref{tab:RESCi} along with the 
relations in Eq.~(\ref{eq:after fffff}) one recovers the predictions 
for $C_{12}$, $ C_{34} $ and $C_{38}$ from
Ref.~\cite{Cirigliano:2005xn}. 

In the resonance Lagrangian approach a new feature arises. The OPE 
expansion for the 
$\langle S P P \rangle$ function demands that its leading term
behaves as 
$\Pi_{SPP} (\lambda^2 s, \lambda^2t ,\lambda^2 u ) 
\sim {\cal O} (1/\lambda^2)$, while one can see that our result
of Eq.~(\ref{eq:sppreso}) is of ${\cal O}(1/\lambda^4)$.
This should come as no surprise since the terms with the potential to 
generate ${\cal O}(1/\lambda^2)$ contributions to the $ \langle 
SPP\rangle$ Green function ($\lambda^S_{8,14}$ and $
\lambda^P_{6,10}$) have been discarded when constructing the minimal 
resonance Lagrangian. It turns out, however, that the attempt to
generate $c_{211},c_{121}  \neq 0$ by retaining those terms is bound 
to fail because the restrictions imposed by Eq.~(\ref{eq:poly})
require $\lambda^S_{8} = \lambda^S_{14} =  \lambda^P_{6} = 
\lambda^P_{10} =0 $ nonetheless. (Note that there arise contributions 
proportional to $\lambda^S_8 (t^3+u^3)$ and $\lambda^P_6 s^2(t+u)$ in
$P_3$.)    

It is relevant to notice that, as already commented in 
Ref.~\cite{Cirigliano:2005xn}, the procedure we are devising does not
apply to all Green functions with the same settings. Contrary to the
$\langle VAP \rangle$ case, higher-order corrections furnish the 
Wilson coefficients of the $\langle S P P \rangle$ function with 
anomalous dimensions and therefore the matching in the 
$N_C \rightarrow \infty$ limit that we intend to enforce (with a finite
number of meson states in the spectrum) is not fully feasible in the
$\langle S P P \rangle$ case. Thus we take the softer approach of 
demanding that asymptotically our Green functions behave no worse
than given by the OPE expansion at leading order, as 
implemented in Ref.~\cite{Cirigliano:2005xn}.

\section{Conclusions}
The LECs of $\chi$PT encode the dynamical information on the massive
hadronic states of QCD, which are not present explicitly in the
effective Goldstone Lagrangian. The most important contributions to
the LECs are expected to originate from the low-lying mesonic
resonances because contributions from heavier hadrons are
suppressed by inverse powers of their masses.

In the large-$N_C$ limit of QCD,
the correlators of colour-singlet quark-antiquark currents are given
by tree-level exchanges of infinite towers of narrow mesons.
Crossing and unitarity imply that these sums correspond to the
lowest-order approximation of some effective mesonic Lagrangian. The
construction of this general Lagrangian, with an infinite number of
hadronic states, is beyond our present abilities. However,
truncating the hadronic spectrum to the lowest-lying multiplets with
$J^{PC}=1^{--},1^{++},0^{-+}$ and $0^{++}$, one obtains a very good
approximation at low energies. The couplings of this resonance
Lagrangian should be determined by imposing that the corresponding
Green functions reproduce the asymptotic behaviour of
QCD${}_\infty$.
Integrating {out} all massive fields at the level of the generating
functional, one obtains the low-energy effective Lagrangian of the
Goldstone modes. Therefore, from the resonance chiral Lagrangian one
can determine the LECs of $\chi$PT as functions of resonance
parameters.

In this paper we have constructed the most general $SU(3)_L \times
SU(3)_R$ invariant Lagrangian, containing Goldstone bosons and the
lowest-lying $J^{PC}=1^{--},1^{++},0^{-+}$ and $0^{++}$ multiplets,
that contributes to ${\cal L}_6^{\rm \chi PT}$ after integrating out
the resonance fields. Since any resonance exchange involves a
suppression factor $1/M_R^2$ from the resonance propagator, we only
need to consider terms involving one, two or three resonance fields,
coupled to chiral monomials of $\cO(p^4)$, $\cO(p^2)$ and
$\cO(p^0)$, respectively. This can also be seen by expanding the
classical equations of motion of the resonance fields in inverse
powers of the heavy masses, which shows that for $p^2\ll M_R^2$ the
resonance fields scale as $R\sim\cO(p^2/M_R^2)$.

The number of possible chiral structures is rather large. Using
partial integration, the lowest-order equations of motion and
algebraic identities to eliminate linearly dependent terms, we have
identified a basis of 70 monomials in $\cL^R_{(4)}$, 38 in
$\cL^{RR}_{(2)}$ and 7 in $\cL^{RRR}_{(0)}$. This basis is, however,
highly redundant for the purpose of determining the LECs
of $\cO(p^6)$ because only certain combinations of the resonance
couplings occur in the LECs. This can be understood through
appropriate redefinitions of the resonance fields that leave
the generating functional invariant. We have found a large number of
linear field redefinitions of the type (\ref{eq:redef1}),
allowing us to eliminate 47 terms of the resonance
Lagrangian $\cL^R_{(4)}$ by transforming them into structures of
higher chiral order that do not contribute to ${\cal L}_6^{\rm \chi
PT}$. We have chosen to eliminate preferentially terms with
$\chi$'s, while keeping those with many derivatives. 
Higher-derivative structures generate a worse short-distance behaviour 
and, therefore, will be most easily eliminated through high-energy QCD
constraints. In fact, six surviving operators can be discarded because
they induce an unacceptable high-energy behaviour of two-point functions.
In addition, there are two
constants in $\cL^R_{(4)}$ that only lead to contributions subleading 
in $1/\nc$. We have finally kept a total
of 15 terms in $\cL^R_{(4)}$: four vector, three axial-vector, five
scalar  and three pseudoscalar operators.

Once the relevant resonance Lagrangian has been determined, we have
performed the functional integration of the resonance fields and
obtained their contribution to the LECs of  $\cO(p^6)$.
The results, given in App.~\ref{app:resultsCI}, still show the presence
of redundant terms; many resonance couplings appear only in definite
combinations. Again, this can be understood through additional
non-linear field redefinitions of the type (\ref{eq:redef3}), which
could be used to eliminate all seven trilinear couplings and six
bilinear operators. One further bilinear coupling can be dismissed on
the basis of large $\nc$. Altogether, 46 independent combinations of
resonance couplings appear in the LECs of $\cO(p^6)$.

We have adopted the usual chiral formulation of spin-1 fields in
terms of antisymmetric tensors. For completeness, we have
added possible contributions to ${\cal L}_6^{\rm \chi PT}$ from
odd-parity vector and axial-vector terms, with $\cO(p^3)$ chiral  
monomials in the Proca field formulation. This introduces three
additional couplings, not present in the antisymmetric formalism at
this order in the momentum expansion. The equivalence of different
formalisms for vector fields, once short-distance constraints are
taken into account, was demonstrated at $\cO(p^4)$ in 
Ref.~\cite{Ecker:1989yg}. One of the three Proca-type couplings
contributing at $\cO(p^6)$ is already known to be nonvanishing
\cite{Ecker:1990in}. In the antisymmetric tensor formulation, 
corresponding local terms of $\cO(p^6)$ would have to be added to
the chiral resonance Lagrangian. For the other two couplings, 
a short-distance analysis remains to be done. 

The large-$N_C$ counting is more transparent in the $U(3)_L \times
U(3)_R$ effective theory, where multiple-trace terms are suppressed
by corresponding powers of $1/N_C$. However, one needs then to
consider the special role of the $U(1)_A$ anomaly, which is a very
important physical effect not present at leading order in $1/N_C$.
This can be incorporated through a more involved counting in powers
of $p^2\sim m_q \sim 1/N_C$. Integrating {out} the singlet Goldstone field,
one recovers the more standard $SU(3)_L \times SU(3)_R$ chiral
framework. 
This procedure is sketched in App.~\ref{sec:multiple}, 
hereby elucidating the role of the multiple-trace terms encountered 
in our resonance Lagrangian. 
A more systematic analysis of the
$\eta'$ contributions to the $\cO(p^6 )$ LECs will be given in
\cite{RK}.

The complete list of resonance contributions to the chiral couplings 
of $\cO(p^6)$ constitutes our main result. It shows which LECs
are not sensitive at all to resonance exchange and, therefore, may
be expected to be negligible. Moreover, there are interesting
relations among different couplings, which can be useful for
phenomenological applications,
e.g., $C_{20}= -3 C_{21}=C_{32}=C_{35}/6=C_{94}/8$ or $C_{24}= 6 C_{28}=
3 C_{30}$, etc.
In a few cases, e.g.,  $C_1 +4 C_3$, $3C_1-4 C_4$, $C_{12}$, 
$C_{1}/12-C_{28}+r_S C_{32}$, $C_{38}$,
$C_{87}$,  $C_{88}- C_{90}$, $C_{91}$,
$C_{93}$,  etc., 
the $\cO(p^6 )$ LECs only depend on resonance couplings already 
present at $\cO(p^4)$.

Our program towards a systematic analysis of resonance contributions
still involves a missing step: the determination of the resonance
couplings through appropriate short-distance QCD constraints. These
couplings could be fixed by enforcing a systematic matching
procedure between the Green functions of QCD${}_\infty$ and their
corresponding correlators in the effective low-energy theory. The
matching cannot be exact, i.e. it is not possible to perform it for
all possible Green functions, because that would require to
introduce an infinite number of hadronic states. Nevertheless, it is
certainly possible to accomplish a matching good enough to correctly
describe a wide set of interesting physical observables. We have
shown two known examples, the $\langle VAP\rangle$
and $\langle SPP\rangle$ three-point functions, which provide very
useful constraints on some low-energy couplings. A thorough study of
other three-point correlators is under way.

\section*{Acknowledgements}
We are grateful to Marc Knecht for a discussion on the relevance
of the Proca terms. We also thank Santiago Peris and Eduardo de Rafael
for a very helpful correspondence. 
J.P. wishes to thank Gabriel Amor\'os for many interesting discussions on the
role of the Resonance Chiral Theory. This work has been
supported in part by
MCYT (Spain) under grant FPA2004-00996, by Generalitat Valenciana
(Grants GRUPOS03/013, GV04B-594 and GV05/015) and by ERDF funds from the
European Commission.
R.K. was supported by the Swiss National Science Foundation.

\appendix
\renewcommand{\theequation}{\Alph{section}.\arabic{equation}}
\renewcommand{\thetable}{\Alph{section}.\arabic{table}}
\setcounter{equation}{0}
\setcounter{table}{0}

\section{Multiple-trace terms and \mbox{\boldmath$\eta^\prime$} exchange}
\label{sec:multiple}

The form of the EOM (\ref{eq:eom2}) implies that the number of traces
is not conserved over the course of constructing the effective
Lagrangian. In principle, this could be circumvented at the cost of 
ignoring the EOM and writing the Lagrangian in terms of 
$\nabla_\mu u^\mu $. More fundamentally, this circumstance reflects 
the fact that the counting of traces in the effective theory is in
general not in direct correspondence with the order in $1/\nc$ of a 
term \footnote{There are cases where such a mismatch is introduced 
artificially by using Cayley-Hamilton identities to trade terms with 
fewer traces for terms with more traces. If desired, this is repaired 
easily and shall not be our concern here.}.

This discrepancy is generated by the occurrence of an intermediate 
singlet pseudoscalar (the $\eta'$) which in the present context is 
treated as massive, as is reflected by the Goldstone manifold being 
$SU(3)$. To arrive at a classification of contributions with respect to 
their order in $1/\nc$ one should instead start from a Lagrangian 
which involves the singlet field as an explicit degree of freedom. In 
this case the leading-order Lagrangian reads \cite{LeffU(3)}
\begin{equation} 
\tilde{{\cal L}} = \frac{{F}_0^2}{4} \langle  \tilde{u}_\mu
\tilde{u}^\mu +   \tilde{\chi}_+  \rangle  +  \frac{{F}_0^2}{3}  M_0^2
\ln^2 (\det \tilde{u})  ,
\label{eq:U(3)}
\end{equation}  
where the tildes refer to building blocks made of an effective field 
$ \tilde{u}(x) \in U(3) $. The additional degree of freedom,  
$\ln(\det \tilde{u})$, comes with a nonzero chiral limit mass $M_0$ 
that prevents it from causing a `$U(1)$ problem' 
\cite{U(1)GoldstoneBoson}. It is well known, however, that this mass 
vanishes in the large-$\nc$ limit, $M_0^2 = {\cal O}(1/\nc)$. The structure
of the above effective Lagrangian implies a balance between the scales 
set by the momenta, quark masses and $1/\nc$  ,
\begin{align}
p^2 \sim m_q \sim M_0^2 \, . 
\end{align}    
The 
validity of the theory relies on the assumption that all of these 
scales are small in comparison to the intrinsic scale of QCD. While 
that is a nontrivial assumption, the benefit lies in the fact that the 
large-$\nc$ counting rules are now `canonical': terms with single
traces are of order $\nc$ while additional traces reduce the order in 
$1/\nc$ by unity. Factors of $\ln (\det \tilde{u})$ also lead to a 
suppression in $1/\nc$, which can be understood in a framework where 
singlet external fields are present 
\cite{Kaiser:2000gs,Kaiser:2005eu,HerreraandCo}. 

Contact with the standard effective theory is established when
treating the mass $ M_0^2$ as large in comparison to the octet meson 
masses and momenta squared, such that by the EOM
\begin{align}
\tilde{u} = u  \{1 + \frac{1}{12 M_0^2} \langle \chi_- \rangle + 
{\cal O}(p^4)  \} , 
\end{align} 
with $\det u = 1$. 
 Proceeding in this manner, one recovers the 
well-known $\eta'$ contribution to the coupling constant
$L_7$~\cite{Gasser:1984gg},
\begin{align}
L_7^{(\eta')} = -\frac{F_0^2}{48 M_0^2} \, .  
\end{align}
As emphasized in Refs.~\cite{Gasser:1984gg,Peris:1994dh}, some care
is needed in performing the large-$\nc$ limit for $L_7$, that is still
an open question. We refer
to Refs.~\cite{Leutwyler:1997yr,RK} for detailed 
expositions in which sense $L_7$ can be counted as ${\cal O}(N_C^2)$
even though it is the coefficient of the double-trace term
$ \langle \chi_- \rangle^2 $.
On the other hand, 
contributions to $L_1- \frac{1}{2} L_2 $, 
$L_4$ and $L_6$ do not occur. These LECs are therefore booked
as ${\cal O}(1)$ at large-$\nc$, in accordance with the double-trace
structure of the corresponding monomials in the chiral $SU(3)$ 
Lagrangian of ${\cal O}(p^4)$~\cite{Gasser:1984gg}.    

Similarly, one also has the possibility to set up a resonance
Lagrangian in the framework where the singlet field is explicitly 
present \cite{Kaiser:2005eu}. Apart from additional terms that
involve factors of $\ln (\det \tilde{u})$ one has terms of the same 
form as those in Eq.~(\ref{eq:R_int}). Proceeding as  above one 
generates the following multiple-trace terms,  
\begin{align}
c_m \langle S \tilde{\chi}_+ \rangle  & \rightarrow c_m \langle S 
{\chi}_+ \rangle - \frac{c_m}{6 M_0^2}  \langle S {\chi}_- \rangle 
\langle \chi_- \rangle + \ldots
\label{eq:mtrace}
\\ \nonumber
c_d \langle S  \tilde{u}_\mu \tilde{u}^\mu  \rangle  & \rightarrow  
c_d \langle S  {u}_\mu {u}^\mu  \rangle  + \frac{c_d}{3 M_0^2} i  
\langle Su_\mu \rangle \partial^\mu \langle \chi_-  \rangle + \ldots
\\ \nonumber 
i \, d_m \langle P \tilde{\chi}_- \rangle  & \rightarrow  i \, d_m \langle P 
{\chi}_- \rangle -\frac{d_m}{6 M_0^2}  i \langle P {\chi}_+ \rangle 
\langle \chi_- \rangle + \ldots~,
\end{align}
which are obviously in direct correspondence to those generated by 
the application of the EOM in Eq.~(\ref{eq:eom2}). Again the
occurrence of the factor $M_0^{-2} = {\cal O}(N_c)$ leads to an enhancement 
in $1/\nc$, which is why we do not dispose of those terms. In an 
accompanying paper \cite{RK}, the $\eta^\prime$ 
contributions to the LECs of $\cO(p^4)$ \cite{Gasser:1984gg} and 
$\cO(p^6)$ are studied in a more systematic manner.

In the following we discuss in more detail how one arrives at our
effective Lagrangian if one starts from a resonance Lagrangian including  
the $\eta'$. In the diagram below, this Lagrangian sits in the upper left  
corner ($ \tilde{\cal L}_{{\rm eff}}^{\cal R}$):
\begin{align}
& \tilde{\cal L}_{{\rm eff}}^{\cal R} \quad \overset{\eta'
\hspace{-.6em}/}{\longrightarrow} \quad {\cal L}_{{\rm eff}}^{\cal 
R(\eta')}
\no
& \downarrow {\cal R\hspace{-.7em}/} \hspace{5em} \downarrow {\cal R 
\hspace{-.7em}/}
\\
& \tilde{\cal L}_{{\rm eff}}^{(\cal R)} \quad \overset{\eta'
\hspace{-.6em}/}{\longrightarrow} \quad {\cal L}_{{\rm eff}}^{\cal 
(\eta',R)}
\nonumber
\end{align}
The notation is the following: a superscript ${\cal R}$ denotes a  
Lagrangian with explicit resonance fields, whereas superscripts in  
brackets  denote contributions from the resonances and/or $\eta'$ in  
the coupling constants. The presence of the $\eta'$ (the chiral group  
being $U(3)$) is indicated by a tilde ($\tilde{\quad}$). A slash (/)  
symbolizes the process of integrating out a field. The final effective
Lagrangian of this work is ${\cal L}_{{\rm eff}}^{\cal 
(\eta',R)}$ in the lower right corner. Here, we want to analyse
the transition from the upper left to the upper right. 

Let us start with the low-energy expansion of $\tilde{\cal L}_{{\rm  
eff}}^{\cal R}$. As indicated above, the relevant expansion is the  
one where  powers of momenta, quark masses and $1/\nc$ are treated as  
small, according to
\begin{align}
p^2 \sim m_q \sim 1/\nc = {\cal O}(\delta) ,
\end{align}
where a counting parameter $\delta$ has been introduced 
\cite{Kaiser:2000gs}. The expansion for our Lagrangian thus takes the
form 
\begin{align}
\tilde{\cal L}_{{\rm eff}}^{\cal R}  = \tilde{\cal L}_{0}^{\cal R} +  
\tilde{\cal L}_{1}^{\cal R}  + \tilde{\cal L}_{2}^{\cal R} +  O 
(\delta^3)  .
\end{align}
The leading-order term $\tilde{\cal L}_{0}^{\cal R}$ is of course  
nothing but the Lagrangian $\tilde{\cal L}$ in (\ref{eq:U(3)}) and it
is in fact  independent of the resonance fields ${\cal R}$.
The term $\tilde{\cal L}_{1}^{\cal R}$ has been given in  
Ref.~\cite{Kaiser:2005eu} and involves several terms of order $\nc p^4 
$ as well as one contribution (in $\tilde{\cal L}_{1}^{P}$) of  
order $\nc^0 p^2 $ proportional to $ \ln (\det \tilde{u}) \, \langle  
P \rangle$, viz.
\begin{align}\label{dzero}
\tilde{\cal L}_{1}^{P} = \ldots -2 i d_0 \ln (\det \tilde{u})  
\, \langle P \rangle .
\end{align}
Here, the resonance fields have implicitly been counted as order $  
\sqrt{\nc} p^2 $. The factor of $ \sqrt{\nc} $ arises simply because  
of the normalization of the resonance fields, whereas the power of  
$p^2$ can be understood when treating the resonance masses as large  
(i.e. of order $1$) and solving the EOM.
In simplifying the expression for the Lagrangian $ 
\tilde{\cal L}_{1}^{\cal R}$, the EOM for the  
Goldstone fields has been used to eliminate terms involving $\nabla_ 
\mu \tilde{u}^\mu$. In the present case it takes the form
\begin{align}\label{eom u tilde}
2 \nabla_\mu \tilde{u}^\mu = i \tilde{\chi}_- - \frac{4i}{3} M_0^2    
\ln \det \tilde{u} 
\end{align}
as one derives from the Lagrangian (\ref{eq:U(3)}).

We now consider the terms of ${\cal O}(\delta^2)$. The Lagrangian $\tilde 
{\cal L}_{2}^{\cal R}$ collects the contributions of order $ \nc p^6 
$, $\nc^0 p^4 $ and $\nc^{-1} p^2 $,
\begin{align}
\tilde{\cal L}_{2}^{\cal R} = \tilde{\cal L}_{2}^{\cal R}|_{ \nc  
p^6}  +\tilde{\cal L}_{2}^{\cal R}|_{\nc^0 p^4 } +\tilde{\cal L}_{2}^ 
{\cal R}|_{\nc^{-1} p^2}  .
\end{align}
The recipe to determine the first of these terms is to consult the  
tables of the present paper and identify all the terms that
only involve a single trace; in those replace the effective  
field $u$ by $\tilde{u}$ and, finally, equip the associated coupling  
constants with tildes, i.e.
\begin{align}
\tilde{\cal L}_{2}^{\cal R}|_{ \nc p^6} = \tilde{\lambda}^{V}_1  
i \langle V_{\mu\nu} \tilde{u}^\mu \tilde{u}_\alpha \tilde{u}^ 
\alpha  \tilde{u}^\nu   \rangle  + \ldots
\end{align}
with all the above $\tilde{\lambda}^{R}_i  = {\cal O}(\sqrt{\nc})$. We  
will not attempt to give the next term explicitly, but simply  
indicate exemplary contributions:
\begin{align}
\tilde{\cal L}_{2}^{\cal R}|_{ \nc^0 p^4} &= \tilde{\lambda}^{S} 
_n  \langle {S} \rangle  \langle \tilde{\chi}_+ \rangle + \ldots  
+ \tilde{\lambda}^{S}_{n'}  \langle {S}  \rangle^2   +  
\ldots + \tilde{\lambda}^{S}_{n''}  \ln \det \tilde{u}  \langle  
{S}  \tilde{\chi}_- \rangle + \ldots ,
\no
\tilde{\lambda}^{S}_{n('('))} &= {\cal O}(\nc^{-\frac{1}{2}}) .
\end{align}
Finally, the last piece of the ${\cal O}(\delta^2) $ Lagrangian consists  
of a single term
\begin{align}
\tilde{\cal L}_{2}^{\cal R}|_{\nc^{-1} p^2}  &= \tilde{\lambda}^S_m  
\ln^2 (\det \tilde{u} )  \langle S \rangle ,
\end{align}
with $\tilde{\lambda}^S_{m} = {\cal O}(\nc^{-\frac{3}{2}}) $. 
Again, the EOM has been  
used to eliminate $\nabla_\mu \tilde{u}^\mu$ terms. The difference to  
the standard framework lies in the fact that the application of the  
EOM does not generate factors of $\langle \chi_-  
\rangle$ but instead factors of $ \ln \det \tilde{u} $ and thereby  
intertwines the three types of contributions to $\tilde{\cal L}_{2}^ 
{\cal R}$.

Let us now turn to the transition from the upper left to the upper  
right of our diagram. There are several contributions to be considered:
\begin{enumerate}

\item When integrating out the $\eta'$ the term $\tilde{\cal L}_{0}^ 
{\cal R} = \tilde{\cal L}$ generates several {\it purely} $\eta'$  
contributions to low-energy constants, $L_7^{(\eta')}$ is an example.  
For these we refer to Ref.~\cite{RK}.

\item In the Lagrangian $\tilde{\cal L}_{1}^{\cal R}$ the first class  
of contributions is generated by replacing the field $\tilde{u} $ by  
$ u$ which produces the Lagrangian (\ref{eq:R_int}). Note that  
the difference of $\tilde{u} $ and  $ u$ is of order $p^2$ when the $ 
\eta'$ mass $M_0$ is treated as large.

\item Retaining terms up to ${\cal O}(p^6)$ only, the three terms  
generated by that difference are those given in
Eq.~(\ref{eq:mtrace}). 

\item These would have been all contributions from $\tilde{\cal L}_{1} 
^{\cal R}$ would it not be for the term $\propto d_0$ in Eq.~(\ref 
{dzero}). Closer inspection reveals, however, that contributions of  
this term can be neglected without significant loss of information  
because it exclusively leads to nonleading contributions in $1/\nc$.  
This can be seen from the solution of the singlet field's EOM,
\begin{align}
\ln \det \tilde{u} = \frac{1}{4 M_0^2 } \{ \langle  \chi_- \rangle +  
\frac{12i d_0 }{F_0^2 } \langle {P}  \rangle \} + {\cal O}(p^4) .
\end{align}
Upon integrating out the pseudoscalar resonance $P$, $d_0$  
generates a contribution $\propto  \langle \chi_- \rangle$,  
relatively suppressed by $d_0 d_m /F_0^2 = {\cal O}(1/\nc)$.

\item It remains to work out the contributions from $\tilde{\cal L}_ 
{2}^{\cal R}$. Again, the singly traced ${\cal O}(\nc p^6)$ terms are in  
direct correspondence with the single-trace terms in the tables of  
the present  work. For these one finds the trivial matching relations
\begin{align}
\lambda^R_i & = \tilde{\lambda}^R_i .
\end{align}
As far as the remaining terms are concerned they either lead to 
genuinely suppressed contributions ($\tilde{\lambda}^{S}_n,  
\tilde{\lambda}^{S}_{n'}$, etc.) or to contributions that are  
suppressed relatively to the leading $\eta'$ exchange  
contributions. For instance, the contribution of $\tilde{\lambda}^{S}_m  
\ln^2 (\det  \tilde{u} )  \langle {S} \rangle $ to the term 
$\langle \chi_- \rangle^2  \langle \chi_+  \rangle $ is of
${\cal O}(\nc)$, whereas the leading-order contributions  to 
this term are of ${\cal O}(N_C^3)$~\cite{RK}. For this reason we will 
simply  neglect these terms altogether.
\end{enumerate}
To summarize, the transition from the upper left to the upper right  
of our diagram has lead to a resonance Lagrangian that consists of  
single-trace terms only, up to three double-trace terms $
\cO^S_4$, $\cO^S_5$ and $\cO^P_3$ as given explicitly in 
Eq.~(\ref{eq:mtrace}).

\setcounter{equation}{0}
\setcounter{table}{0}

\section{Field redefinitions}
\label{app:lfr}
In this appendix we report some details of our analysis of linear
field redefinitions. Let us start by listing the allowed redefinitions
for resonance fields.

\subsection*{Redefinitions for vector meson fields}
\begin{eqnarray}
1) \qquad V_{\mu\nu} & \longrightarrow & 
V_{\mu\nu} \, + \, g \, \left[ \, A_{\mu\nu} \, , \, \chi_- \, \right] \; , 
\nonumber \\
2) \qquad V_{\mu\nu} & \longrightarrow & 
V_{\mu\nu} \, + \, i \, g \, \left( \,\left[ \, A_{\mu\alpha} \, ,  
\,f_{-\beta\nu} \, \right] \, - \, \left[ \, A_{\nu\alpha} \, , \, 
f_{-\beta\mu} \, \right] \, 
\right) \, g^{\alpha\beta} \; , \nonumber  \\
3) \qquad V_{\mu\nu} & \longrightarrow & V_{\mu\nu} \, + \, i \, 
g \, \left( \,\left[ \, A_{\nu\alpha} \, 
, h_{\mu}^{\alpha} \, \right] \, - \, \left[ \, A_{\mu\alpha} \, , 
\, h_{\nu}^{\alpha} \, \right] \, \right) \,\; , \nonumber \\
4) \qquad V_{\mu\nu} & \longrightarrow & V_{\mu\nu} \, + \, i \, g \, 
\left( \, \left[ \, \nabla_{\mu} P \,
, \, u_{\nu} \, \right] \, - \, \left[ \, \nabla_{\nu} P \,, \,
u_{\mu} \, \right] \, \right) \; , \nonumber 
\\
5) \qquad V_{\mu\nu} & \longrightarrow & V_{\mu\nu} \, + \, i \, g \, 
\left[ \, P \, , \, f_{-\mu\nu} \, \right] \, , \nonumber \\
6) \qquad V_{\mu\nu} & \longrightarrow & V_{\mu\nu} \, + \, i \, g \, 
\left\lbrace \, S \, , \, 
\left[ \, u_{\mu} \, , \, u_{\nu} \, \right] \, \right\rbrace \; , 
\nonumber \\
7) \qquad V_{\mu\nu} & \longrightarrow & V_{\mu\nu} \, + \, g \, 
\left\lbrace \, S \, , \, f_{+\mu\nu} \, 
\right\rbrace \; , \nonumber \\
8) \qquad V_{\mu\nu} & \longrightarrow & V_{\mu\nu} \, + \, i g \, 
\left( u_\mu \, S \, u_\nu \, 
-  u_\nu \, S \, u_\mu \right) \; , \nonumber \\
9) \qquad V_{\mu\nu} & \longrightarrow & V_{\mu\nu}  \, + \, g \, 
\left\lbrace \, V_{\mu \nu} \, , \,  u_{\alpha} \, u^{\alpha} \, 
\right\rbrace \ , 
 \\
10) \qquad V_{\mu\nu} & \longrightarrow & V_{\mu\nu}  \, + \, g \, 
\left\lbrace  \, V_{\mu \nu} \, , \chi_+ \, \right\rbrace \,  \; , 
\nonumber \\
11) \qquad V_{\mu\nu} & \longrightarrow & V_{\mu\nu}  \, + \, 
g \ u_\alpha \, V_{\mu \nu} \, u^\alpha \,  \; , \nonumber  \\
12) \qquad V_{\mu\nu} & \longrightarrow & V_{\mu\nu}  \, + \, i \, g  \,
\left( \, \left[ f_{+ \mu \alpha} \, , \, V_{\beta \nu} \right] \, - \, 
\left[ f_{+ \nu \alpha} \, , \, V_{\beta \mu} \right] \, \right) \, 
g^{\alpha \beta} 
\,  \; , \nonumber \\
13) \qquad V_{\mu\nu} & \longrightarrow & V_{\mu\nu}  \, + \, g  \,
\left( \, u^\alpha \, V_{\mu \alpha} \, u_\nu \, + \, u_\nu \, 
V_{\mu \alpha} \, u^\alpha  
\, - \, \, u^\alpha \, V_{\nu \alpha} \, u_\mu \, - \, u_\mu \, 
V_{\nu \alpha} \, u^\alpha  
\, \right) 
\,  \; , \nonumber \\
14) \qquad V_{\mu\nu} & \longrightarrow & V_{\mu\nu}  \, + \, g  \, 
\left( \, \left[ \, \left[u_\mu , u^\alpha \right] \, , \, 
V_{\nu \alpha} \, \right] \, - \, 
\left[ \, \left[u_\nu , u^\alpha \right] \, , \, V_{\mu \alpha} 
\, \right] \, \right) 
\,  \; , \nonumber \\
15) \qquad V_{\mu\nu} & \longrightarrow & V_{\mu\nu}  \, + \, g  \, 
\left( \, \lbrace \, \lbrace u_\mu , u^\alpha \rbrace \, , \, 
V_{\nu \alpha} \, \rbrace \, - \, 
\lbrace \, \lbrace u_\nu , u^\alpha \rbrace \, , \, V_{\mu \alpha} 
\, \rbrace \, \right) 
\,  \; , \nonumber \\
16) \qquad V_{\mu\nu} & \longrightarrow & V_{\mu\nu}  \, + \, i \,  g  \, 
\left( \, \left[  \, \nabla_\mu \, A_{\nu \alpha} \, - \, 
\nabla_{\nu} \, A_{\mu \alpha} \, , \, 
u^\alpha \, \right] \, \right) 
\,  \; , \nonumber \\
17) \qquad V_{\mu\nu} & \longrightarrow & V_{\mu\nu}  \, + \, i \,  g  \, 
 \left[  \, \nabla_\alpha \, A_{\mu \nu} \, , \, u^\alpha \, \right]  
\,  \; , \nonumber \\
18) \qquad V_{\mu\nu} & \longrightarrow & V_{\mu\nu}  \, + \, i \,  g  \, 
\left( \, \left[  \, \nabla^\alpha \, A_{\alpha \nu} \, , \, 
u_\mu \, \right] \, - \, 
\left[ \,  \nabla^{\alpha} \, A_{\alpha \mu} \, , \, 
u_\nu \, \right] \, \right) 
\,  \; . \nonumber
\end{eqnarray}

\subsection*{Redefinitions for scalar fields}
\begin{eqnarray}
1) \qquad S & \longrightarrow & S \, +  \, g \,  \left\lbrace \, 
\nabla_{\mu} A^{\mu\nu} \, 
, \, u_{\nu} \, \right\rbrace \,  \; , \nonumber \\
2) \qquad S & \longrightarrow & S \, +  \, g \, \left\lbrace \, 
A^{\mu\nu} \, , \, f_{-\mu\nu} \,
\right\rbrace \; , \nonumber \\
3) \qquad S & \longrightarrow & S \, + \, i \, g \, \left\lbrace \, 
V^{\mu\nu} \, , \, 
 u_{\mu}  \, u_{\nu}  \, \right\rbrace \; , \nonumber \\
4) \qquad S & \longrightarrow & S \, + \, g \, \left\lbrace \, 
V^{\mu\nu} \, , \, f_{+\mu\nu} \, 
\right\rbrace \; , \nonumber \\
5) \qquad S & \longrightarrow & S \, + \, i \, g \, u_\mu \, 
V^{\mu \nu} \, u_\nu  \; , \nonumber \\
6) \qquad S & \longrightarrow & S  \, + \, g \, \left\lbrace \, 
\nabla_{\mu} P \, , \, u^{\mu} \, \right\rbrace \; ,  \\
7) \qquad S & \longrightarrow & S  \, + \, i \, g \, \left\lbrace \,  
P \, , \, \chi_{-} \, \right\rbrace \; , \nonumber \\
8) \qquad S & \longrightarrow & S \, + \, i \, g \, P \, \left\langle 
\, \chi_{-} \, \right\rangle \; , \nonumber \\
9) \qquad S & \longrightarrow & S \, + \, g \, 
\left\lbrace \, S \, , \,  u_{\alpha} \, u^{\alpha} \, \right\rbrace 
  \; , \nonumber \\
10) \qquad S & \longrightarrow & S \, + \, g \, \left\lbrace  \, 
S \, , \chi_+ \, \right\rbrace \,
  \; , \nonumber \\
11) \qquad S & \longrightarrow & S \, + \, g \ u_\alpha \, S \, 
u^\alpha \,  \; . 
\nonumber
\end{eqnarray} 

\subsection*{Redefinitions for axial-vector fields}
\begin{eqnarray}
1) \qquad A_{\mu\nu} & \longrightarrow & A_{\mu\nu} \, + \, g \, 
\left[ \, V_{\mu\nu} \, , \, \chi_- \, \right] 
\; , \nonumber \\
2) \qquad A_{\mu\nu} & \longrightarrow & A_{\mu\nu} \, + \, i \, g 
\, \left( \,\left[ \, V_{\mu\alpha} \, 
,  \,f_{-\beta\nu} \, \right] \, - \, \left[ \, V_{\nu\alpha} \, , 
\, f_{-\beta\mu} \, \right] \, \right) \, g^{\alpha\beta} \; , \nonumber \\
3) \qquad A_{\mu\nu} & \longrightarrow & A_{\mu\nu} \, + \, i \, g 
\, \left( \,\left[ \, V_{\nu\alpha} \, 
, h_{\mu}^{\alpha} \, \right] \, - \, \left[ \, V_{\mu\alpha} \, , 
\, h_{\nu}^{\alpha} \, \right] \, \right) \,\; , \nonumber \\
4) \qquad A_{\mu\nu} & \longrightarrow & A_{\mu\nu} \, + \, i \, g 
\, \left[ \, P \, , \, f_{+\mu\nu} \, \right] \; , \nonumber \\
5) \qquad A_{\mu\nu} & \longrightarrow & A_{\mu\nu} \, +  \, g \, 
\left[ \, P \, , \, \left[ \, u_{\mu} \,
, \, u_{\nu} \, \right] \, \right] \; , \nonumber \\
6) \qquad A_{\mu\nu} & \longrightarrow & A_{\mu\nu} \, +  \, g \, 
\left( \, \left\lbrace \, \nabla_{\mu} S \, 
, \, u_{\nu} \, \right\rbrace \, - \, \left\lbrace \, \nabla_{\nu} 
S \, , \, u_{\mu} \, \right\rbrace \, \right) \; , \nonumber \\
7) \qquad A_{\mu\nu} & \longrightarrow & A_{\mu\nu} \, +  \, g \, 
\left\lbrace \, S \, , \, f_{-\mu\nu} \,
\right\rbrace \; , \nonumber \\
8) \qquad A_{\mu\nu} & \longrightarrow & A_{\mu\nu}  \, + \, g \, 
\left\lbrace \, A_{\mu \nu} \, , \,  u_{\alpha} \, u^{\alpha} \, 
\right\rbrace \; , \nonumber \\
9) \qquad A_{\mu\nu} & \longrightarrow & A_{\mu\nu}  \, + \, g \, 
\left\lbrace  \, A_{\mu \nu} \, , \chi_+ \, \right\rbrace \,
\; , \\
10) \qquad A_{\mu\nu} & \longrightarrow & A_{\mu\nu}  \, + \, g \ 
u_\alpha \, A_{\mu \nu} \, u^\alpha \,  \; , \nonumber  \\
11) \qquad A_{\mu\nu} & \longrightarrow & A_{\mu\nu}  \, + \, i \, g  \,
\left( \, \left[ f_{+ \mu \alpha} \, , \, A_{\beta \nu} \right] \, - \, 
\left[ f_{+ \nu \alpha} \, , \, A_{\beta \mu} \right] \, \right) \, 
g^{\alpha \beta} 
\,  \; , \nonumber \\
12) \qquad A_{\mu\nu} & \longrightarrow & A_{\mu\nu}  \, + \, g  \,
\left( \, u^\alpha \, A_{\mu \alpha} \, u_\nu \, + \, u_\nu \, 
A_{\mu \alpha} \, u^\alpha  
\, - \, \, u^\alpha \, A_{\nu \alpha} \, u_\mu \, - \, u_\mu \, 
A_{\nu \alpha} \, u^\alpha  
\, \right) 
\,  \; , \nonumber \\
13) \qquad A_{\mu\nu} & \longrightarrow & A_{\mu\nu}  \, + \, g  \, 
\left( \, \left[ \, \left[u_\mu , u^\alpha \right] \, , \, 
A_{\nu \alpha} \, \right] \, - \, 
\left[ \, \left[u_\nu , u^\alpha \right] \, , \, A_{\mu \alpha} 
\, \right] \, \right) 
\,  \; , \nonumber \\
14) \qquad A_{\mu\nu} & \longrightarrow & A_{\mu\nu}  \, + \, g  \, 
\left( \, \lbrace \, \lbrace u_\mu , u^\alpha \rbrace \, , \, 
A_{\nu \alpha} \, \rbrace \, - \, 
\lbrace \, \lbrace u_\nu , u^\alpha \rbrace \, , \, A_{\mu \alpha} 
\, \rbrace \, \right) 
\,  \; ,  \nonumber \\
15) \qquad A_{\mu\nu} & \longrightarrow & A_{\mu\nu}  \, + \, i \,  g  \, 
\left( \, \left[  \, \nabla_\mu \, V_{\nu \alpha} \, - \, 
\nabla_{\nu} \, V_{\mu \alpha} \, , \, 
u^\alpha \, \right] \, \right) 
\,  \; , \nonumber \\
16) \qquad A_{\mu\nu} & \longrightarrow & A_{\mu\nu}  \, + \, i \,  g  \, 
 \left[  \, \nabla_\alpha \, V_{\mu \nu} \, , \, u^\alpha \, \right]  
\,  \; , \nonumber \\
17) \qquad A_{\mu\nu} & \longrightarrow & A_{\mu\nu}  \, + \, i \,  g  \, 
\left( \, \left[  \, \nabla^\alpha \, V_{\alpha \nu} \, , \, u_\mu 
\, \right] \, - \, 
\left[ \,  \nabla^{\alpha} \, V_{\alpha \mu} \, , \, 
u_\nu \, \right] \, \right) 
\,  \; . \nonumber 
\end{eqnarray} 

\subsection*{Redefinitions for pseudoscalar fields}
\begin{eqnarray}
1) \qquad P & \longrightarrow & P \, + \, i \, g \, \left[ \, 
A^{\mu\nu} \, , \, f_{+\mu\nu} \, \right] \; , \nonumber \\
2) \qquad P & \longrightarrow & P \, +  \, g \, \left[ \, A^{\mu\nu} 
\, , \, u_{\mu}  \, 
u_{\nu}  \, \right] \; , \nonumber \\
3) \qquad P & \longrightarrow & P  \, + \, i \, g \,  \left[ \, 
\nabla_{\mu} V^{\mu\nu} \,, \, u_{\nu} \, \right] \; , \nonumber 
\\
4) \qquad P & \longrightarrow & P\, + \, i \, g \, \left[ \, 
V^{\mu\nu} \, , \, f_{-\mu\nu} \, \right] \, , \nonumber 
\\
5) \qquad P & \longrightarrow & P  \, + \, g \, \left\lbrace \, 
\nabla_{\mu} S \, , \, u^{\mu} \, \right\rbrace \; , \\
6) \qquad P & \longrightarrow & P  \, + \, i \, g \, \left\lbrace 
\,  S \, , \, \chi_{-} \, \right\rbrace \; , \nonumber \\
7) \qquad P & \longrightarrow & P \, + \, i \, g \, S \, 
\left\langle \, \chi_{-} \, \right\rangle \; , \nonumber \\
8) \qquad P & \longrightarrow & P  \, + \, g \, 
\left\lbrace \, P \, , \,  u_{\alpha} \, u^{\alpha} \, \right\rbrace 
  \; , \nonumber \\
9) \qquad P & \longrightarrow & P  \, + 
\, g \, \left\lbrace  \, P \, , \chi_+ \, \right\rbrace \,
\; , \nonumber \\
10) \qquad P & \longrightarrow & P \, + \, g \ u_\alpha \, P \, 
u^\alpha \,  \; . 
\nonumber
\end{eqnarray} 

Applying the redefinitions of the resonance fields $R_i$ to 
${\cal L}_{(2)}^{R}$ generates monomials $\cO_n^{R_j}$ of 
${\cal L}_{(4)}^{R}$.
The results are reported in Tables~\ref{tab:rdfV}, \ref{tab:rdfA},
\ref{tab:rdfS}.  Note that when considering redefinitions of $V$ and
$S$, we cannot eliminate both entries in a given line in the tables at
the same time because the ratios $F_V/G_V$ and $c_d/c_m$ are fixed.   


\begin{table}
\begin{center}
\hspace*{-1.5cm} 
\renewcommand{\arraystretch}{1.5}
\begin{tabular}{|c|c||c|} 
\hline 
& &  \\[-.5cm] 
\multicolumn{1}{|c|}{i} &
\multicolumn{1}{|c||}{ 
$\displaystyle\frac{F_{V}}{2\sqrt{2}}\; \langle V^{\mu\nu} 
f_{+ \, \mu\nu}\rangle $ generates
} & 
\multicolumn{1}{|c|}{
$\displaystyle\frac{i\, G_{V}}{\sqrt{2}} \,\,\langle V^{\mu\nu} 
u_\mu u_\nu\rangle$
generates
}  \\[.3cm] 
\hline
\hline
 1 & $\cO^{A}_4$ & $\cO^{A}_5$ \\
\hline 
 2 & $\cO^{A}_{14}$ & $\cO^{A}_9 - \cO^{A}_{10}$ \\
\hline
 3 & $\cO^{A}_{12}$ & $\cO^{A}_1 - \cO^{A}_2$ \\
\hline
 4 & $2 \cO^{P}_9 - \cO^{P}_{11}$ & $\cO^{P}_1 - 2 \cO^{P}_2 -
\cO^{P}_4 + 2 \cO^{P}_5 - 3 \cO^{P}_8$ \\
\hline
 5 & $\cO^{P}_{11}$ & $\cO^{P}_8$ \\
\hline
 6 & $\cO^{S}_{10}$ & $\cO^{S}_2 - \cO^{S}_3$ \\
\hline
 7 & $\cO^{S}_{15}$ & $\cO^{S}_{10}$ \\
\hline
 8 & $\cO^{S}_{11}$ & $\cO^{S}_1 - \cO^{S}_3$ \\
\hline
 9 & $\cO^{V}_{11}$ & $\cO^{V}_4$ \\
\hline
 10 & $\cO^{V}_{6}$ & $\cO^{V}_8$ \\
\hline
 11 & $\cO^{V}_{12}$ & $\cO^{V}_2$ \\
\hline
 12 & $\cO^{V}_{7}$ & $\cO^{V}_{14} - \cO^{V}_{15}$ \\
\hline
 13 & $\cO^{V}_{13}$ & $\cO^{V}_{4} - \cO^{V}_{3}$ \\
\hline
 14 & $\cO^{V}_{14} - \cO^{V}_{15}$ & $\cO^{V}_{1} + \cO^{V}_2 - 
\cO^{V}_{3}$ \\
\hline
 15 & $\cO^{V}_{14} + \cO^{V}_{15}$ & $\cO^{V}_{1} - \cO^{V}_2$ \\
\hline
 16 & $\cO^{A}_{12} + \cO^{A}_{14} - 2 \cO^{A}_{15}$ & $\cO^{A}_{2} - 
\cO^{A}_3 + \cO^{A}_5 + \cO^{A}_{10} -
 \cO^{A}_{11}$ \\
\hline
 17 & $\cO^{A}_{4} - 2 \cO^{A}_{13}$ & $\cO^{A}_{2} - \cO^{A}_3 + 
\cO^{A}_5 + \cO^{A}_{10} - \cO^{A}_{11}$ \\
\hline
 18 & $\cO^{A}_{7} - 2 \cO^{A}_{8} + 2 \cO^{A}_{10} - 2 \cO^{A}_{11}$  
  & $\cO^{A}_{1} - 2 \cO^{A}_2 + \cO^{A}_3 + \cO^{A}_7$ \\ 
 & $ + 2 \cO^{A}_{12} + 2 \cO^{A}_{13} - 2
 \cO^{A}_{14}$ & 
$ - 2 \cO^{A}_8 - \cO^{A}_9 + 2 \cO^{A}_{10} - \cO^{A}_{11}$ \\
\hline
\end{tabular} 
\end{center}
\caption{
Results of applying redefinitions of $V$ fields 
to ${\cal L}_{(2)}^V$. 
\label{tab:rdfV}
}
\end{table}


\begin{table}
\begin{center}
\hspace*{-1.5cm} 
\renewcommand{\arraystretch}{1.5}
\begin{tabular}{|c|c||c|} 
\hline
& &  \\[-.5cm] 
\multicolumn{1}{|c|}{i} &
\multicolumn{1}{|c||}{
$\displaystyle\frac{F_{A}}{2\sqrt{2}} \; \langle A^{\mu\nu} 
f_{- \, \mu\nu} \rangle$
generates
} & 
\multicolumn{1}{|c|}{
$d_{m}\;\langle P\, \chi_- \rangle $
generates
}  \\[.3cm] 
\hline
\hline
 1 & $\cO^{V}_{20}$ & $\cO^{A}_4$ \\
\hline 
 2 & $\cO^{V}_{5}$ & $\cO^{A}_5$ \\
\hline
 3 & $\cO^{V}_{19}$ & $2 \cO^{V}_{10} +  \cO^{V}_{20}$ \\
\hline
 4 & $\cO^{P}_{11}$ & $\cO^{V}_{20}$ \\
\hline
 5 & $\cO^{P}_{8}$ & $\cO^{S}_8 - \cO^{S}_{14} + \cO^{S}_{5}/3$ \\
\hline
 6 & $\cO^{S}_{12} + \cO^{S}_{16}$ & $\cO^{S}_{14}$ \\
\hline
 7 & $\cO^{S}_{16}$ & $\cO^{S}_{5}$ \\
\hline
 8 & $\cO^{A}_{7}$ & $\cO^{P}_4$ \\
\hline
 9 & $\cO^{A}_{16}$ & $\cO^{P}_{10}$ \\
\hline
 10 & $\cO^{A}_{8}$ & $\cO^{P}_5$ \\
\hline
 11 & $\cO^{A}_{14}$ &  \\
\hline
 12 & $\cO^{A}_{11}$ &  \\
\hline
 13 & $\cO^{A}_{9} - \cO^{A}_{10}$ &  \\
\hline
 14 & $\cO^{A}_{9} + \cO^{A}_{10}$ &  \\
\hline
 15 & $2 \cO^{V}_{5} - 2 \cO^{V}_{18} + \cO^{V}_{19}$ &  \\
\hline
 16 & $2 \cO^{V}_{17} - \cO^{V}_{20}$ &  \\
\hline
 17 & $2 \cO^{V}_{5} + 2 \cO^{V}_{16} - \cO^{V}_{19}$ &  \\
\hline
\hline
\end{tabular} 
\end{center}
\caption{
Results of applying redefinitions of $A$ and $P$ fields 
to ${\cal L}_{(2)}^A$ and ${\cal L}_{(2)}^P$. 
\label{tab:rdfA}
}
\end{table}


\begin{table}
\begin{center}
\hspace*{-1.5cm} 
\renewcommand{\arraystretch}{1.5}
\begin{tabular}{|c|c||c|} 
\hline
\multicolumn{1}{|c|}{i} &
\multicolumn{1}{|c||}{
$c_{d} \; \langle S\, u^\mu u_\mu\rangle$
generates
} & 
\multicolumn{1}{|c|}{
$c_{m} \; \langle S\, \chi_+ \rangle $
generates
}  \\[.1cm] 
\hline
\hline
 1 & $\cO^{A}_1 + \cO^{A}_3 + \cO^{A}_7 - \cO^{A}_9 - \cO^{A}_{11}$ 
& $2 \cO^{A}_6 + \cO^{A}_{16}$ \\
\hline 
 2 & $\cO^{A}_{7}$ & $\cO^{A}_{16}$ \\
\hline
 3 & $\cO^{V}_{4}$ & $\cO^{V}_{8}$ \\
\hline
 4 & $\cO^{V}_{11}$ & $\cO^{V}_{6}$ \\
\hline
 5 & $\cO^{V}_{1}$ & $\cO^{V}_9$ \\
\hline
 6 & $\cO^{P}_{1} + 2 \cO^{P}_2 - 2 \cO^{P}_7/3 + \cO^{P}_4 + 
\cO^{P}_{8}$ & $2 \cO^{P}_6 + \cO^{P}_{10} - 2 \cO^{P}_{3}/3$ \\
\hline
 7 & $\cO^{P}_{4}$ & $\cO^{P}_{10}$ \\
\hline
 8 & $\cO^{P}_{7}$ & $\cO^{P}_{3}$ \\
\hline
 9 & $\cO^{S}_{1}$ & $\cO^{S}_{6}$ \\
\hline
 10 & $\cO^{S}_{6}$ & $\cO^{S}_{13}$ \\
\hline
 11 & $\cO^{S}_{2}$ & $\cO^{S}_7$  \\
\hline
\hline
\end{tabular} 
\end{center}
\caption{
Results of applying redefinitions of $S$  fields 
to ${\cal L}_{(2)}^S$. 
\label{tab:rdfS}
}
\end{table}
\vspace*{14cm}
\setcounter{equation}{0}
\setcounter{table}{0}

\section{Integrating out resonance fields}
\label{app:defintout}
To arrive at the resonance exchange Lagrangian (\ref{eq:intout}),
we first perform the linear field redefinitions of
Sec.~\ref{sec:minimal} and then rewrite the interaction Lagrangians
(\ref{eq:R_int}), (\ref{eq:l4r}) and (\ref{eq:l2rr}) as follows:	
\begin{eqnarray}
{\cal L}_{(2)}^{R} \, + \, {\cal L}_{(4)}^{R} & = & 
 \sum_{R=S,P} \left\langle R \, (g_2^R + g_4^R) 
\right\rangle +
\sum_{R=V,A} \left\langle R_{\mu\nu} \, (g_2^R + g_4^R)^{\mu\nu} 
\right\rangle \; , \\
& & \nonumber \\
{\cal L}_{(2)}^{RR} & = & 
\sum_{R=S,P} \left[  \left\langle \, R^2 \, h_2^R \, \right\rangle + 
\lambda_{2}^{RR} \left\langle \,  R \, u_{\mu} \, R \, u^{\mu} \,
\right\rangle \right] \nonumber \\ & & \nonumber \\
& &  \,+ \, \lambda_{1}^{SP} \, \left\langle \left\lbrace \,
\nabla_{\mu}S, P \,\right\rbrace \, u^{\mu} \, \right\rangle \, + i \,
\lambda_{2}^{SP} \, 
\left\langle \,\left\lbrace S, \,P\right\rbrace  \,\chi_- \,\right\rangle 
\,
\nonumber \\
 & & \nonumber \\
 & & 
+\,  \sum_{R=V,A} \left[ \, \left\langle R_{\mu\nu} \,  R_{\alpha\beta} \, 
\Theta_{2R}^{\mu\nu\alpha\beta} \, \right\rangle + 
\left\langle \, R_{\mu\nu}\,  u_{\gamma} \, R_{\alpha\beta} \, u_{\delta} \,
\Lambda_{2R}^{\mu\nu\alpha\beta\gamma\delta}\, \right\rangle \right]  
 \nonumber \\
& & \nonumber \\
& & \qquad \; \; + \, \left\langle \,  \left[ \, 
V_{\mu\nu}, A_{\alpha\beta}\, \right] \, 
\Theta_{VA}^{\mu\nu\alpha\beta} \, \right\rangle \ + \, 
\left\langle \, \left[ \, \nabla_{\gamma} V_{\mu\nu}, A_{\alpha\beta} \, 
\right] \,
\Upsilon_{VA}^{\mu\nu\alpha\beta\gamma} \, \right\rangle \,  \nonumber \\
& & \nonumber \\
& & \qquad \; \; + \, \left\langle \, \left\lbrace \,S, V_{\mu\nu} \, 
\right\rbrace \, 
\Omega_{2SV}^{\mu\nu} \,
\right\rangle \, + \left\langle \, \left[ \, P, A_{\mu\nu}\, \right] \, 
\Omega_{2PA}^{\mu\nu}
\, \right\rangle \, + \, i \lambda_2^{SV} \,\left\langle \, S \, u_{\mu} 
 \, V^{\mu\nu}
\, u_{\nu} \, \right\rangle  \nonumber \\
& & \nonumber \\
& & \qquad \; \; + \, \lambda_{1}^{SA} \, \left\langle \, \left\lbrace \, 
\nabla_{\mu} S, A^{\mu\nu} \, \right\rbrace \, u_{\nu} \, \right\rangle \,+ \, 
\lambda_{2}^{SA} \, \left\langle \, \left\lbrace \, S, A_{\mu\nu} \, \right
\rbrace \,
f_{-}^{\mu\nu} \, \right\rangle \, 
\nonumber \\
& & \nonumber \\
& & \qquad \; \; + \, i \, \lambda_{1}^{PV} \, 
\left\langle \, \left[ 
\nabla^{\mu} P, V_{\mu\nu} \, \right]  \, u^{\nu} \,  \right\rangle  
+ \, i \, \lambda_{2}^{PV} \, \left\langle \, 
\left[ P, V_{\mu \nu} 
\, \right] \, f_{-}^{\mu\nu} \,
\right\rangle \,\, \; .
\end{eqnarray}
Finally, ${\cal L}_{(0)}^{RRR}$ in Eq.~(\ref{eq:l0rrr}) must be
included. The explicit expressions for $g_n^R$, $(g_n^R)_{\mu \nu}$, 
$h_2^R$,... can be read off from Tables~\ref{tab:lagV} --
\ref{tab:PVPAVA}: 
\begin{eqnarray}
g_2^S \, & = &  \, c_d \, u_{\mu} u^{\mu} \, + \, c_m \, \chi_{+} \, 
\; , \nonumber \\
g_2^P \, & = & \, i \, d_m \, \chi_{-} \, \; , \nonumber \\
\left( g_2^V \right)_{\mu\nu} \, & = & \, \frac{F_V}{2\sqrt{2}} \, 
f_{+\mu\nu} \, + \, i\frac{G_V}{2\sqrt{2}} \, 
\left[ u_{\mu}, u_{\nu} \right] \, \; , \nonumber \\
\left( g_2^A\right)_{\mu\nu}  \, & = & \, \frac{F_A}{2\sqrt{2}} \, 
f_{-\mu\nu} \, \; ,\nonumber \\
g_4^S  & = & \lambda_{1}^{S} \, u_{\mu}u^{\mu}u_{\nu}u^{\nu} \, + \,
\lambda_{2}^{S} \, u_{\mu}u_{\nu}u^{\nu}u^{\mu}  \, + \, 
\lambda_{3}^{S} \, u_{\mu}u_{\nu}u^{\mu}u^{\nu} \, + \, 
i \,\lambda_{4}^{S} \, u_{\mu} \, \left\langle \nabla^{\mu} \chi_- 
\right\rangle  \, + \,\lambda_{5}^{S} \, \chi_- \, 
\left\langle \chi_- \right\rangle 
\; , \nonumber \\
g_4^P & = & \lambda_{1}^{P} \, \left\lbrace h_{\mu\nu}, 
u^{\mu}u^{\nu}\right\rbrace \, + \,
\lambda_{2}^{P} \, u_{\mu} h^{\mu\nu} u_{\nu} \, + \, 
i \, \lambda_{3}^{P} \, \chi_{+} \left\langle \chi_{-} \right\rangle \, 
\; ,  \nonumber \\
\left( g_{4}^{V} \right)_{\mu\nu} & = & 
i \, \lambda_{1}^{V} \, u_{\mu}u_{\alpha}u^{\alpha}u_{\nu} \, + \, 
i \, \lambda_{2}^{V} \, u_{\alpha}u_{\mu}u_{\nu}u^{\alpha} \, + \, 
i \, \lambda_{3}^{V} \, \left\lbrace u^{\alpha},u_{\mu}u_{\alpha}
u_{\nu}\right\rbrace \, + \,
i \, \lambda_{4}^{V} \, \left\lbrace u_{\mu}u_{\nu},u_{\alpha}
u^{\alpha}\right\rbrace  \,
  \, \; ,  \nonumber \\
\left( g_{4}^{A} \right)_{\mu\nu} & = & 
\lambda_{1}^{A} \, \left( u_{\mu}u^{\alpha}h_{\nu\alpha} + 
h_{\nu\alpha} u^{\alpha} u_{\mu} \right) 
\, + \,
\lambda_{2}^{A} \, \left( u^{\alpha}u_{\mu}h_{\nu\alpha} + 
h_{\nu\alpha} u_{\mu} u^{\alpha} \right) 
\, \nonumber \\ & & + \, 
\lambda_{3}^{A} \, \left( u_{\mu}h_{\nu\alpha}u^{\alpha} + 
u^{\alpha} h_{\nu\alpha} u_{\mu} \right) 
\,
\, \; ,  \nonumber
\end{eqnarray} 
\begin{eqnarray} 
h_{2}^{R} & = & \lambda_{1}^{RR} \, u_{\mu} u^{\mu} \, + \, 
\lambda_{3}^{RR} \chi_{+} \; ,
\; \; \; \; \qquad \qquad R = S,P \; ,  \nonumber \\
\Theta_{2R}^{\mu\nu\alpha\beta} & = & \lambda_{1}^{RR} \, 
g^{\mu\alpha} g^{\nu\beta} u^{\gamma}u_{\gamma} \, + \, 
\lambda_{3}^{RR} \, g^{\nu\beta} u^{\mu} u^{\alpha} \, + \, 
\lambda_{4}^{RR} \, g^{\nu\beta}
u^{\alpha}u^{\mu} 
\, + \, \lambda_{6}^{RR} \, g^{\mu\alpha} 
g^{\nu\beta} \chi_{+} \,  \nonumber \\ & &  + \, 
i \, \lambda_{7}^{RR}  \, g^{\nu\alpha} f_{+}^{\mu\beta} \;, 
\qquad\qquad \qquad\qquad
\qquad\qquad \qquad\qquad\quad \; \; \; \; \; \;R = V,A , 
 \nonumber \\
\Lambda_{2R}^{\mu\nu\alpha\beta\gamma\delta} & = & 
\lambda_{2}^{RR} \, g^{\mu\alpha} g^{\nu\beta} g^{\gamma\delta} \, + \, 
\lambda_{5}^{RR} \, g^{\mu\alpha} \left( g^{\nu\gamma} g^{\beta\delta} + 
g^{\beta\gamma} g^{\nu\delta} \right)  \; , \qquad\qquad  R = V,A ,
 \nonumber \\
\Theta_{VA}^{\mu\nu\alpha\beta} & = & 
\lambda_{1}^{VA} \, g^{\mu\alpha} g^{\nu\beta} \chi_{-} \, + \, 
i \, \lambda_{2}^{VA} \, g^{\nu\alpha} h^{\beta\mu} \, + \, 
i \, \lambda_{6}^{VA} \, g^{\mu\alpha} f_{-}^{\beta\nu} \; , 
\nonumber \\
\Upsilon_{VA}^{\mu\nu\alpha\beta\gamma} & = & 
i \, \lambda_{3}^{VA} \, g^{\gamma\mu} g^{\nu\alpha} u^{\beta} \, + \, 
i \, \lambda_{4}^{VA} \, g^{\gamma\alpha} g^{\nu\beta} u^{\mu} \, + \, 
i \, \lambda_{5}^{VA} \, g^{\mu\alpha} g^{\nu\beta} u^{\gamma} \, , 
\nonumber \\
\Omega_{2SV}^{\mu\nu} & = & i \, \lambda_{1}^{SV} \, u^{\mu} u^{\nu} \, + \, 
\lambda_{3}^{SV} \, f_{+}^{\mu\nu} \, \; ,  
\nonumber\\
\Omega_{2PA}^{\mu\nu} & = & i \, \lambda_{1}^{PA} \, f_{+}^{\mu\nu} \, + \, 
\lambda_{2}^{PA} \, u^{\mu} u^{\nu} \; \; . 
\end{eqnarray} 

The final result of integrating out the resonance fields up to
$\cO(p^6)$ is contained in the Lagrangian of Eq.~(\ref{eq:intout}),
with  
\begin{eqnarray}
{\cal L}_{SP} & = & \sum_{R=S,P} \, \left[ \frac{1}{2 M_{R}^{4}} \, 
\left\langle \, \nabla_{\mu} g_2^R \, 
\nabla^{\mu} g_2^R \, \right\rangle \, + \, \frac{1}{2 M_{R}^{2}} \, 
\left\langle \,
g_2^R \, g_2^R \, \right\rangle \, + \, \frac{1}{M_{R}^{2}} \, 
\left\langle \, g_2^R \, g_4^R \, \right\rangle \, \right. \nonumber \\
& & \nonumber \\
& & \qquad \quad \left. + \frac{1}{M_R^4} \, \left\langle \, g_2^R \, 
g_2^R \, h_2^R \, \right\rangle \, + \, 
\frac{\lambda_{2}^{RR}}{M_R^4} \, \left\langle \, g_2^R \, u_{\mu} \, 
 g_2^R \, u^{\mu} \,
\right\rangle \, \right] \; \nonumber \\ & & \nonumber \\
& & + \, \frac{\lambda_{1}^{SP}}{M_S^2 \, M_P^2} \, 
\left\langle  \, \left\lbrace \, \nabla_{\mu} g_2^S, g_2^P \, \right\rbrace 
\, u^{\mu} \,
\right\rangle \, + \, \frac{i \, \lambda_{2}^{SP}}{M_S^2 \, M_P^2} \, 
\left\langle \, \left\lbrace \, g_2^S, g_2^P \, \right\rbrace \,
\chi_- \, \right\rangle \, \; , 
\end{eqnarray} 
\begin{eqnarray}
{\cal L}_{VA} & = & - \, \sum_{R=V,A} \, \left[ \, \frac{2}{M_R^4} \, 
\left\langle \, \nabla^{\lambda} \left( g_2^R\right)_{\lambda\mu} \, 
\nabla_{\nu} 
\left( g_2^R \right)^{\nu\mu} \, \right\rangle \, + \, \frac{1}{M_R^2} 
\, \left\langle 
\left( g_2^R \right)_{\mu\nu}  \, \left( g_2^R \right)^{\mu\nu} 
 \right\rangle  \, \right. \nonumber \\
& & \nonumber \\
& & \qquad \quad \; \; \left. + \, \frac{2}{M_R^2} \, \left\langle \, 
\left( g_2^R \right)_{\mu\nu}  \, \left( g_4^R \right)^{\mu\nu} \, 
\right\rangle \,  -  \,
\frac{4}{M_R^4} \, \left\langle \, \left( g_2^R \right)_{\mu\nu} \, 
\left( g_2^R \right)_{\alpha\beta} \, \Theta_{2R}^{\mu\nu\alpha\beta} 
\, \right\rangle \, \right. \nonumber \\
& & \nonumber \\
& & \qquad \quad \; \; \left. - \, \frac{4}{M_R^4} \, 
\left\langle \, \left( g_2^R \right)_{\mu\nu} \, u_{\gamma} \, 
\left( g_2^R \right)_{\alpha\beta} \, u_{\delta} \, 
\Lambda_{2R}^{\mu\nu\alpha\beta\gamma\delta} \, \right\rangle \, \right] 
\, \nonumber \\
& & \nonumber \\
& & + \, \frac{4}{M_V^2 \, M_A^2} \, \left\langle \, \left[ \, 
\left( g_2^V\right)_{\mu\nu},
\left( g_2^A\right)_{\alpha\beta} \, \right] \, 
\Theta_{VA}^{\mu\nu\alpha\beta} \, 
\right\rangle \, \nonumber \\
& & \nonumber \\ 
& & + \, \frac{4}{M_V^2 \, M_A^2} \, \left\langle \, 
\left[ \, \nabla_{\gamma} 
\left( g_2^V\right)_{\mu\nu},\left( g_2^A\right)_{\alpha\beta} \, 
\right] \, 
\Upsilon_{VA}^{\mu\nu\alpha\beta\gamma}  \, \right\rangle \; , 
\end{eqnarray} 
\begin{eqnarray}
{\cal L}_{SPVA} & = & - \, \frac{2}{M_S^2 \, M_V^2} \, 
\left\langle \, \left\lbrace \, g_2^S, \left( g_2^V\right)_{\mu\nu} 
\, \right\rbrace \, 
\Omega_{2SV}^{\mu\nu} \, \right\rangle \, - \, \frac{2}{M_P^2 \, M_A^2} \, 
\left\langle \, \left[ \, g_2^P, \left( g_2^A\right)_{\mu\nu} \, 
\right] \, 
\Omega_{2PA}^{\mu\nu} \, \right\rangle \; \nonumber \\
& & \nonumber \\
& & -i \, \frac{2 \, \lambda_{2}^{SV}}{M_S^2 \, M_V^2} \, \left\langle 
\, g_2^S \, u_{\mu} \,
\left( g_2^V\right)^{\mu\nu} \, u_{\nu} \, \right\rangle \, - \,
\frac{2 \, \lambda_{1}^{SA}}{M_S^2 \, M_A^2} \, \left\langle \, 
\left\lbrace \,
\nabla_{\mu} g_2^S, \left( g_2^A\right)^{\mu\nu} \, \right\rbrace 
\, u_{\nu} \, \right\rangle 
\nonumber \\
& & \nonumber \\
& & - \, \frac{2 \, \lambda_{2}^{SA}}{M_S^2 \, M_A^2} \, \left\langle
 \, \left\lbrace \,
g_2^S, \left( g_2^A\right)^{\mu\nu} \, \right\rbrace \, f_{-\mu\nu} \, 
\right\rangle  \, - i
\, \frac{2 \, \lambda_{1}^{PV}}{M_P^2 \, M_V^2} \, \left\langle \, 
\left[ \, \nabla_{\mu} g_2^P,
\left( g_2^V\right)^{\mu\nu} \right] \, u_{\nu} \, \right\rangle \, 
\nonumber \\
& & \nonumber \\
& & -i \, \frac{2 \, \lambda_{2}^{PV}}{M_P^2 \, M_V^2} \, \left\langle 
\, \left[ \,
g_2^P, \left( g_2^V\right)^{\mu\nu} \, \right] \, f_{-\mu\nu} \, 
\right\rangle  \;  \nonumber \\
& & \nonumber \\
& & + \, \frac{4 \, \lambda^{SVV}}{M_S^2 \, M_V^4} \, \left\langle 
\, g_2^S \, \left( g_2^V \right)_{\mu\nu} \,
\left( g_2^V \right)^{\mu\nu} \, \right\rangle \, + \, \frac{4 \, 
\lambda^{SAA}}{M_S^2 \, M_A^4} \,
\left\langle \, g_2^S \, \left( g_2^A \right)_{\mu\nu} \,
\left( g_2^A \right)^{\mu\nu} \, \right\rangle \, \nonumber \\
& & \nonumber \\
& & + \, \frac{\lambda^{SSS}}{M_S^6} \, \left\langle \, g_2^S \, 
g_2^S \, g_2^S \, \right\rangle \,
+ \, \frac{\lambda^{SPP}}{M_S^2 \, M_P^4} \, \left\langle \, g_2^S 
\, g_2^P \, g_2^P \, \right\rangle \,
\nonumber \\
& & \nonumber \\
& & - \, i \, \frac{8 \, \lambda^{VVV}}{M_V^6} \, g_{\rho\sigma} \,
\left\langle \, \left( g_2^V \right)_{\mu\nu} \, \left( g_2^V 
\right)^{\mu\rho} \,
\left( g_2^V \right)^{\nu\sigma} \, \right\rangle \, + \, i \, 
\frac{4 \, \lambda^{VAP}}{M_P^2 \, M_V^2 \, M_A^2} \, \left\langle  
\, \left[ \, \left( g_2^V 
\right)_{\mu\nu} , \left( g_2^A \right)^{\mu\nu} \, \right] \, g_2^P 
\, \right\rangle \nonumber \\
& & \nonumber \\
& & - \, i \, \frac{8 \, \lambda^{VAA}}{M_V^2 \, M_A^4} \, g_{\rho\sigma} \,
\left\langle \, \left( g_2^V \right)_{\mu\nu} \, \left[ \, 
\left( g_2^A\right)^{\mu\rho} , 
\left( g_2^A \right)^{\nu\sigma} \, \right] \, \right\rangle \; . 
\end{eqnarray} 
%

\setcounter{equation}{0}
\setcounter{table}{0}

\section{Resonance contributions to the LECs of  $\mathbf{{\cal
      O}(p^6)}$} 
\label{app:resultsCI}

\renewcommand{\arraystretch}{1.9}
\setlength{\LTcapwidth}{\textwidth}
In Sec.~\ref{sec:integratingout} we integrated out the resonance
fields up to ${\cal O}(p^6)$ in the chiral Lagrangian. By identifying 
the result at ${\cal O}(p^6)$ with the $\chi$PT Lagrangian
\begin{equation}
{\cal L}_6^{\rm \chi PT} \, = \, \sum_{i=1}^{94} \, C_i \; {\cal O}_i^{(6)}
\, ,
\end{equation}
we extract the LECs $C_i$. Notice that we find contributions for 
only 64 of the couplings, reflecting the absence of genuine 
multiple-trace terms in the resonance Lagrangian. Moreover, the 
couplings $C_{91}$, \dots, 
$C_{94}$ correspond to local operators that involve
external fields only (contact terms).
The results are given in Table~\ref{tab:RESCi}.
{\small 
\begin{longtable}[c]{|r|l|}
\hline
$ i $ & $C_i^{\cal R}$\\
\hline
\hline
\endhead
\hline
\endfoot
\hline
\caption[]{\label{tab:RESCi}
\rule{0cm}{2em} Explicit expressions for the nonvanishing resonance 
contributions to the $ C_i$. For ease of notation, the $\lambda$--type 
couplings have been rescaled:  $\overline{\lambda}^{R,R',\ldots}_i = 
\lambda^{R,R',\ldots}_i/(\mu_R \mu_{R'}\cdots)$, with
$\mu_S = \MS^2/\cm$, $\mu_P = \MP^2/d_m$, $\mu_V = \MV^2/\FV$ and 
$\mu_A = \MA^2/\FA$.  Further, $\rS = \cd/\cm$, $\rV = \GV/\FV$. A
number of coupling constants are conveniently written in terms of 
$\Cs$ and $\Css$, which is manifestly displayed in the table. 
For additional symbols see text.} 
\endlastfoot

1&$
- \displaystyle\frac{\cd^2}{4\,\MS^4} + 
\displaystyle\frac{\GV^2}{8\,\MV^4} - {\sqrt{2}}\,{\lb{A}_2} + 
  {\rV}\,\left( \lVA_2 - \displaystyle\frac{{\lVA_4}}{2} - {\lVA_5} \right)
$ \\

3&$
\displaystyle\frac{{\lb{A}_2}}{2\,{\sqrt{2}}} + {\rV}\,\left( 
\displaystyle\frac{-\lVA_2}{4} + 
\displaystyle\frac{\lVA_4}{8} + 
     \displaystyle\frac{\lVA_5}{4} \right)
$ \\

4&$
\displaystyle\frac{\GV^2}{8\,\MV^4} - \displaystyle\frac{3\,
{\lb{A}_2}}{2\,{\sqrt{2}}} + 
  {\rV}\,\left( \displaystyle\frac{3\,\lVA_2}{4} - 
\displaystyle\frac{3\,\lVA_4}{8} - 
     \displaystyle\frac{3\,\lVA_5}{4} \right)
$ \\

5&$
\displaystyle\frac{{\cd}\,{\cm}}{2\,\MS^4} + {\lb{S}_1} + 
\displaystyle\frac{{\rV}\,{\lb{SV}_2}}{{\sqrt{2}}} + 
  2\,{\rS}\,({\lb{SS}_1})' + \rS^2\,({\lb{SS}_3})'
$ \\

8&$
\displaystyle\frac{{\cd}\,{\cm}}{2\,\MS^4} + {\lb{S}_2} + 
2\,{\rS}\,{\lb{SS}_2} + \rV^2\,({\lb{VV}_6})'
$ \\

10&$
- \displaystyle\frac{{\cd}\,{\cm}}{\MS^4}  + {\lb{S}_3} - 
\displaystyle\frac{{\rV}\,{\lb{SV}_2}}{{\sqrt{2}}} - 
  \rV^2\,({\lb{VV}_6})'
$ \\

12&$
\displaystyle\frac{- {\cd}\,{\cm} }{2\,\MS^4}
$ \\

14&$
\displaystyle\frac{-\dm^2}{4\,\MP^4} + ({\lb{SS}_1})' + 
2\,{\rS}\,({\lb{SS}_3})'
$ \\

17&$
\displaystyle\frac{-\dm^2}{4\,\MP^4} + {\lb{SS}_2}$ \\

19&$
\displaystyle\frac{-2\,\Css}{3} + ({\lb{SS}_3})'
$ \\

20&$
\Css
$ \\

21&$
\displaystyle\frac{-\Css}{3}
$ \\

22&$
\displaystyle\frac{\cd^2}{8\,\MS^4} + 
\displaystyle\frac{\GV^2}{16\,\MV^4} + {\lb{P}_1} + 
\displaystyle\frac{{\rV}\,{\lb{PV}_1}}{2\,{\sqrt{2}}} + 
  \displaystyle\frac{{\rS}\,{\lb{SP}_1}}{2}
$ \\

24&$
6\,\Cs
$ \\

25&$
\displaystyle\frac{\cd^2}{4\,\MS^4} - 
\displaystyle\frac{3\,\GV^2}{8\,\MV^4} + 
\displaystyle\frac{3\,{\lb{A}_2}}{{\sqrt{2}}} + {\lb{P}_2} + 
  {\rS}\,{\lb{SP}_1} + {\rV}\,\left( -
\displaystyle\frac{{\lb{PV}_1}}{{\sqrt{2}}}   - 
     \displaystyle\frac{3\,{\lVA_2}}{2} + 
\displaystyle\frac{3\,{\lVA_4}}{4} + 
     \displaystyle\frac{3\,{\lVA_5}}{2} \right)
$ \\

26&$
-4\,\Cs - \displaystyle\frac{\dm^2}{2\,\MP^4} - 
\displaystyle\frac{\cd^2}{4\,\MS^4} - 
\displaystyle\frac{{\cd}\,{\cm}}{2\,\MS^4} - 
\displaystyle\frac{\cm^2}{4\,\MS^4} + 
  \displaystyle\frac{\GV^2}{4\,\MV^4} - {\sqrt{2}}\,{\lb{A}_2} - {\lb{SP}_1} 
  \extraline
  + 
  {\rV}\,\left( \displaystyle\frac{{\lb{PV}_1}}{{\sqrt{2}}} + {\lVA_2} - 
     \displaystyle\frac{{\lVA_4}}{2} - {\lVA_5} \right)  - ({\lb{PP}_1})' - 
  2\,{\rS}\,({\lb{SP}_2})'
$ \\

27&$
-2\,\Cs - 6\,\Css + {\rS}\,\left( \displaystyle\frac{-{\lb{S}_4}}{2} 
+ {\lb{S}_5} \right)
$ \\

28&$
-\displaystyle\frac{\cd^2}{36\,\MS^4} + 
\displaystyle\frac{\GV^2}{72\,\MV^4} - 
\displaystyle\frac{{\lb{A}_2}}{6\,{\sqrt{2}}} - 
   \displaystyle\frac{{\rS}\,{\lb{S}_4}}{6} + 
{\rV}\,\left( \displaystyle\frac{{\lVA_2}}{12} - 
      \displaystyle\frac{{\lVA_4}}{24} - 
\displaystyle\frac{{\lVA_5}}{12} \right)
$ \\

29&$
-2\,\Cs - \displaystyle\frac{\dm^2}{2\,\MP^4} - 
\displaystyle\frac{{\cd}\,{\cm}}{2\,\MS^4} - 
\displaystyle\frac{\cm^2}{4\,\MS^4} - \displaystyle\frac{\GV^2}{8\,\MV^4} + 
  \displaystyle\frac{{\lb{A}_2}}{2\,{\sqrt{2}}} - 
{\lb{PP}_2} - {\lb{SP}_1}
 \extraline
   + 
  {\rV}\,\left( - \displaystyle\frac{{\lb{PV}_1}}{{\sqrt{2}}}   
- \displaystyle\frac{{\lVA_2}}{4} + 
     \displaystyle\frac{{\lVA_4}}{8} + \displaystyle\frac{{\lVA_5}}{4} \right)
$ \\

30&$
2\,\Cs
$ \\

31&$
-2\,\Css - \displaystyle\frac{\dm^2}{2\,\MP^4} - 
\displaystyle\frac{{\cd}\,{\cm}}{2\,\MS^4} - 2\,({\lb{SP}_2})'
$ \\

32&$
- \displaystyle\frac{ {\cd}\,{\cm} }{18\,\MS^4} - 
\displaystyle\frac{{\lb{S}_4}}{6}
$ \\

33&$
-4\,\Css + \displaystyle\frac{\dm^2}{6\,\MP^4} - {\lb{P}_3} - 
\displaystyle\frac{{\lb{S}_4}}{2} + {\lb{S}_5}
$ \\

34&$
\displaystyle\frac{\dm^2}{2\,\MP^4} + 
\displaystyle\frac{{\cd}\,{\cm}}{2\,\MS^4} + 
\displaystyle\frac{\cm^2}{2\,\MS^4} + {\lb{SP}_1}
$ \\

35&$
6\,\Css
$ \\

38&$
\displaystyle\frac{-\dm^2}{2\,\MP^4} + 
\displaystyle\frac{\cm^2}{2\,\MS^4}
$ \\

40&$
\displaystyle\frac{\cd^2}{4\,\MS^4} - 
\displaystyle\frac{\GV^2}{8\,\MV^4} + {\sqrt{2}}\,{\lb{A}_2} + 
  {\rS}\,\left( {\lb{S}_1} + 
\displaystyle\frac{{\rV}\,{\lb{SV}_2}}{{\sqrt{2}}} \right)  + 
  {\rV}\,\left( - \displaystyle\frac{{\lb{V}_1}}{{\sqrt{2}}}  - {\lVA_2} + 
     \displaystyle\frac{{\lVA_4}}{2} + {\lVA_5} \right) 
     \extraline
      + 
  \rS^2\,({\lb{SS}_1})' + \displaystyle\frac{\rV^2\,({\lb{VV}_4})'}{2}
$ \\

42&$
\displaystyle\frac{\cd^2}{4\,\MS^4} - 
\displaystyle\frac{\GV^2}{8\,\MV^4} + 
\displaystyle\frac{{\lb{A}_2}}{{\sqrt{2}}} + {\rS}\,{\lb{S}_2} + 
  \rS^2\,{\lb{SS}_2} + {\rV}\,\left( - {\sqrt{2}}\,{\lb{V}_4}  - 
     \displaystyle\frac{{\lVA_2}}{2} + 
\displaystyle\frac{{\lVA_4}}{4} + \displaystyle\frac{{\lVA_5}}{2}
     \right) 
     \extraline 
      + \rV^2\,\left( -{\lb{VV}_5} + ({\lb{VV}_1})' \right) + 
      \displaystyle\frac{(g_1^V)^2}{2 M_V^2}
$ \\

44&$
\displaystyle\frac{-\cd^2}{2\,\MS^4} + 
\displaystyle\frac{\GV^2}{4\,\MV^4} - 2\,{\sqrt{2}}\,{\lb{A}_2} + 
  {\rS}\,\left( {\lb{S}_3} - 
\displaystyle\frac{{\rV}\,{\lb{SV}_2}}{{\sqrt{2}}} \right)
  + \extraline
    + 
  {\rV}\,\left( \displaystyle\frac{{\lb{V}_1}}{{\sqrt{2}}} - 
{\sqrt{2}}\,{\lb{V}_3} + {\sqrt{2}}\,{\lb{V}_4} + 
     2\,{\lVA_2} - {\lVA_4} - 2\,{\lVA_5} \right)
     \extraline 
       + 
  \rV^2\,\left( 2\,{\lb{VV}_5} - ({\lb{VV}_1})' + 
     \displaystyle\frac{({\lb{VV}_3})'}{2} - ({\lb{VV}_4})' \right) - 
     \displaystyle\frac{(g_1^V)^2}{M_V^2}
$ \\

46&$
- \displaystyle\frac{{\lb{A}_2}}{{\sqrt{2}}}   + {\rV}\,
   \left( \displaystyle\frac{{\lb{V}_2}}{{\sqrt{2}}} + 
\displaystyle\frac{{\lVA_2}}{2} - \displaystyle\frac{{\lVA_4}}{4} - 
     \displaystyle\frac{{\lVA_5}}{2} \right)  + 
  \rV^2\,\left( -{\lb{VV}_2} + \displaystyle\frac{({\lb{VV}_3})'}{2} \right) -
  \displaystyle\frac{(g_1^V)^2}{2 M_V^2}
$ \\

47&$
{\sqrt{2}}\,{\lb{A}_2} + {\rV}\,\left( -
\displaystyle\frac{{\lb{V}_2}}{{\sqrt{2}}}  + {\sqrt{2}}\,{\lb{V}_3} - 
     {\lVA_2} + \displaystyle\frac{{\lVA_4}}{2} + {\lVA_5} \right) 
     \extraline 
      + 
  \rV^2\,\left( {\lb{VV}_2} - {\lb{VV}_5} - ({\lb{VV}_3})' + 
     \displaystyle\frac{({\lb{VV}_4})'}{2} \right) + 
     \displaystyle\frac{(g_1^V)^2}{M_V^2}
$ \\

48&$
\displaystyle\frac{\cd^2}{4\,\MS^4} - 
\displaystyle\frac{\GV^2}{8\,\MV^4} + 
\displaystyle\frac{{\lb{A}_2}}{{\sqrt{2}}} - 
\displaystyle\frac{{\lb{V}_4}}{{\sqrt{2}}} - 
  {\sqrt{2}}\,{\rS}\,{\rV}\,({\lb{SV}_3})' 
  \extraline
  + 
  {\rV}\,\left( \displaystyle\frac{-{\lVA_2}}{2} + 
\displaystyle\frac{{\lVA_4}}{4} + 
     \displaystyle\frac{{\lVA_5}}{2} - {\lb{VV}_5} + ({\lb{VV}_1})' \right) +
     2 \, \displaystyle\frac{g_1^V g_2^V}{M_V^2}
$ \\

50&$
\displaystyle\frac{{\FV}\,{\GV}}{4\,\MV^4} + 
{\sqrt{2}}\,{\lb{A}_2} + \displaystyle\frac{{\lb{PV}_1}}{{\sqrt{2}}} - 
  \displaystyle\frac{{\lb{V}_2}}{{\sqrt{2}}} 
  \extraline
  + {\rV}\,\left( -{\lVA_2} + \displaystyle\frac{{\lVA_4}}{2} + 
     {\lVA_5} + 2\,{\lb{VV}_2} - ({\lb{VV}_3})' \right)  - 
  \displaystyle\frac{\rV^2\,({\lb{VV}_7})'}{2} + 
  4 \, \displaystyle\frac{g_1^V g_2^V}{M_V^2}
$ \\

51&$
\displaystyle\frac{\cd^2}{2\,\MS^4} + 
\displaystyle\frac{{\FV}\,{\GV}}{4\,\MV^4} - 
\displaystyle\frac{\GV^2}{4\,\MV^4} + 2\,{\sqrt{2}}\,{\lb{A}_2} + 
  \displaystyle\frac{{\lb{PV}_1}}{{\sqrt{2}}} + 
\displaystyle\frac{{\rS}\,{\lb{SV}_2}}{{\sqrt{2}}} - 
  \displaystyle\frac{{\lb{V}_1}}{{\sqrt{2}}} 
  \extraline 
  + {\rV}\,\left( -2\,{\lVA_2} + {\lVA_4} + 
     2\,{\lVA_5} + ({\lb{VV}_4})' \right)  - 
  \displaystyle\frac{\rV^2\,({\lb{VV}_7})'}{2}
$ \\

52&$
\displaystyle\frac{- {\FV}\,{\GV} }{4\,\MV^4} - 
\displaystyle\frac{{\lb{A}_2}}{{\sqrt{2}}} - 
  \displaystyle\frac{{\lb{PV}_1}}{{\sqrt{2}}} - 
\displaystyle\frac{{\lb{V}_3}}{{\sqrt{2}}}    
  \extraline 
  + 
  {\rV}\,\left( \displaystyle\frac{{\lVA_2}}{2} - 
\displaystyle\frac{{\lVA_4}}{4} - 
     \displaystyle\frac{{\lVA_5}}{2} + {\lb{VV}_5} + 
\displaystyle\frac{({\lb{VV}_3})'}{2} - 
     \displaystyle\frac{({\lb{VV}_4})'}{2} \right)  
     + \displaystyle\frac{\rV^2\,({\lb{VV}_7})'}{2} -
     2 \, \displaystyle\frac{g_1^V g_2^V}{M_V^2}
$ \\ 

53&$
\displaystyle\frac{\FA^2}{16\,\MA^4} - 
\displaystyle\frac{3\,\FV^2}{16\,\MV^4} - 
\displaystyle\frac{{\FV}\,{\GV}}{8\,\MV^4} - 
  \displaystyle\frac{{\lb{PV}_1}}{2\,{\sqrt{2}}} + 
\displaystyle\frac{{\lVA_4}}{4} + 
  \displaystyle\frac{{\lVA_5}}{2} - {\sqrt{2}}\,{\rS}\,({\lb{SV}_3})' + 
  \displaystyle\frac{({\lb{VV}_1})'}{2} 
  \extraline 
 \;  + \, 2 \, \displaystyle\frac{(g_2^V)^2}{M_V^2}
$ \\

55&$
\displaystyle\frac{-\FA^2}{16\,\MA^4} + 
\displaystyle\frac{3\,\FV^2}{16\,\MV^4} + 
\displaystyle\frac{{\FV}\,{\GV}}{8\,\MV^4} + 
  \displaystyle\frac{{\lb{PV}_1}}{2\,{\sqrt{2}}} - 
\displaystyle\frac{{\lVA_4}}{4} - 
  \displaystyle\frac{{\lVA_5}}{2} + \displaystyle\frac{{\lb{VV}_2}}{2} + 
   2 \, \displaystyle\frac{(g_2^V)^2}{M_V^2}
$ \\

56&$
\displaystyle\frac{-\FA^2}{8\,\MA^4} + 
\displaystyle\frac{3\,\FV^2}{8\,\MV^4} - 
\displaystyle\frac{{\FV}\,{\GV}}{4\,\MV^4} - 
  \displaystyle\frac{{\lb{PV}_1}}{{\sqrt{2}}} - 
\displaystyle\frac{{\lVA_4}}{2} - {\lVA_5} + 
  \displaystyle\frac{({\lb{VV}_3})'}{2} + {\rV}\,({\lb{VV}_7})' -
   4 \, \displaystyle\frac{(g_2^V)^2}{M_V^2}
$ \\

57&$
\displaystyle\frac{\FV^2}{8\,\MV^4} + 
\displaystyle\frac{{\FV}\,{\GV}}{2\,\MV^4} + {\sqrt{2}}\,{\lb{PV}_1} + 
  \displaystyle\frac{({\lb{VV}_4})'}{2} - {\rV}\,({\lb{VV}_7})'
$ \\

59&$
\displaystyle\frac{\FA^2}{16\,\MA^4} - 
\displaystyle\frac{\FV^2}{4\,\MV^4} - 
\displaystyle\frac{{\FV}\,{\GV}}{8\,\MV^4} - 
  \displaystyle\frac{{\lb{PV}_1}}{2\,{\sqrt{2}}} + 
\displaystyle\frac{{\lVA_4}}{4} + 
  \displaystyle\frac{{\lVA_5}}{2} + \displaystyle\frac{{\lb{VV}_5}}{2} -
   2 \, \displaystyle\frac{(g_2^V)^2}{M_V^2}
$ \\

61&$
- {\sqrt{2}}\,({\lb{SV}_3})'  + \displaystyle\frac{({\lb{VV}_6})'}{2}
$ \\

63&$
\displaystyle\frac{{\cd}\,{\cm}}{2\,\MS^4} + {\rV}\,\left( - 
{\sqrt{2}}\,({\lb{SV}_3})'  + 
     ({\lb{VV}_6})' \right)
$ \\

65&$
\displaystyle\frac{{\cd}\,{\cm}}{\MS^4} + 
\displaystyle\frac{{\lb{SV}_2}}{{\sqrt{2}}}
$ \\

66&$
\displaystyle\frac{\cd^2}{4\,\MS^4} + 
\displaystyle\frac{\GV^2}{8\,\MV^4} - 
\displaystyle\frac{{\lb{A}_1}}{{\sqrt{2}}} - 
\displaystyle\frac{{\lb{A}_2}}{{\sqrt{2}}} + 
  \displaystyle\frac{{\rS}\,{\lb{SA}_1}}{2\,{\sqrt{2}}} + 
  {\rV}\,\left( \displaystyle\frac{{\lVA_3}}{4} - 
\displaystyle\frac{{\lVA_4}}{4} - 
     \displaystyle\frac{{\lVA_5}}{2} \right)
$ \\

69&$
\displaystyle\frac{\cd^2}{4\,\MS^4} - 
\displaystyle\frac{\GV^2}{8\,\MV^4} - 
\displaystyle\frac{{\lb{A}_3}}{{\sqrt{2}}} + 
  \displaystyle\frac{{\rS}\,{\lb{SA}_1}}{2\,{\sqrt{2}}} + 
  {\rV}\,\left( \displaystyle\frac{-{\lVA_3}}{4} + 
\displaystyle\frac{{\lVA_4}}{4} + 
     \displaystyle\frac{{\lVA_5}}{2} \right)
$ \\

70&$
\displaystyle\frac{\cd^2}{4\,\MS^4} + 
\displaystyle\frac{\FV^2}{8\,\MV^4} - 
\displaystyle\frac{{\FV}\,{\GV}}{8\,\MV^4} - 
\displaystyle\frac{\GV^2}{8\,\MV^4} + 
  {\sqrt{2}}\,{\lb{A}_2} - 
\displaystyle\frac{{\lb{PV}_1}}{2\,{\sqrt{2}}} - 
\displaystyle\frac{{\lVA_4}}{4} 
  \extraline
  + 
  {\rV}\,\left( -{\lVA_2} + {\lVA_4} + {\lVA_5} \right)  + 
  \displaystyle\frac{({\lb{AA}_1})'}{2} +
   2 \, \displaystyle\frac{(g_1^A)^2}{M_A^2}
$ \\

72&$
\displaystyle\frac{-\FV^2}{8\,\MV^4} + 
\displaystyle\frac{{\FV}\,{\GV}}{8\,\MV^4} - 
\displaystyle\frac{{\lb{A}_2}}{2\,{\sqrt{2}}} + 
  \displaystyle\frac{{\lb{AA}_2}}{2} + 
\displaystyle\frac{{\lb{PV}_1}}{2\,{\sqrt{2}}} + 
  \displaystyle\frac{{\lVA_4}}{4} + {\rV}\,\left( 
\displaystyle\frac{{\lVA_2}}{4} - 
     \displaystyle\frac{5\,{\lVA_4}}{8} - 
\displaystyle\frac{{\lVA_5}}{4} \right)
     \extraline 
     + \, 2 \, \displaystyle\frac{(g_1^A)^2}{M_A^2}
$ \\

73&$
\displaystyle\frac{-\FA^2}{8\,\MA^4} - 
\displaystyle\frac{\FV^2}{8\,\MV^4} + 
\displaystyle\frac{{\FV}\,{\GV}}{4\,\MV^4} - {\sqrt{2}}\,{\lb{A}_2} + 
  \displaystyle\frac{{\lb{PV}_1}}{{\sqrt{2}}} + {\rV}\,
   \left( {\lVA_2} - {\lVA_4} \right)  + \displaystyle\frac{{\lVA_4}}{2} + 
  \displaystyle\frac{({\lb{AA}_3})'}{2}
  \extraline
  - \, 4 \, \displaystyle\frac{(g_1^A)^2}{M_A^2}
$ \\

74&$
\displaystyle\frac{\FA^2}{8\,\MA^4} + 
\displaystyle\frac{\cd^2}{2\,\MS^4} - 
\displaystyle\frac{\GV^2}{4\,\MV^4} + {\sqrt{2}}\,{\lb{A}_2} + 
  \displaystyle\frac{{\rS}\,{\lb{SA}_1}}{{\sqrt{2}}} 
  \extraline
  + 
  {\rV}\,\left( -{\lVA_2} - \displaystyle\frac{{\lVA_3}}{2} + 
     \displaystyle\frac{{\lVA_4}}{2} + {\lVA_5} \right)  + 
\displaystyle\frac{({\lb{AA}_4})'}{2}
$ \\ 

76&$
\displaystyle\frac{\cd^2}{4\,\MS^4} + 
\displaystyle\frac{\FV^2}{16\,\MV^4} - 
\displaystyle\frac{{\FV}\,{\GV}}{8\,\MV^4} + 
  \displaystyle\frac{3\,{\lb{A}_2}}{2\,{\sqrt{2}}} + 
\displaystyle\frac{{\lb{AA}_5}}{2} - 
  \displaystyle\frac{{\lb{PV}_1}}{2\,{\sqrt{2}}} + 
\displaystyle\frac{{\rS}\,{\lb{SA}_1}}{2\,{\sqrt{2}}} - 
  \displaystyle\frac{{\lVA_4}}{4} 
  \extraline
  + {\rV}\,\left( \displaystyle\frac{-3\,{\lVA_2}}{4} + 
     \displaystyle\frac{{\lVA_3}}{4} + 
\displaystyle\frac{5\,{\lVA_4}}{8} + \displaystyle\frac{{\lVA_5}}{4}
     \right) - \, 2 \, \displaystyle\frac{(g_1^A)^2}{M_A^2}
$ \\

78&$
\displaystyle\frac{\FV^2}{4\,\MV^4} + 
\displaystyle\frac{{\FV}\,{\GV}}{8\,\MV^4} + 
\displaystyle\frac{{\lb{PV}_1}}{2\,{\sqrt{2}}} + 
  \displaystyle\frac{{\lVA_2}}{2} - 
\displaystyle\frac{{\lVA_4}}{4} - \displaystyle\frac{{\lVA_5}}{2}
$ \\

79&$
\displaystyle\frac{-\FA^2}{8\,\MA^4} + 
\displaystyle\frac{\FV^2}{8\,\MV^4} - 
\displaystyle\frac{{\FV}\,{\GV}}{8\,\MV^4} - 
  \displaystyle\frac{{\lb{PV}_1}}{2\,{\sqrt{2}}} + 
\displaystyle\frac{{\lVA_4}}{4} + 
  \displaystyle\frac{{\lVA_5}}{2} + \displaystyle\frac{({\lb{AA}_7})'}{4} - 
  \displaystyle\frac{3\,({\lb{VV}_7})'}{4}
$ \\

80&$
\displaystyle\frac{{\cd}\,{\cm}}{2\,\MS^4} + 
\displaystyle\frac{({\lb{AA}_6})'}{2}
$ \\

82&$
\displaystyle\frac{-\FV^2}{16\,\MV^4} - 
\displaystyle\frac{{\FV}\,{\GV}}{16\,\MV^4} + 
\displaystyle\frac{{\lb{PV}_1}}{4\,{\sqrt{2}}} + 
  \displaystyle\frac{{\lVA_4}}{8} + \displaystyle\frac{{\lVA_5}}{4} - 
  \displaystyle\frac{({\lb{PA}_1})'}{{\sqrt{2}}} + 
\displaystyle\frac{({\lb{PV}_2})'}{{\sqrt{2}}}
$ \\

83&$
\displaystyle\frac{-\cd^2}{8\,\MS^4} - 
\displaystyle\frac{{\cd}\,{\cm}}{2\,\MS^4} + 
\displaystyle\frac{3\,\GV^2}{16\,\MV^4} - 
\displaystyle\frac{{\lb{A}_2}}{{\sqrt{2}}} - 
  \displaystyle\frac{{\lb{SA}_1}}{2\,{\sqrt{2}}} - 
\displaystyle\frac{{\rS}\,{\lb{SP}_1}}{2} 
  \extraline
  + 
  {\rV}\,\left( \displaystyle\frac{3\,{\lb{PV}_1}}{2\,{\sqrt{2}}} + 
\displaystyle\frac{{\lVA_2}}{2} + 
     \displaystyle\frac{{\lVA_3}}{4} - 
\displaystyle\frac{{\lVA_4}}{4} - \displaystyle\frac{{\lVA_5}}{2} + 
     {\sqrt{2}}\,({\lb{PV}_2})' \right)
$ \\

85&$
- \displaystyle\frac{{\cd}\,{\cm}}{\MS^4}   - 
\displaystyle\frac{{\lb{SA}_1}}{{\sqrt{2}}}
$ \\

87&$
\displaystyle\frac{-\FA^2}{8\,\MA^4} + \displaystyle\frac{\FV^2}{8\,\MV^4}
$ \\

88&$
\displaystyle\frac{- {\FV}\,{\GV}  }{4\,\MV^4} - 
\displaystyle\frac{{\lb{PV}_1}}{{\sqrt{2}}}
$ \\

89&$
\displaystyle\frac{\FV^2}{2\,\MV^4} + 
\displaystyle\frac{{\FV}\,{\GV}}{4\,\MV^4} + 
\displaystyle\frac{{\lVA_3}}{2} - 
  \displaystyle\frac{{\lVA_4}}{2} - {\lVA_5}
$ \\

90&$
- \displaystyle\frac{{\lb{PV}_1}}{{\sqrt{2}}}
$ \\

91&$
\displaystyle\frac{2\,\dm^2}{\MP^4}
$ \\

92&$
\displaystyle\frac{\FV^2}{\MV^4} - 2\,({\lb{VV}_7})'
$ \\

93&$
\displaystyle\frac{-\FV^2}{4\,\MV^4}
$ \\

94&$
8\,\Css
$

\end{longtable}
}

The following abbreviations have been used:
\begin{align} \label{eq:RelevComb}
({\lb{SS}_1})' &= {\lb{SS}_1} + {\rS}\,\lb{SSS} \, ,
\no
({\lb{SS}_3})' &= {\lb{SS}_3} + \lb{SSS} \, ,
\no
({\lb{SP}_2})' &= \frac{{\lb{PP}_3}}{2} + {\lb{SP}_2} + 
   \frac{\lb{SPP}}{2} \, , 
   \no
   ({\lb{SV}_3})' &= \frac{-{\lb{SV}_1}}{2\,{\rV}} + {\lb{SV}_3} \, , 
   \no
({\lb{PP}_1})' &= {\lb{PP}_1} - {\rS}\,{\lb{PP}_3} \, , 
\no
   ({\lb{PV}_2})' &= \frac{{\lb{PA}_2}}{2\,{\rV}} + {\lb{PV}_2} + 
   \frac{{\lVA_1}}{{\sqrt{2}}} - \frac{\lb{VAP}}{{\sqrt{2}}} \, , 
   \no
   ({\lb{PA}_1})' &= {\lb{PA}_1} + \frac{{\lb{PA}_2}}{2\,{\rV}} \, , 
   \no
   ({\lb{VV}_1})' &= {\lb{VV}_1} + 
   {\rS}\,\left( - \frac{{\sqrt{2}}\,{\lb{SV}_1}}{\rV}  + 
   \lb{SVV} \right) \, , 
   \no
({\lb{VV}_3})' &= {\lb{VV}_3} + \frac{{\rV}\,\lb{VVV}}{{\sqrt{2}}} \, , 
\no
({\lb{VV}_4})' &= {\lb{VV}_4} - \frac{{\rV}\,\lb{VVV}}{{\sqrt{2}}} \,  ,
\no
   ({\lb{VV}_6})' &= -\frac{{\sqrt{2}}\,{\lb{SV}_1}}{\rV}  + 
   {\lb{VV}_6} + \lb{SVV} \, , 
   \no
({\lb{VV}_7})' &= {\lb{VV}_7} + \frac{\lb{VVV}}{{\sqrt{2}}} \, , 
\no
({\lb{AA}_1})' &= {\lb{AA}_1} + 
   {\rS}\,\left( -2\,{\sqrt{2}}\,{\lb{SA}_2} + \lb{SAA} \right) \, , 
   \no   
   ({\lb{AA}_3})' &= {\lb{AA}_3} + 
   {\rV}\,\left( 2\,{\lVA_6} + \frac{\lb{VAA}}{{\sqrt{2}}} \right) \, , 
   \no
   ({\lb{AA}_4})' &= {\lb{AA}_4} - 
   {\rV}\,\left( 2\,{\lVA_6} + \frac{\lb{VAA}}{{\sqrt{2}}} \right) \, , 
   \no
({\lb{AA}_6})' &= {\lb{AA}_6} - 2\,{\sqrt{2}}\,{\lb{SA}_2} + 
   \lb{SAA} \,  ,
\no
   ({\lb{AA}_7})' &= {\lb{AA}_7} + 2\,{\lVA_6} + 
   \frac{\lb{VAA}}{{\sqrt{2}}} \, .
%
\end{align}
{Eq. (\ref{eq:RelevComb}) shows explicitly that several couplings
appear always together in the given combinations. This redundancy
can be understood through non-linear redefinitions of the resonance
fields of the type given in Eq. (\ref{eq:redef3}). For instance, the
field redefinition
\begin{equation}
S\;\longrightarrow\; S \, +\,\frac{\lambda^{SSS}}{M_S^2}\; SS
\end{equation}
generates a contribution from the scalar mass term, which eliminates
the operator $\cO^{SSS}$. At the same time, it generates
contributions to $\cO_1^{SS}$ and $\cO_3^{SS}$ originating in the
$c_d$ and $c_m$ terms of $\cL^S_{(2)}$ in (\ref{eq:R_int}). As a
result, the couplings $\lambda^{SSS}$, $\lambda_1^{SS}$ and
$\lambda_3^{SS}$ can only appear through the combinations
$\left(\bar\lambda_1^{SS}\right)'$ and
$\left(\bar\lambda_3^{SS}\right)'$.

Similarly, the field redefinitions
\begin{equation}
S\;\longrightarrow\; S \, -\,\frac{\lambda^{PP}_3}{c_m}\; PP
\qquad , \qquad
P\;\longrightarrow\; P \, +\,\left(\frac{\lambda^{SPP}}{2 M_P^2}
+ \frac{\lambda^{PP}_3}{2 c_m}\,\frac{M_S^2}{M_P^2}
\right)\;\left\{P,S\right\} \, , 
\end{equation}
eliminate the operator $\cO^{SPP}$ through the contributions generated
from the scalar and pseudoscalar mass terms. The contributions originating
in the $c_d$, $c_m$ and $d_m$ terms in (\ref{eq:R_int}) eliminate also
the operator $\cO^{PP}_3$ and modify the couplings $\lambda_1^{PP}$ and
$\lambda_2^{SP}$. The net result is that $\lambda_1^{PP}$, $\lambda_3^{PP}$,
$\lambda_2^{SP}$ and $\lambda^{SPP}$ can only contribute to the
$\cO(p^6)$ couplings $C_{i}^{\cal R}$ through the combinations
$\left(\bar\lambda_1^{PP}\right)'$ and $\left(\bar\lambda_2^{SP}\right)'$.

All trilinear couplings $\lambda^{R_iR_jR_k}$ and 6 bilinear ones 
$\lambda^{R_iR_j}_k$
can be eliminated with this type of field redefinitions. One gets then the
combinations of couplings in (\ref{eq:RelevComb}).}

\end{document}